\documentclass[aps,prd,nofootinbib,twocolumn,superscriptaddress,preprintnumbers,balancelastpage,longbibliography]{revtex4-1}
\usepackage{aas_macros}
\usepackage{placeins}
\usepackage{amsmath,amssymb,mathtools,bm}
\usepackage{lipsum}
\usepackage{comment}
\usepackage{ulem}

\usepackage{graphicx, color, hepunits}
\usepackage[dvipsnames]{xcolor}
\usepackage{float}
\usepackage{filecontents}
\usepackage{multirow}
\usepackage{slashed}

\usepackage{tabularx}

\usepackage[colorlinks=true 
,urlcolor=purple 
,anchorcolor=blue
,citecolor=blue 
,filecolor=magenta  
,linkcolor=blue 
,menucolor=blue
,linktocpage=true
,pdfproducer=medialab
,pdfa=true
]{hyperref}
\usepackage[utf8]{inputenc}
\usepackage[english]{babel}
\let\vec\mathbf

\newcommand{\D}{{\rm d}}

\newcommand{\es}[2] {\begin{equation} \label{#1} \begin{split} #2 \end{split} \end{equation}}

\begin{document}

\title{Signatures of Primordial Energy Injection from 
Axion Strings}

\author{Joshua N. Benabou}
\affiliation{Berkeley Center for Theoretical Physics, University of California, Berkeley, CA 94720, USA}
\affiliation{Theoretical Physics Group, Lawrence Berkeley National Laboratory, Berkeley, CA 94720, USA}

\author{Malte Buschmann}
\affiliation{Department of Physics, Princeton University, Princeton, NJ 08544, USA}

\author{Soubhik Kumar}
\affiliation{Berkeley Center for Theoretical Physics, University of California, Berkeley, CA 94720, USA}
\affiliation{Theoretical Physics Group, Lawrence Berkeley National Laboratory, Berkeley, CA 94720, USA}

\author{Yujin Park}
\affiliation{Berkeley Center for Theoretical Physics, University of California, Berkeley, CA 94720, USA}
\affiliation{Theoretical Physics Group, Lawrence Berkeley National Laboratory, Berkeley, CA 94720, USA}

\author{Benjamin R. Safdi}
\affiliation{Berkeley Center for Theoretical Physics, University of California, Berkeley, CA 94720, USA}
\affiliation{Theoretical Physics Group, Lawrence Berkeley National Laboratory, Berkeley, CA 94720, USA}

\date{\today}

\begin{abstract}
Axion strings are horizon-size topological defects that may be produced in the early Universe. 
Ultra-light axion-like particles may form strings that persist to temperatures below that of big bang nucleosynthesis.
Such strings have been considered previously as sources of gravitational waves and cosmic microwave background (CMB) polarization rotation. In this work we show, through analytic arguments and dedicated adaptive mesh refinement cosmological simulations, that axion strings deposit a sub-dominant fraction of their energy into high-energy Standard Model (SM) final states, for example, by the direct production of heavy radial modes that subsequently decay to SM particles. 
  This high-energy SM radiation is absorbed by the primordial plasma, leading to novel signatures 
in precision big bang nucleosynthesis, the CMB power spectrum, and gamma-ray surveys.  
In particular, we show that CMB power spectrum data 
constrains axion strings with decay constants $f_a \lesssim 10^{12}$ GeV, up to model dependence on the ultraviolet completion, for 
axion masses $m_a \lesssim 10^{-29}$~eV; future CMB surveys could find striking evidence of axion strings with lower decay constants.   
\end{abstract}
\maketitle

\tableofcontents

\section{Introduction}

\begin{figure*}[!htb]
    \centering
    \includegraphics[width=1\textwidth]{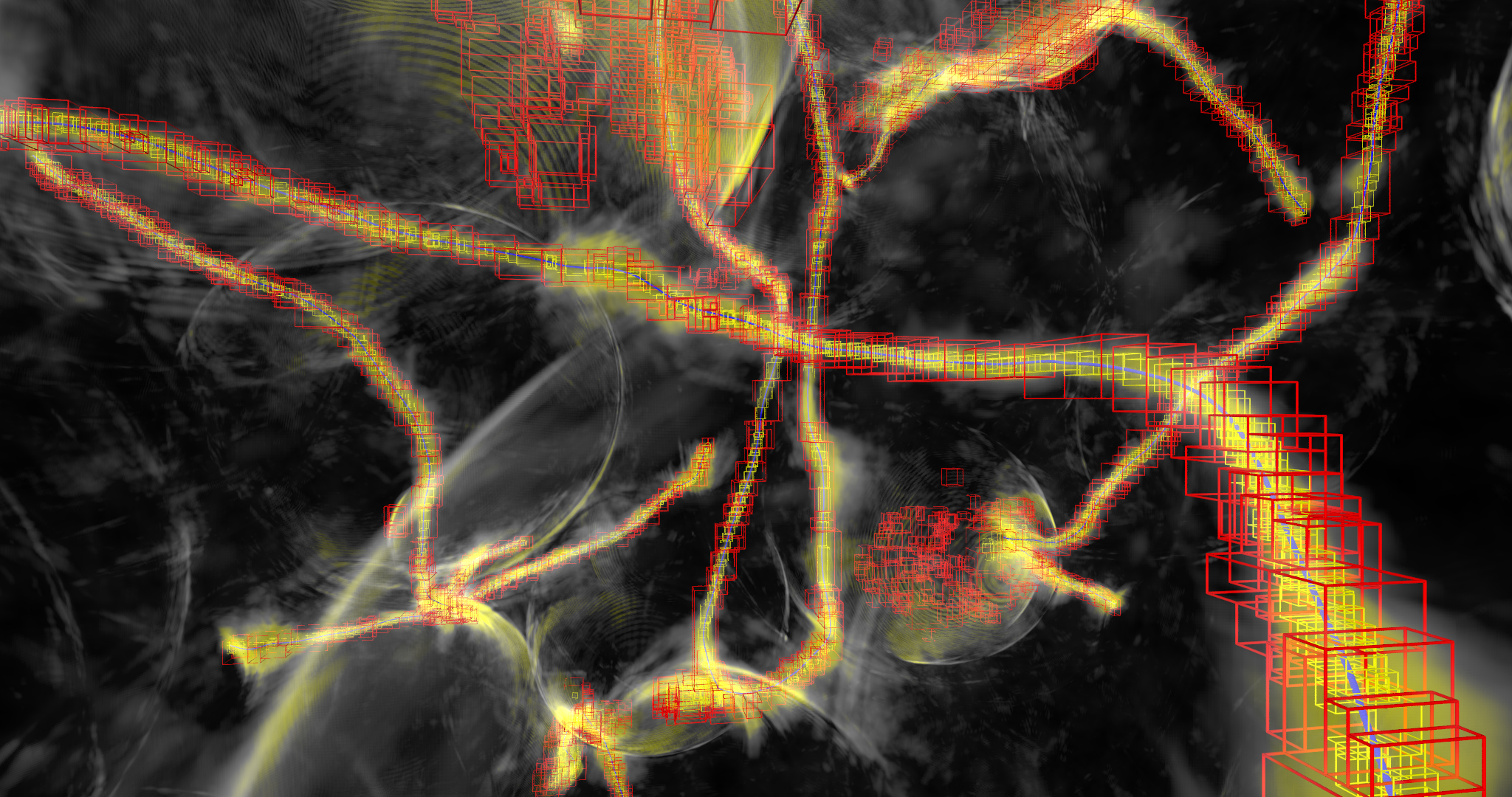}
    \caption{A zoom-in of the axion-string network as realized in a PQ-Higgs field simulation, which is discussed in Sec.~\ref{sec:axion-Higgs strings}.  We show the energy density in axion radiation in a 3D volume rendering enclosing approximately 1.9 Hubble volumes at $\log(m_s / H) \approx 7.2$; the string network evolves to maintain the scaling solution by emitting energy into relativistic axion modes. As we discuss in this work, however, axion strings can also efficiently produce high-energy SM radiation through the production and subsequent decay of heavy radial modes and through the direct production of SM Higgs bosons. Inset in this figure is the outline of the AMR grid structure, showing the different refinement level locations at this snapshot. The finest refinement level (yellow boxes) is nested within the intermediate level (red boxes) and is mostly localized around the string cores. The intermediate level covers a larger area and ensures proper numerical convergence of 
    the outgoing radiation wherever necessary. The blue lines correspond to the string cores where the width of the line is 
    matched to the string core width.
  Note that due to the periodic nature of the simulation box, some strings seem to end suddenly but are in reality continuing on the other side of the volume.  Animations available \href{https://goo.by/qHk9d}{here}.
    }
    \label{fig:ani}
\end{figure*}

Axion strings are extended topological defects stretching over cosmological distances that may develop for high post-inflationary reheat temperatures. For example, if the axion arises as the pseudo-Goldstone boson of a spontaneously broken global $U(1)$ Peccei-Quinn (PQ) symmetry, then axion strings develop so long as the reheat temperature is above the temperature of $U(1)$ symmetry restoration. The axion strings are characterized by the property that the axion field, which is a periodic field, undergoes a full field excursion when traversing a circle encompassing the string core; to resolve the singularity at the string core, the heavy radial mode of the PQ complex scalar deviates from its vacuum expectation value (VEV) and sends the full complex scalar field to zero at the core center.  Apart from at the string cores, the radial mode is otherwise frozen at its VEV for temperatures well below that of PQ symmetry breaking. The axion-string network evolves to maintain an approximate scaling solution, where there is roughly one string per Hubble patch at any time (see~\cite{Safdi:2022xkm} for a review).  Fig.~\ref{fig:ani} illustrates a snapshot of an axion-string network in the context of a cosmological adaptive mesh refinement (AMR) simulation performed in this work. 

Axion strings were originally discussed in the context of the quantum chromodynamics (QCD) axion, which was  introduced to solve the strong-{\it CP} problem~\cite{Peccei:1977hh,Peccei:1977ur,Weinberg:1977ma,Wilczek:1977pj}. It was later realized that the QCD axion may also make up the observed dark matter (DM) abundance~\cite{Preskill:1982cy,Abbott:1982af,Dine:1982ah}.  If the PQ symmetry is broken after inflation, then QCD axion DM is predominantly produced through horizon-scale axion radiation produced by the axion-string network just prior to its collapse at the QCD phase transition.  Numerical simulations of the axion-string network predict that in order for the QCD axion to produce the correct relic abundance from axion strings, it should have a mass on the order of 10's to 100's of $\mu$eV~\cite{Vilenkin:1982ks,Sikivie:1982qv,Davis:1986xc,Harari:1987ht,Shellard:1987bv,Davis:1989nj,Hagmann:1990mj,Battye:1993jv,Battye:1994au,Yamaguchi:1998gx,Klaer:2017ond,Gorghetto:2018myk,Vaquero:2018tib,Buschmann:2019icd,Gorghetto:2020qws,Dine:2020pds,Buschmann:2021sdq}.

Axion strings may also develop in axion-like particle models; for ultralight axion masses $m_a$ below that of the QCD axion, the resulting string networks can persist near or below the epoch of big bang nucleosynthesis (BBN), leading to a number of observable signatures.  These include, for example, gravitational wave production, contribution to the effective number of neutrino degrees of freedom $N_{\rm eff}$, and the polarization rotation of cosmic microwave background (CMB) photons.  The string network persists until $H \sim m_a$, with $H$ the Hubble parameter.  If the axion domain wall number $N_{\rm dw}$ is larger than unity then stable domain walls develop for $H \lesssim m_a$, which could themselves lead to novel cosmological signatures.  If $N_{\rm dw} = 1$ then the string-domain-wall network collapses around $H \sim m_a$. In this work, we restrict to axion masses low enough at a given cosmological epoch that we do not have to consider domain walls or specifiy $N_{\rm dw}$, for simplicity. 

It has been shown that axion strings with axion masses $m_a$ roughly less than $10^{-18}$ eV, with decay constants  $f_a \gtrsim 10^{14}$ GeV, could produce detectable gravitational wave signatures at next-generation observatories~\cite{Gorghetto:2021fsn} (see also~\cite{Hindmarsh:1994re,Saikawa:2017hiv,Chang:2021afa,Gelmini:2021yzu}).    The gravitational waves are sourced by the energy density in the evolving string network.  If the network persists until after the CMB decouples ($m_a \lesssim 10^{-29}$ eV), then axion strings may rotate the polarization of CMB photons by an amount that is proportional to the electromagnetic anomaly coefficient; polarization observations of the CMB already constrain this scenario~\cite{Agrawal:2019lkr,Jain:2021shf,Yin:2021kmx,Jain:2022jrp,Yin:2023vit, Hagimoto:2023tqm}.  Interestingly, the polarization signatures are independent of $f_a$ and only depend on the electromagnetic anomaly coefficient of the axion.

In this work, we point out that axion-like particle strings, which we refer to as axion strings for simplicity, also leave novel energy injection signatures in the primordial plasma. It is well established that the axion-string network evolves by emitting axions to maintain the scaling solution. We show, however, that the string network releases a subdominant fraction of its energy into heavy radial modes (see also~\cite{Drew:2019mzc,Saurabh:2020pqe}) that promptly decay to Standard Model (SM) final states and, under certain circumstances that we enumerate, high-energy SM Higgs particles.  This high-energy SM radiation is absorbed by the primordial plasma and can undo the success of BBN, modify the CMB power spectrum, or even lead to observable gamma-ray signatures today. Understanding how to translate cosmological measurements of these observables to constraints on or evidence for axion strings requires a detailed understanding of how axion strings emit heavy radial modes.

While this work focuses on global strings, local string networks are closely related and result from gauging the $U(1)_{\rm PQ}$ symmetry that gives rise to the axion as a Goldstone mode.  The axion is then ``eaten" by the abelian gauge field, which acquires a mass of order the scale of symmetry breaking.  Thus, unlike for global strings local strings do not emit massless radiation, except gravitational wave emission and perhaps other massless radiation that they may couple to indirectly.  High-energy SM radiation from cosmic string networks has been studied before in the context of local strings (see, {\it e.g.},~\cite{Vincent:1997cx,Hindmarsh:2008dw,Vachaspati:2009jx,Vachaspati:2009kq,Hyde:2013fia,Long:2014mxa,Long:2014lxa,SantanaMota:2014xpw}). It is debated whether or not local strings can directly emit heavy modes with masses of order the symmetry breaking scale (see, {\it e.g.},~\cite{Vincent:1997cx,Hindmarsh:2008dw,Hindmarsh:2017qff,Blanco-Pillado:2023sap}), but if they can then they would be constrained through analogous probes to those studied in this work. Higgs condensates surrounding local strings have also been studied previously~\cite{Vachaspati:2009kq,SantanaMota:2014xpw, Hyde:2013fia}, in analogy with the Higgs configurations around global strings studied in this work. 

The remainder of this article is organized as follows. In Sec.~\ref{sec:existing} we review constraints on and future probes of axion-like particle strings from the axion contribution to $N_{\rm eff}$, gravitational waves, CMB distortions, and CMB polarization rotation. In Sec.~\ref{sec:prompt_radial_mode_emission}, through analytic arguments and AMR simulations, we compute the amount of heavy radial mode emission from an axion string network.
The emitted radial modes can, in turn, decay into SM degrees of freedom and inject energy into the primordial plasma.
To compute this effect, in Sec.~\ref{sec:radial_decay}, we derive the branching ratios of the radial mode into SM particles.
Among such SM final states, the radial mode can also decay into a pair of SM Higgs via a PQ scalar-Higgs quartic coupling.
In particular, such a coupling would give rise to non-winding classical Higgs configurations surrounding axion strings.
In Sec.~\ref{sec:axion-Higgs strings} we compute the properties of such Higgs `sheaths' analytically and using AMR simulations.
Using these results, in Sec.~\ref{sec:obs_cons} we obtain constraints on axion strings from the energy injection signatures they would leave in the primordial plasma at the epochs of BBN and CMB decoupling, in addition to constraints arising from present-day gamma-ray surveys.  We conclude in Sec.~\ref{sec:discuss} with a discussion of how some of our results may apply to local strings.

\section{Existing probes of axion strings}
\label{sec:existing}

We describe some of the current and future probes of axion strings that have been previously discussed in order of decreasing axion mass $m_a$, as the constraints typically become stronger as one allows the string network to persist to later times.  
Note that throughout this work we consider only the so-called ``field theory axions," which are those that emerge as the pseudo-Goldstone bosons of spontaneously broken $U(1)$ PQ symmetries. Axion-like particles are also motivated by the framework of the string axiverse~\cite{Svrcek:2006yi, Arvanitaki:2009fg}, where the axions arise not from global symmetry breaking but rather as the zero modes of higher-dimensional gauge fields integrated over the compact manifolds in string theory compactifications~\cite{Witten:1984dg,Arkani-Hamed:2003xts,Svrcek:2006yi,Cicoli:2012sz,Demirtas:2018akl,Halverson:2019cmy,Demirtas:2021gsq}.  Axion string production for string theory axions is more subtle than for field theory axions and is not considered here; we discuss the formation and signatures of string theory axion strings in~\cite{inpress} (see also~\cite{March-Russell:2021zfq}). 

To produce axion strings in field theory realizations, the Universe must reheat after inflation to a temperature $T_{\rm RH} > f_a$, large enough to restore the PQ symmetry.  This is because finite-temperature corrections to the effective potential for the PQ field restore the PQ symmetry for temperatures $T \gtrsim \sqrt{3} f_a$, where the numerical pre-factor is somewhat model dependent  
(see, {\it e.g.},~\cite{Wantz:2009it, Hiramatsu:2012gg}). The maximum reheat temperature, given a Hubble parameter during inflation $H_{I}$, is determined by hypothesizing that the inflaton decays promptly at the end of inflation to the SM, such that the energy density after reheating $\rho_{{\rm RH}} \sim T_{\rm RH}^4$ is equal to the energy density directly before reheating, which is roughly $3 M_{\rm pl}^2 H_{I}^2$, with $M_{\rm pl}$ the reduced Planck mass. If the inflaton does not decay promptly then the reheat temperature may be lower. The scale of Hubble during inflation is constrained by the tensor-to-scalar ratio as measured by CMB anisotropy measurements. A combination of Planck data and data from the BICEP2/Keck Array constrain $H_I \lesssim 5 \times 10^{13}$ GeV at 95\% confidence level~\cite{Planck:2018jri}, which implies $T_{\rm RH} \lesssim 10^{16}$ GeV. Thus, we conclude that $f_a \lesssim 6 \times 10^{15}$ GeV, regardless of the axion mass $m_a$, in order to produce axion strings.  

The string network is relatively unconstrained at present until BBN.  However, this may change in the future with the next-generation gravitational wave observatories.  Axion strings evolve primarily through axion emission. As we discuss in this work, a sub-dominant fraction of the energy density is also emitted in the form of heavy radial modes, though the energy density in this fraction is suppressed relative to that in axions by an amount $\sim$$\log(f_a / H) \sim 10^2$, where $H$ is evaluated at late times, such as during BBN or CMB decoupling.  On the other hand, the axion string network also sources gravitational waves but with a rate heavily suppressed relative to axion emission by an amount $\propto (f_a / M_{\rm pl})^2$.  Gravitational wave emission thus has a negligible effect on the dynamics of axion strings for the decay constants of interest ($f_a \lesssim 10^{15}$ GeV).  However, the spectrum of gravitational waves emitted by the string network may be detectable through next-generation  low-frequency gravitational wave observatories such as LISA and the square kilometer array (SKA) with pulsar timing; in particular, Ref.~\cite{Gorghetto:2021fsn} concluded that string networks with $f_a \gtrsim {\rm few} \times 10^{14}$ GeV for $m_a \lesssim 10^{-18}$ eV may be detectable.  Note that the recently-detected gravitational wave signal at pulsar timing array experiments, including EPTA and NANOGrav, is not compatible with axion string emission because of the $N_{\rm eff}$ bound discussed below~\cite{NANOGrav:2023gor,NANOGrav:2023hvm,Antoniadis:2023rey,Servant:2023mwt,Madge:2023cak}.  

At BBN, existing constraints on axion strings arise from the contribution of relativistic axions to $N_{\rm eff}$ (see, {\it e.g.},~\cite{Gorghetto:2021fsn}).  The string network evolves by emitting relativistic axions; the energy density in axions is~\cite{Gorghetto:2018myk} $\rho_a \approx {4 \over 3} H^2 c_1 \pi f_a^2 \log_*^3$, where $\log_* \equiv \log(m_s / H)$, with $H$ the Hubble parameter at the epoch of interest and where $c_1$ is an ${\cal O}(1)$ coefficient that we discuss later in this article.  We may divide this energy density by the energy density in one species of neutrino to compute the contribution to $N_{\rm eff}$ during radiation domination, in particular, at 1~MeV: 
\es{eq:Delta_Neff}{
\Delta N_{\rm eff}\bigg\rvert_{\rm BBN} \approx &1.1 \left( {g_{*} \over 10.75 }\right)\left( {c_1 \over 0.25 }\right)\left( {\log_* \over 90} \right)^3 \\
&\times \left({f_a\over 10^{15}~{\rm GeV}} \right)^2 \,,
}
with $g_{*}$ the effective number of degrees of freedom.
At BBN, $\Delta N_{\rm eff}$ is constrained at 95\% to be less than $0.46$, which implies $f_a \lesssim 7 \times 10^{14}$ GeV for $m_a \lesssim 10^{-18}$ eV. Note that it is more appropriate to account for the change in $g_{*}$ around the epoch of BBN, and a more detailed calculation incorporating this effect can be found in~\cite{Gorghetto:2021fsn}, which finds an approximately similar upper bound $f_a \lesssim 9\times 10^{14}$~GeV.  
Note that the BBN constraints leave a narrow range of decay constants that may be detectable with gravitational wave observations. In this work, we further constrain and narrow the parameter space that may be detectable in gravitational waves through radial mode emission.

If the string network persists to the epoch of CMB decoupling then additional probes arise from the gravitational effects of the string network imprinted on CMB anisotropies~\cite{Zeldovich:1980gh,Vilenkin:1981iu,Kaiser:1984iv}.   
No analyses of these anisotropies have been performed to-date that account for the scaling violation of the string network. 
However, extrapolating from existing results from global strings~\cite{Lopez-Eiguren:2017dmc}, to account for the larger values of $\xi$ expected at the CMB decoupling epoch, suggest that $f_a \lesssim 2 \times 10^{14}$ GeV are likely in tension with CMB anisotropy data~\cite{Gorghetto:2021fsn}. 
Axions emitted from strings contribute to $\Delta N_{\rm eff}$ during the CMB epoch as well, 
\es{eq:Delta_Neff_cmb}{
\Delta N_{\rm eff}\bigg\rvert_{\rm CMB} \approx 0.8 \left( {c_1 \over 0.25 }\right)\left( {\log_* \over 120} \right)^3\left({f_a\over 5\times 10^{14}~{\rm GeV}} \right)^2 \,,
}
and the bound $\Delta N_{\rm eff} < 0.34$~\cite{Planck:2018vyg} leads to $f_a \lesssim 3\times 10^{14}$~GeV, with $\rm{log}_*$ evaluated at recombination.  Note that the CMB limit is stronger than that from BBN in part because of the larger $\log_*$ value at the CMB epoch and in part because for the CMB  $N_{\rm eff}$ is measured in matter domination, at $z \sim 1100$, while matter-radiation equality is at $z \sim 3400$.

Axion strings persisting until after recombination ($m_a \ll 10^{-29}$ eV) may also rotate the polarization of CMB photons~\cite{Agrawal:2019lkr,Jain:2021shf,Yin:2021kmx, Jain:2022jrp, Hagimoto:2023tqm}.  This effect involves the axion-electromagnetic interaction, which may be parameterized by
\es{}{
{\mathcal L} \supset {{\mathcal A} \alpha_{\rm EM} \over 4 \pi f_a} a F_{\mu\nu} \tilde F^{\mu\nu} \,,
}
where $F$ is the electromagnetic field strength (with $\tilde F$ its dual), $\alpha_{\rm EM}$ the fine-structure constant, and ${\mathcal A}$ the mixed PQ-electromagnetic anomaly coefficient.  If a photon propagates along a trajectory over which the background axion field changes by an amount $\Delta a$, then the polarization angle of the photon will change by $\Delta \Phi = {\mathcal A} \alpha_{\rm EM} / (2 \pi f_a) \Delta a$.  Given that complete loops enclosing strings cores are characterized by $\Delta a = 2 \pi f_a$ field excursions, we expect that photons propagating through a background of cosmic axion strings will undergo polarization angle rotations on the order of $\Delta \Phi \sim {\mathcal A} \alpha_{\rm EM}$.  Ref.~\cite{Yin:2021kmx} performed a dedicated search for this effect in CMB polarization data as measured by Planck and set the upper limit ${\mathcal A} \xi_0 \lesssim 0.93$, 
with $\xi_0 \sim 10$ the expected number of strings per Hubble patch at recombination, though in detail this upper limit likely depends on the morphology of the string network assumed in~\cite{Yin:2021kmx}.  Interestingly, the polarization probes are independent of $f_a$ and close to probing theoretically motivated parameter space for which ${\mathcal A} \gtrsim 0.1$~\cite{Agrawal:2022lsp}.  In contrast, the signatures that we develop in this work are independent of ${\mathcal A}$ but directly probe $f_a$.

\section{Prompt radial mode emission from axion strings}
\label{sec:prompt_radial_mode_emission}

We now turn to our calculation of the SM radiation produced by evolving axion strings. We start by considering SM particles generated by the decay of heavy radial modes that are generated by the evolving strings.  First, we review the standard picture for axion string evolution in the context of a field theory UV completion with a PQ complex scalar field that undergoes spontaneous symmetry breaking. 

\subsection{PQ axion strings 
}

Let us first recall the dynamics of axion strings in the scaling regime, at temperatures $T \ll f_a$ but $H(t) \gg m_a$ (see~\cite{Gorghetto:2018myk,Safdi:2022xkm} for modern reviews).  Note that the dynamics we describe here are valid both for axion-like particle strings and for the QCD axion string network at temperatures above the QCD phase transition.  This is because the QCD axion mass is temperature dependent in the early Universe; it only becomes relevant at the QCD epoch, where it rises rapidly and quickly exceeds the Hubble parameter. Thus, at temperatures well above a GeV the dynamics of the QCD axion string are the same as those of axion-like particle strings with $H(t) \gg m_a$. Furthermore, in this limit, the domain wall number also does not affect the dynamics, since the domain wall number only plays a role in the evolution when the axion potential becomes important.

In the standard field theory UV completion the axion $a$ arises as the (pseudo-)~Goldstone boson of global PQ symmetry breaking of a complex scalar field $\Phi$, which for $T \ll f_a$ we represent as 
\es{Phi_def}{
\Phi = {(f_a + s) \over \sqrt{2}} e^{-i a / f_a}  \,,
}  
with $s$ the radial mode.\footnote{For simplicity we assume a KSVZ-type UV completion~\cite{Kim:1979if,Shifman:1979if}.} 
In vacuum, the field $\Phi$ is subject to the Lagrangian 
\es{eq:PQ}{
{\mathcal L} = \partial_\mu \Phi \partial^\mu \Phi^\dagger - \lambda_\Phi \left(|\Phi|^2 - \frac{f_a^2}{2}\right)^2 \,.
}
In the thermal Universe the field $\Phi$ also has a thermal mass term $V_{\rm therm}(\Phi) =  m_{\rm therm}^2 |\Phi|^2$, with $m_{\rm therm}^2 \approx {\lambda_\Phi T^2 / 3}$ (see, {\it e.g.},~\cite{Hiramatsu:2012gg}). The thermal potential restores the PQ symmetry at $T \gg f_a$.  
After PQ symmetry breaking the radial mode acquires a mass $m_s = \sqrt{2 \lambda_\Phi} f_a$ while the axion is massless.

After PQ symmetry breaking the radial mode abundance redshifts like matter until it decays to lighter states everywhere except at the location of the axion string cores. 
Axion strings are topologically protected solutions to the PQ equations of motion for which the axion acquires a $2 \pi f_a$ phase shift when traversing a path that encloses a string. 
The radial mode, which takes values $s \approx 0$ far from the strings, reaches the value $s = - f_a$, so that $\Phi = 0$, at the string core.  
As we discuss more below, semi-analytic solutions are available for infinitely straight strings (see, {\it e.g.},~\cite{Safdi:2022xkm} for a review), though in the cosmological context, the strings move, bend, combine, and disappear dynamically, thus requiring numerical simulations (see~\cite{Gorghetto:2018myk,Gorghetto:2020qws,Buschmann:2021sdq} for the current cutting-edge simulations).  
The string network primarily loses energy by radiating axions, though -- importantly for this work -- a small fraction of the dissipated energy goes into heavy radial modes.

The axion strings have tension $\mu_{\rm eff} \approx \pi f_a^2 \log(m_s / H)$, to leading order in large $\log (m_s / H)$. 
The $\log$ arises because the axion configuration that surrounds the string has an energy density that falls off slowly with distance, leading to a logarithmic divergence in the string tension; the large-distance cut-off is $\sim$$H^{-1}$, which is approximately the distance to the nearest string in the scaling regime.  
The energy density in the string network is $\rho_s \approx 4 \xi H^2 \mu_{\rm eff}$, where $\xi$ is the average number of strings per Hubble patch.  Note that $\xi$ is formally defined by $\xi \equiv \ell t^2 / {\mathcal V}$ at time $t$, where $\ell$ is the string length within a large volume ${\mathcal V}$.  
In the scaling regime, $\xi$ is approximately constant, regardless of whether the Universe is, {\it e.g.}, matter or radiation dominated.  
Logarithmic derivations to the scaling solution are now understood to arise~\cite{Gorghetto:2020qws,Buschmann:2021sdq}, with $\xi \approx c_1 \log(m_s / H)$ at large $\log$, with $c_1 \approx 0.25$ in the radiation dominated epoch~\cite{Buschmann:2021sdq}. In App.~\ref{strings:matter_domm} we show that $c_1 \approx 0.06$ in matter domination.

At large $\log(m_s / H)$ the rate of axion production is ({\it e.g.},~\cite{Gorghetto:2018myk})
\es{eq:gamma_a}{
\Gamma_a \approx 8 H^3 \xi \mu_{\rm eff} \,.
}
Recent simulations suggest that the momentum-space distribution of radiated modes is nearly conformal: $\partial \Gamma_a / \partial k \propto (H / k)^q$, with $q =1.02 \pm 0.03$, for $1 \ll k/H \ll m_s / H$~\cite{Buschmann:2021sdq} with $k$ being the  momentum of radiated axion.   

\subsection{Radial mode emission: general expectations}
\label{sec:general_expectations}

In addition to radiating axions, the strings may also radiate radial modes.  In the next subsection, we compute the radial mode emission rate by performing dedicated simulations, but in this section, we discuss our general expectations for this emission rate.  At momenta $k \gtrsim m_s$ we do not expect that the PQ theory differentiates between axion emission and radial mode emission, since at these high energies the radial mode emission is mildly relativistic.  On the other hand, radial mode emission is disallowed at frequencies less than $m_s$. 
Since $d \Gamma_a / d k \propto 1 /k$, as observed in simulations for $H \ll k \ll m_s$, we conjecture that $\Gamma_a^{k \gtrsim m_s} / \Gamma_a^{k \lesssim m_s} \sim c / \log(m_s / H)$, for some constant $c$, where $\Gamma_a^{k \gtrsim m_s}$ denotes the axion emission with $k \gtrsim m_s$ and $\Gamma_a^{k \lesssim m_s}$ is that with $k \lesssim m_s$. Since for high-$k$ the theory should not differentiate axion versus radial mode emission, we then conjecture the radial-mode emission rate $\Gamma_s \approx \Gamma_a^{k \gtrsim m_S}$:
\es{eq:Gamma_s}{
\Gamma_s \approx 8 c H^3 \xi \pi f_a^2 \,,
}
to leading order in large $\log(m_s/ H)$, where $c$ is an undetermined constant expected to be of order unity.

A deeper understanding of the relation between~\eqref{eq:gamma_a} and~\eqref{eq:Gamma_s} is found through the distribution of string loops. As discussed in~\cite{Buschmann:2021sdq}, we may understand the spectrum $d\Gamma_a / dk \sim 1 / k$ observed in simulations through the observation, in the same simulations, that $\ell d n_\ell / d \ell \approx {\rm const}$. Here, $n_\ell$ represents the number density of string loops with lengths less than $\ell$ at any given time.  A loop of length $\ell$ radiates axions at a characteristic wavelength $k \sim 1/\ell$.\footnote{More precisely, axion string loops of size $\ell$ appear to radiate axions with a conformal instantaneous spectrum $\propto 1/k$ for \mbox{${1 \over \ell} \lesssim k \lesssim m_s$}~\cite{Saurabh:2020pqe}; in App.~\ref{sec:single_loop} we perform numerical simulations of collapsing string loops to verify this scaling and discuss in more detail how it leads to a conformal emission spectrum for the network as a whole.}  Moreover, it has been shown that string loops (and also kinks in long strings) radiate energy at a constant rate $dE / dt$ regardless of the loop (kink) size~\cite{Hagmann:1990mj,Davis:1986xc,Davis:1985pt,Vilenkin:1986ku}.  This implies a conformal spectrum of axion emission $d\Gamma_a / dk \sim 1 / k$, with the high-$k$ modes being emitted by loops and kinks with large curvature; if $\ell \lesssim 1/m_s$, then these loops and kinks also emit radial modes with similar efficiency to axions. Indeed, in the numerical simulations described below, we find that radial modes are dominantly produced in regions of large string curvature.

\subsubsection{Analytic estimate for radial mode emission rate from string tension}

In the following subsections, we simulate the string-network evolution to measure the constant $c$. First, however, we present a rough but insightful analytic argument that suggests a value $c \sim 0.1$, which we later verify and refine with the numerical simulations.  The general idea behind our approach below is that as the string network evolves it radiates energy into axions and radial modes, but that energy must come from the stored energy in the string tension.
Thus, it is plausible that the fraction of radiated energy with $k \gtrsim m_s$ will be proportional to the fraction of the string tension that, in Fourier space, also has $k \gtrsim m_s$.  For $k \gtrsim m_s$ the axion and the radial mode are not qualitatively different, so we hypothesize that the modes emitted with $k \gtrsim m_s$ are split democratically between axions and radial modes.  Below, we make this argument precise.

We parameterize the PQ profile describing a long string oriented along the $z$ axis as
\begin{align}
    \Phi = \frac{f_a}{\sqrt{2}}g(m_s r) \, e^{i\theta} \,,
\end{align}
where $g(m_s r)$ is a dimensionless function and $\{r,\theta, z\}$ describe a 3D cylindrical coordinate system.
For a string solution, $g\sim m_s r$ for small $m_s r$, and $g\sim 1- 1/(m_s r)^2$ for large $m_s r$.
To estimate the emission of radial modes, we can Fourier transform the gradient of the position space profile to see which momentum modes are supported with $k \gtrsim m_s$.
The position space expression for the energy density is given by,
\es{}{
  \rho_{\rm str}(r,\theta,z) &=   |\vec{\nabla} \Phi|^2 + \lambda_\Phi \left( |\Phi|^2 - {f_a^2 \over 2} \right)^2 \\
  &= \frac{m_s^2 f_a^2}{2}\left[ \left(g'^2 + \frac{g^2}{x^2}\right) + {1 \over 4} (g^2 - 1)^2 \right] \,,
}
where $x = m_s r$ and $'$ denotes a derivative with respect to $x$.
The string tension $\mu$ is then given by
\es{}{
    \mu &= \int  d\theta dr \,r \rho_{\rm str} \,.\\
}
Restricting to a given plane orthogonal to the infinite string, the 2D Fourier transform of the energy density is
\es{eq:string_prof_k_2D}{
\tilde{\rho}(k) \equiv \int d^2 \mathbf{x} \, e^{-i\mathbf{k}\cdot  \mathbf{x}} \rho_\mathrm{str}(\mathbf{x})  = 2\pi \int_{0}^{\infty} dr \, r \rho_{\rm str}(r) J_0(kr) \,,
}
where $J_0$ is the $0^\mathrm{th}$ Bessel function of the first kind. At small $k$ we have $\tilde{\rho}(k)\propto \log(k)$, which gives rise to the IR divergence in the string tension $\mu = \tilde{\rho}(0)$.
\begin{figure}
    \centering
    \includegraphics[width=0.48\textwidth]{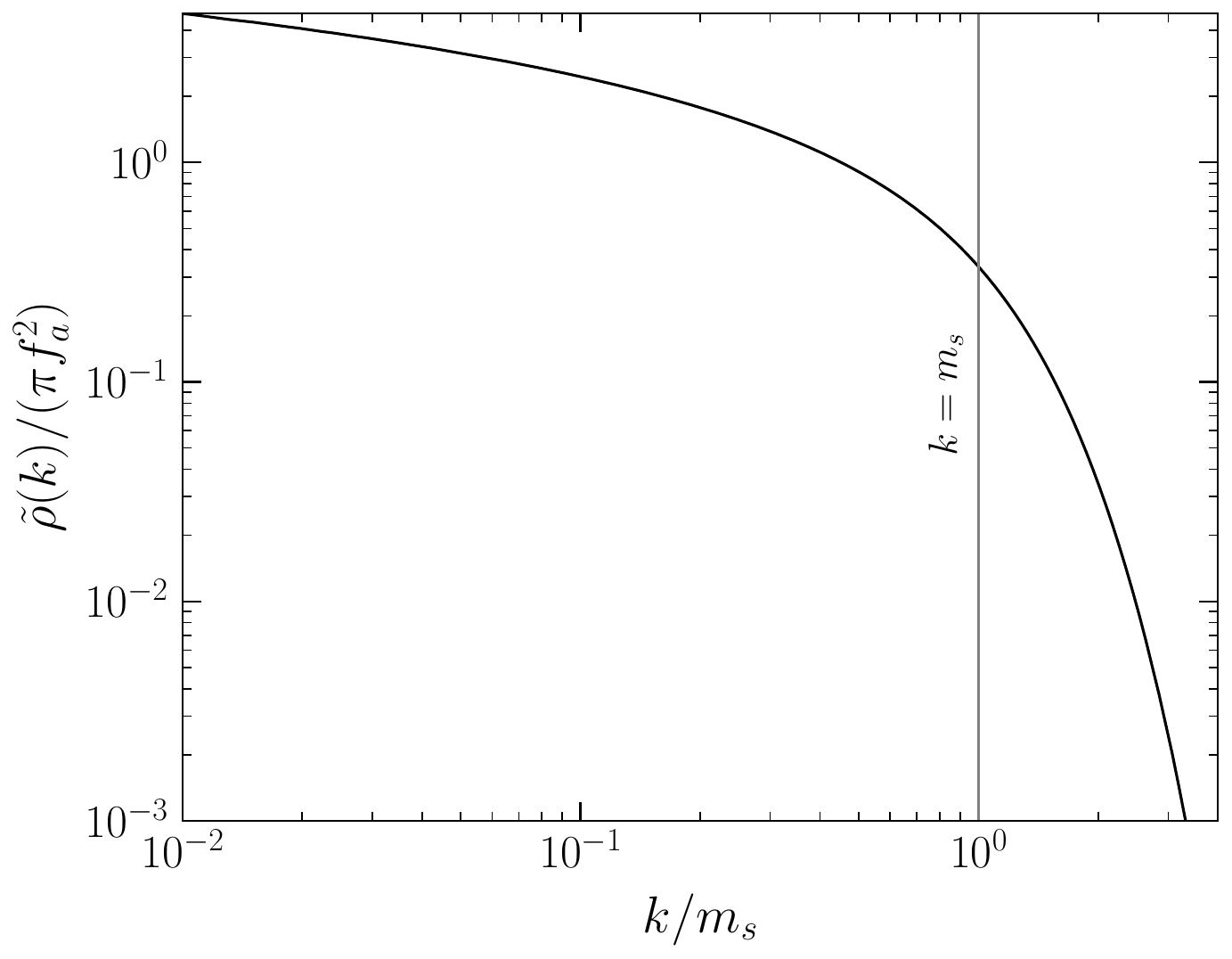}
    \caption{2D Fourier transform of the string energy density for an infinite, straight string, from~\eqref{eq:string_prof_k_2D}.
    }
    \label{fig:string_prof_2D}
\end{figure} 
Physically this divergence is cut off by Hubble providing the largest, relevant length scale at $\tilde{k}_{\rm IR} \sim H/m_s$. (Note that we define $\mu_{\rm eff}$ to be the effective tension computed with the IR cut-off $k_{\rm IR}$.) At large $k$,  $\tilde{\rho}(k)$ falls exponentially, as in Fig.~\ref{fig:string_prof_2D}. 

We estimate the part of the string tension relevant for axion emission as that with $k > m_s$; we compute this contribution as 
$\mu_\mathrm{UV} \equiv \tilde{\rho}(m_s) = 2\pi c_\mathrm{UV}f_a^2$, where the constant $c_{\rm UV}$ is defined with the specific normalization because we show in the following paragraph that it is related to the constant $c$ in~\eqref{eq:Gamma_s}. 
Numerically, we find $c_{\rm UV} \approx 0.16$.

Let us now discuss the relation between $\mu_{\rm UV}$ and $\Gamma_s$.  By comparing the evolution of the energy density of the string network in the scaling solution to that of the free-string network one may infer that the string network must emit energy with rate $\Gamma_{\rm tot} = 8 \xi \mu H^3$~\cite{Gorghetto:2018myk}.  Let us assume that the modes with $k > m_s$ emitted from the strings are split equally between axions and radial modes; while this is almost certainly not completely true, it allows us to make an ${\mathcal O}(1)$ estimate for the radial mode emission rate. Then, we estimate that $\Gamma_s \approx 8 c_{\rm UV} H^3 \xi \pi f_a^2$.  Comparing with~\eqref{eq:Gamma_s} we thus estimate that $c \approx c_{\rm UV} \approx 0.16$. 
As we show in the following subsections, this estimate for $c$ is similar to that we find in dedicated numerical simulations of the string network.

\subsection{PQ simulations for radial mode radiation: setup}

We simulate the evolution of axion strings with the Lagrangian as in~\eqref{eq:PQ} along with the thermal mass term for $\Phi$. We fix $\lambda_\Phi = 1$ throughout this work for definiteness, in all of our simulations, such that the radial-mode mass is $m_s = \sqrt{2} f_a$. 
We follow the basic procedure outlined in~\cite{Buschmann:2021sdq} and the equations of motion, along with common technical details pertaining to the base code, can be found therein.
Ref.~\cite{Buschmann:2021sdq} performed simulations of the axion string network to measure the axion radiation and infer the axion mass that gives rise to the observed DM abundance for the QCD axion. For our purposes, however, we are interested in the ratio of emission rates of radial-mode emission relative to axion emission.

Our code is based on the AMReX framework~\cite{zhang2020amrex} and is capable of AMR.  AMR allows for a dynamical grid with multiple refinement levels that track given spatial locations at higher spatial and temporal resolution (see Fig.~\ref{fig:ani} for an illustration of an AMR grid). While our setup is largely identical to that of~\cite{Buschmann:2021sdq}, our different objective forces two major changes.  First, because the radial mode spectrum is expected to peak at short wavelengths which propagate throughout the entire simulation volume, we do not employ any of the AMR capabilities ({\it i.e.}, we use a single refinement level).  
This is because the AMR setup would miss some of the short-wavelength radiation of the radial mode since its amplitude is often too small compared to that of axion radiation to trigger our refinement criteria. (Note, however, that we do use multiple AMR refinement levels in our simulations including the Higgs field in Sec.~\ref{sec:axion-Higgs strings}.)
Instead, the simulation is performed on a static lattice containing initially $512^3$ grid sites to ensure that the radial mode is resolved as well. As the co-moving width of axion strings decreases over time, the number of grid sites is increased by a factor of $2^3$ through quartic interpolation every time the string width would be resolved by less than four grid sites, up to a final size of $4096^3$ grid sites.

The second change we make relative to the simulation setup in~\cite{Buschmann:2021sdq} involves our initial conditions. Ref.~\cite{Buschmann:2021sdq} started with thermal initial conditions at $\eta_i$ for the complex scalar field $\Phi$. Here, since we are unable to evolve to as large of $\log(m_s / H)$ values as in~\cite{Buschmann:2021sdq} --- since we are not using multiple refinement levels --- we pre-evolve the state following the procedure in~\cite{Gorghetto:2018myk}. The pre-evolution mitigates the effects of transient oscillations to the string cores that are relevant at small values of $\log(m_s / H)$ and which would otherwise contaminate our emission measurement.
The pre-evolution procedure is described in more detail in App.~\ref{sec:pre-evolution} and involves modified equations of motion that support a constant string width and a moderate amount of Hubble friction. 
The pre-evolved state is designed to be close to the attractor solution at our starting time $\log(m_s/H)=2$. To avoid the reintroduction of transient radial-mode excitations when starting from a pre-evolved state due to the sudden change in the underlying physics we introduce a short adiabatic regime between $\log(m_s/H)=2$ and $\log(m_s/H)\sim 3$. In this regime, we smoothly interpolate between the two sets of equations of motion using a logistic function. (See App.~\ref{sec:pre-evolution} for more details.)

The simulation volume is a periodic box with comoving side length $L=33/(R_1H_1)$, where $R_1$ is the scale factor of the Friedmann–Robertson–Walker (FRW) metric at a reference time $t_1$ such that the Hubble parameter is $H(t_1)=f_a\equiv H_1$.
Our simulation is evolved in conformal time $\eta$ defined by $R(t) / R(t_1) \equiv \eta$. The simulation begins at $\eta_i \approx 2.3$ ($\log(m_s/H)=2$), and we evolve until $\eta_f \approx 21.7$ ($\log(m_s/H)\approx 6.5$).   We use a low-storage strong-stability preserving third-order Runge-Kutta algorithm to advance the field with a time step size given by the Courant–Friedrichs–Lewy condition through $\Delta \eta = 0.33 \Delta x$. The Laplacian operator is computed through a finite-difference 7-point stencil. At the end of the simulation, our box contains approximately 3.5 Hubble volumes.

The simulations are performed at the NERSC Perlmutter supercomputer. Each simulation runs for approximately  30 minutes on 256 AMD EPYC 7763 CPUs ({\it i.e.}, 128 total nodes, 256 total CPUs, and 16,384 total CPU cores).

\subsection{PQ simulations for radial mode radiation: results}
\label{sec:PQ_sim_results}

\begin{figure*}[!htb]
    \centering
    \includegraphics[width=1\textwidth]{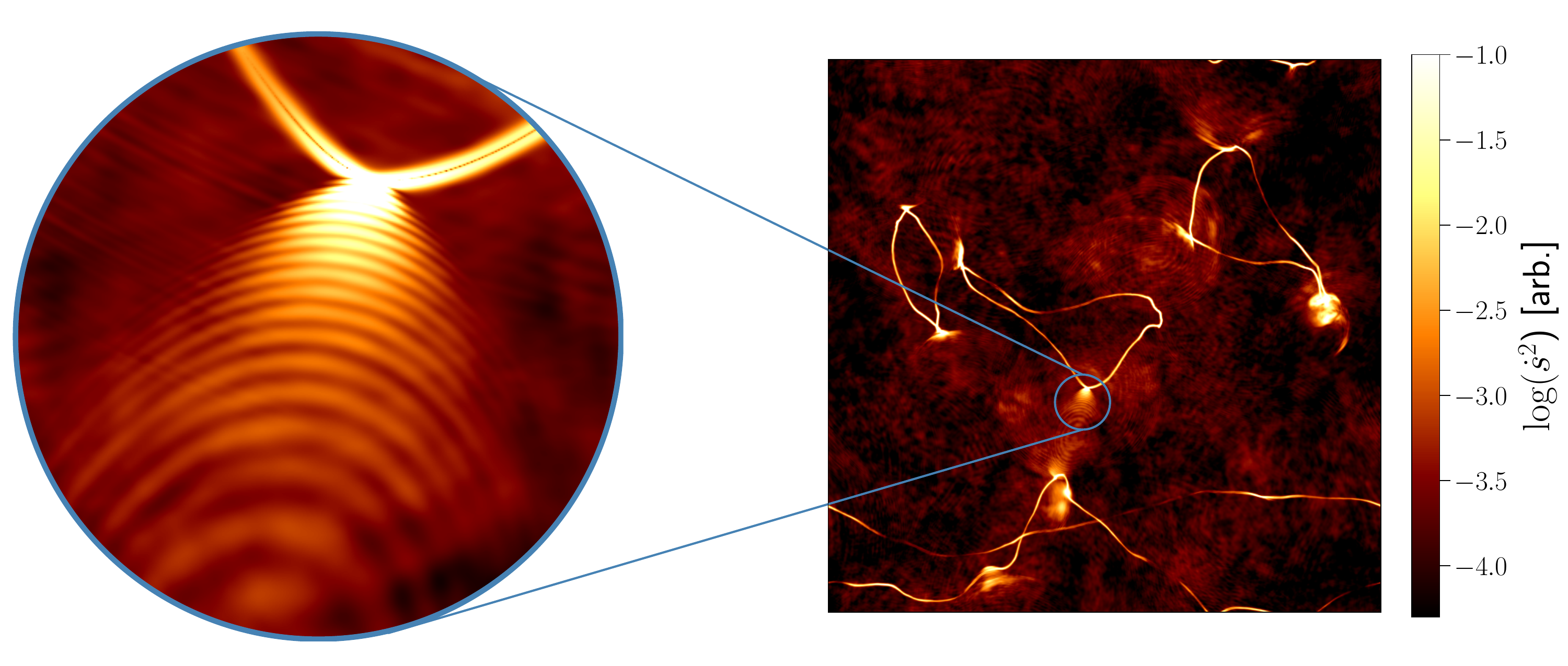}
    \includegraphics[width=1\textwidth]{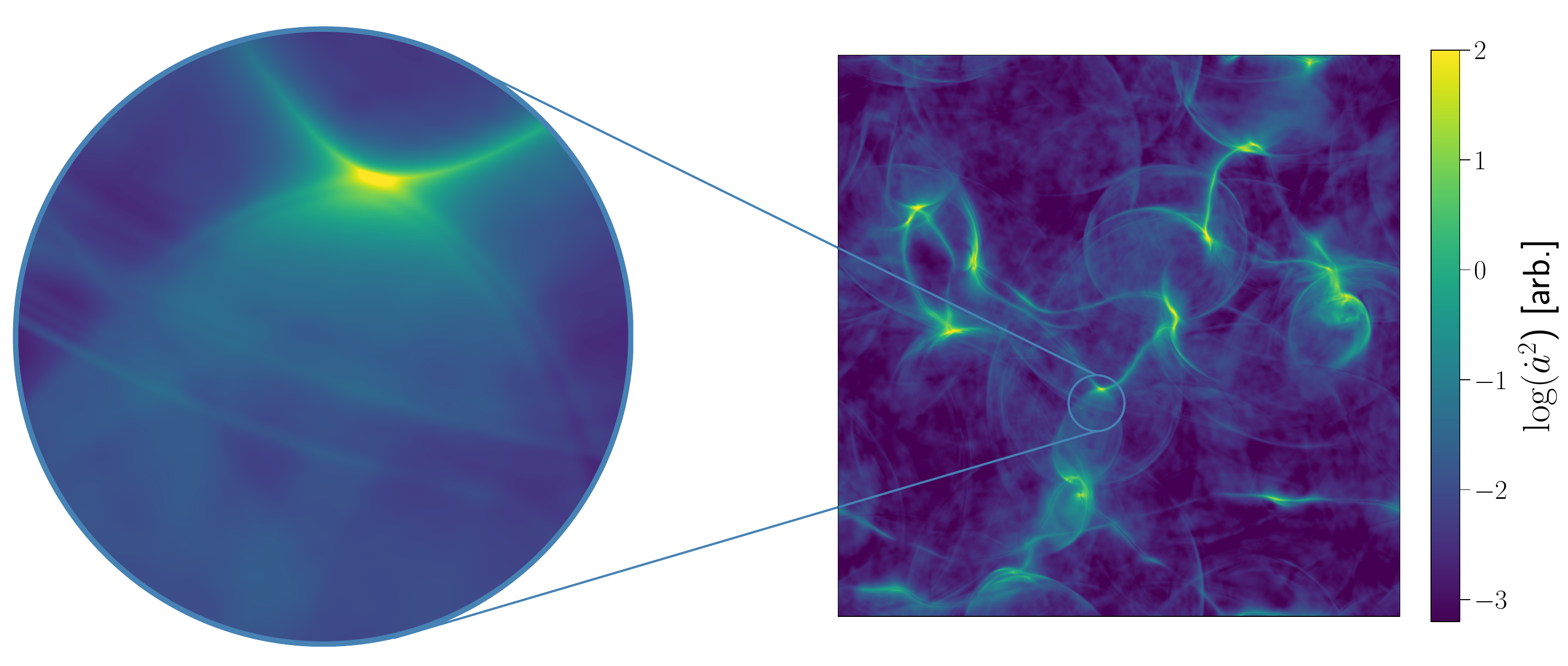}
    \caption{(Top panels) 2D projection of the radial mode energy $\dot s^2$ at the end of our 3D simulation investigating radial mode emission around $\log(m_s/H)\sim 6.5$. The full simulation box, spanning $\sim$1.5 Hubble lengths, is shown on the right with a detailed view shown on the left. Axion strings stand out as bright closed loops with strong emissions in particular around kinks and recent string re-connections. (Bottom panels) The same state of the string network but illustrated for the axion energy density $\dot a^2$ instead of that of the radial mode.  The axion emission has more support at long wavelengths relative to that of the radial mode. 
    }
    \label{fig:RadialModeEmission}
\end{figure*}

We are interested in measuring the amount of axion and radial mode radiation that is emitted from the string network over time.  In Fig.~\ref{fig:RadialModeEmission} we illustrate the radial mode and axion radiation from a snapshot near the end of the simulation, at $\log(m_s/H)\approx 6.5$.  In the top panel, we show the time-derivative of the radial mode squared ($\dot s^2$), in logarithmic units, which is a proxy for the radial mode energy density.  The string network is clearly visible in the top right panel; the bright regions away from the strings are regions of significant axion radiation. The zoom-in on the top left panel shows a string region producing large amounts of radial mode emission. That region of the string is characterized by its high curvature, which suggests that radial modes are predominantly produced from regions of the strings with high curvature of order the radial mode mass itself (see also App.~\ref{sec:single_loop}).  In contrast, the lower panel shows the axion time derivative squared ($\dot a^2$) for the same state as in the left panel. 
The axion radiation has support at longer wavelengths relative to radial mode radiation. 
Thus while the high-curvature region also produces significant axion radiation, the contrast versus the rest of the string regions is not as large. 

\begin{figure}[!t]
    \centering
    \includegraphics[width=1\columnwidth]{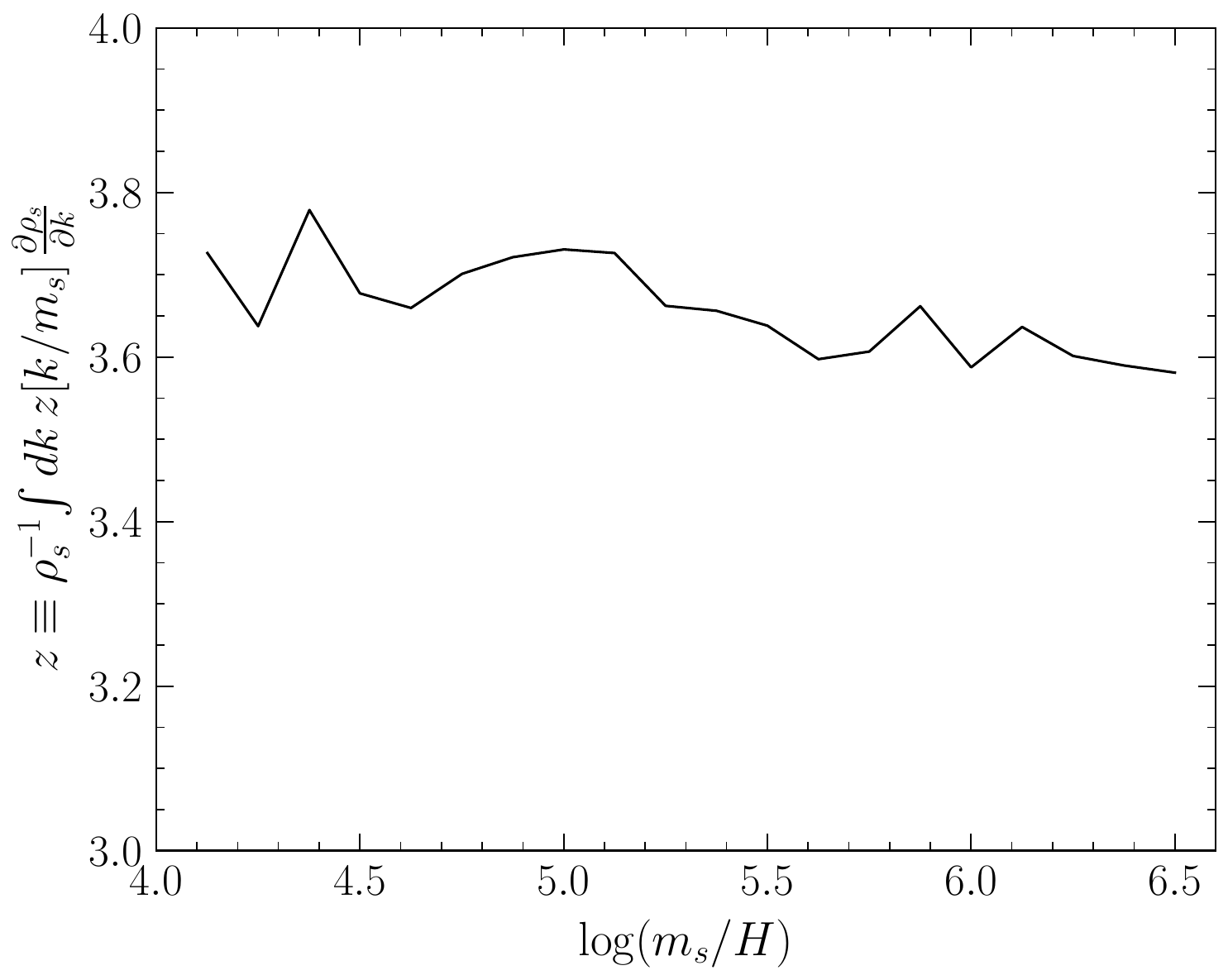}
    \caption{The quantity $z$ computed for the radial mode, defined in~\eqref{eq:Gamma_X_sim} and~\eqref{eq:z_deff}, describes how the instantaneous emission red-shifts at production. Completely non-relativistic (relativistic) radiation has $z = 3$ ($z = 4$), with the free particles scaling with the scale factor like $\rho_s \propto R^{-z}$. The emitted spectrum of radial modes is semi-relativistic, giving the intermediate $z$ shown. 
    }
    \label{fig:z_plot}
\end{figure}

To compute the energy densities more precisely we use the fact that away from the string cores both the axions and radial modes are free fields. At a given point $x$ the energy density of a real, free scalar field $X$, which solves its classical equations of motion, is 
\es{}{
\rho_X(x) &= {1 \over 2} \dot X^2 + {1 \over 2} (\nabla X)^2 + {1 \over 2} m_X^2 X^2 \\
&= \dot X^2 \,,
}
where $m_X$ is the field's mass and where we have applied the equation of motion to arrive at the second line.  
This implies that we can compute the average energy density over the simulation box, $\rho_X \equiv {1 \over L^3} \int d^3 x \rho(x)$, by 
\begin{equation}
    \rho_X = \frac{1}{L^3}\int d^3x \dot{X}^2(x) = \frac{1}{L^3}\int \frac{d^3k}{(2\pi)^3}|\tilde{\dot{X}}(k)|^2 \,.
    \label{eq:rho_X_sim}
\end{equation}
We may take $X(x)$ to be either the axion $a(x)$ or the radial mode $s(x)$. $\tilde{\dot{X}}(k)$ is the Fourier transform of $\dot{X}(x)$ as extracted from the simulation, where $\dot{X}(x)$ is screened to avoid contributions from the strings itself~\cite{Gorghetto:2018myk,Buschmann:2021sdq}. 
Explicitly, we use $\dot{X}(x) \rightarrow \dot{X}(x)(s(x)/f_a+1)^2$, since $(s/f_a+1)^2\approx 1$ away from strings but $\approx 0$ near the core. The exact form of this screening has little effect on the result~\cite{Buschmann:2021sdq}. 
\begin{figure}[!t]
    \centering
    \includegraphics[width=1\columnwidth]{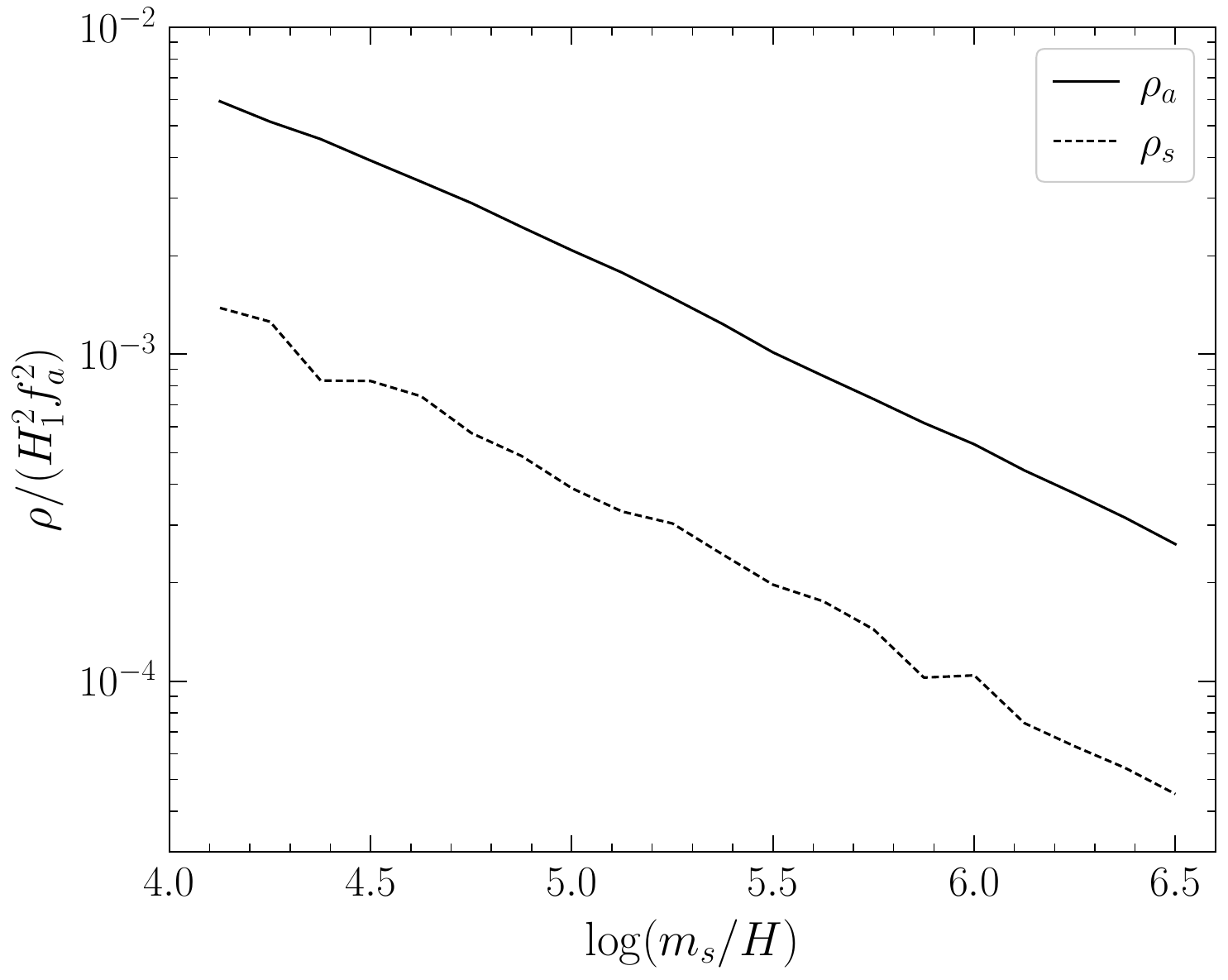}
    \caption{Energy densities $\rho$ for the axion (black) and radial mode (grey) as extracted from the simulation. }
    \label{fig:energy_densities}
\end{figure}
In Fig.~\ref{fig:energy_densities} we show the energy density as a function of time, displayed as $\log(m_s / H)$, for both the axion and the radial mode.  The energy density emitted by the string network is dominated by axion radiation; to compute the emission rates we need to take the appropriate time derivatives: 
\begin{equation}
    \Gamma_X=R^{-z}\frac{d}{dt}\left(R^z\rho_X \right) \,.
     \label{eq:Gamma_X_sim}
\end{equation}
Here, $z$ characterizes how the average energy density of the field $X$ red-shifts, with $z = 3$ for non-relativistic modes and $z = 4$ for relativistic modes. 
The axions have $z=4$ since they are massless. The redshift factor $z$ for the radial mode can be computed by~\cite{Gorghetto:2018myk,Gorghetto:2020qws,Gorghetto:2021fsn} 
\es{eq:z_deff}{
z \equiv {\int dk \, z[k/m_s] \,{\partial \rho_s \over \partial k} \over \rho_s}  \,,
}
with 
\es{}{
z[k/m_s]\equiv 3+{(k/m_s)^2 \over (k/m_s)^2+1}
}
and 
\es{}{
{\partial\rho_s \over \partial k}={|k|^2 \over (2\pi L)^3} \int d\Omega_k |\tilde{\dot{s}}(k)|^2 \,.
}
In Fig.~\ref{fig:z_plot} we show our results for $z$ computed over time from the simulation output for the radial mode. In general, we find $z \approx 3.7$, though it appears that $z$ slightly decreases over time.  Note that $z$ can be no smaller than $z = 3$ and physically we expect $z > 3$ since the radial modes are produced semi-relativistically.  The small decrease in $z$ over time may be expected since as we go to larger $\log(m_s / H)$ values we have a larger dynamical range between Hubble, which provides an IR cut-off, and the UV cut-off provided by $m_s$; this may account for an increased weight for the low-$k$ part of the spectrum at later times.  In our fiducial analysis below we take $z$ as measured at each $\log(m_s/H)$ step, though we show that our results are robust to changes in $z$. For example. we consider $z$ as small as $3.3$ and as large as $3.8$.  

\begin{figure*}[!t]
    \centering
    \includegraphics[width=1\columnwidth]{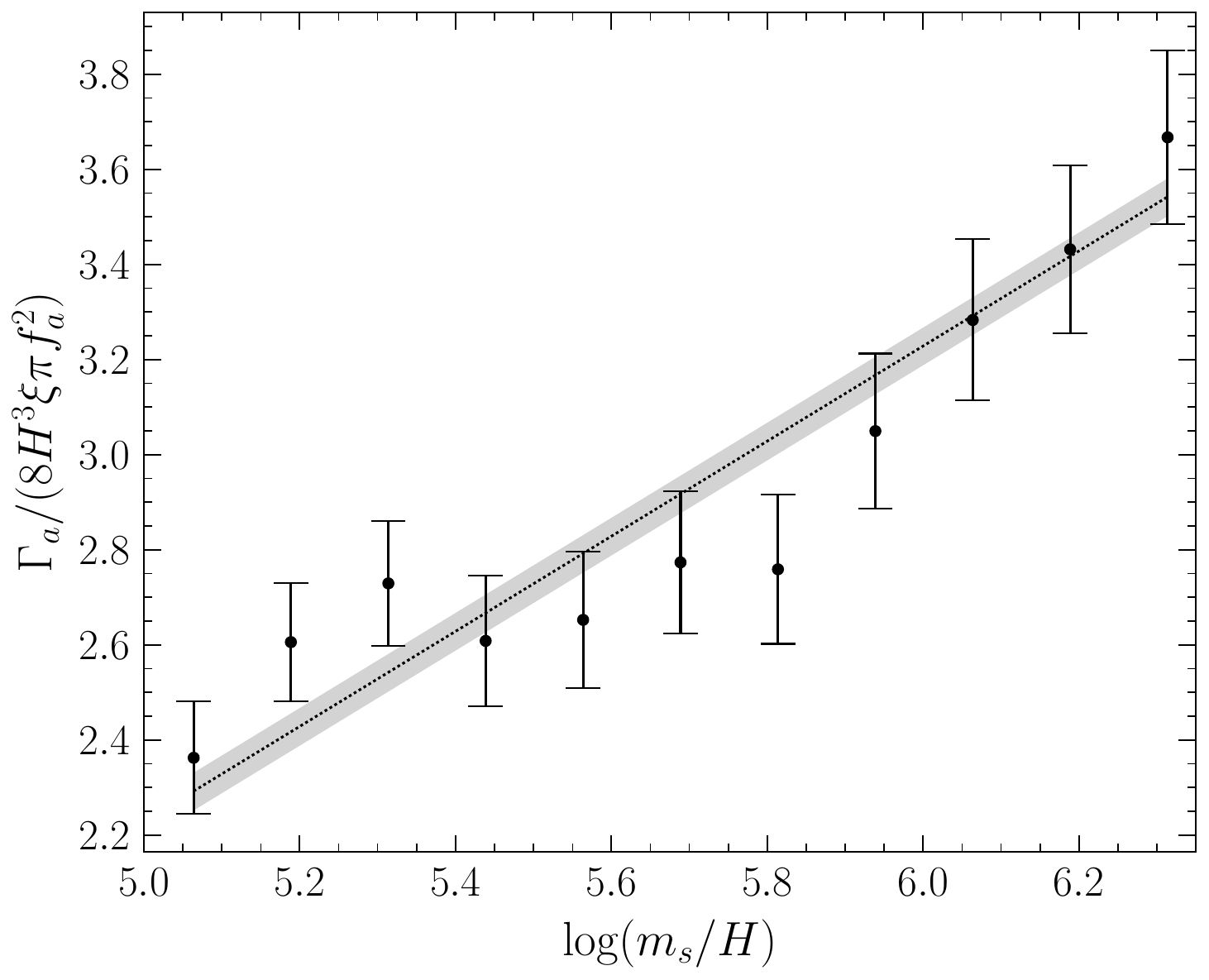}
    \includegraphics[width=1\columnwidth]{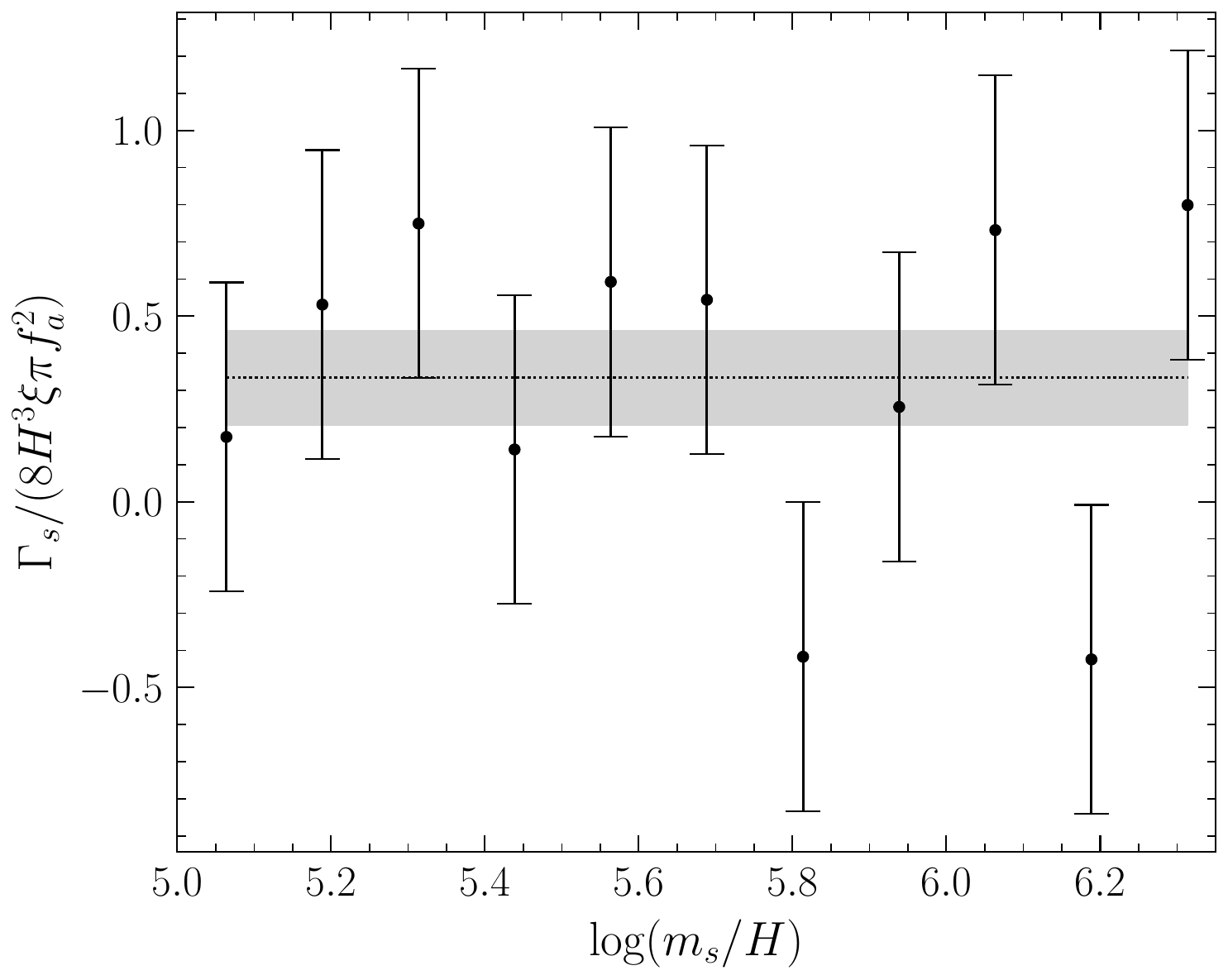}
    \caption{(Left panel) The normalized axion emission rate as measured in our AMR axion string simulation as a function of $\log(m_s / H)$. This quantity is expected to evolve with time as $\log(c_a m_s / H)$, for some constant $c_a$.  We fit this expectation to the data to determine $c_a \approx 0.063 \pm 0.002$ (see text for details). Importantly, the normalized axion emission rate rises logarithmically with time.  The dashed curve and grey band show the best-fit model expectation and the associated uncertainty band, respectively. (Right panel) As in the left panel but for the normalized radial mode emission rate. Unlike the axion emission rate, the normalized radial mode emission rate is expected to be constant with time (see~\eqref{eq:Gamma_s}).  By fitting a constant to the normalized radial-mode emission rate data we determine the constant $c$ in~\eqref{eq:Gamma_s} to be $c \approx 0.33 \pm 0.13$.}
    \label{fig:gamma_ratio}
\end{figure*}

In the left panel of Fig.~\ref{fig:gamma_ratio} we show the instantaneous axion spectrum, divided by $8 H^3 \xi \pi f_a^2$; referring to~\eqref{eq:gamma_a} we expect $\Gamma_a / (8 H^3 \xi \pi f_a^2) \approx \log( c_a m_s / H)$, where $c_a$ is a constant of order unity that accounts for finite contributions to the string tension and the precise form of the IR cut-off to the tension. We fit the model expectation to the data (for $\log(m_s / H) \geq 5$) to get the best-fit curve shown in dashed black, which has $c_a \approx 0.063 \pm 0.002$. (The grey band illustrated the 1$\sigma$ band on $c_a$.)  We perform the fit to the $\Gamma_a$ data assuming that the data points for $\Gamma_a$ have Gaussian uncertainties $\sigma = \alpha \tilde \Gamma_a(c_a)$, where $\alpha$ is a hyperparameter and where $\tilde \Gamma_a(c_a)$ is the model prediction for $\Gamma_a$ for a given choice of $c_a$. To reduce statistical and systematic noise we perform a total of nine simulations with different initial states; see Appendix~\ref{sec:pre-evolution} for more details. The results of these simulations are similar and hence we combine them by averaging $\Gamma_{a}$, $\Gamma_{s}$, and $\xi$.
We profile over $\alpha$ to compute the confidence interval for $c_a$. The best-fit value for $\alpha$ at the best-fit value for $c_a$ is then used to construct the 1$\sigma$ error bars on the data points shown in Fig.~\ref{fig:gamma_ratio}.

The quantity $\Gamma_a / (8 H^3 \xi \pi f_a^2)$ is observed to rise linearly with $\log(m_s / H)$, as expected. In contrast, we expect $\Gamma_a / (8 H^3 \xi \pi f_a^2) = c$, referring to~\eqref{eq:Gamma_s}, to be constant with $\log(m_s / H)$.  In the right panel of Fig.~\ref{fig:gamma_ratio} we show the $\Gamma_s$ data from our fiducial choice of $z$, which is extracted directly from the simulation at each time-step. The best-fit value of $c$ is shown in dashed black with the grey band indicating the 1$\sigma$ confidence interval; we find $c \approx 0.33 \pm 0.13$.  We use this value of $c$ in our analyses below looking at the effect of radial-mode induced energy injection, though keep in mind that different UV completions may give slightly different values for $c$.  

Our best-fit value of $c$ has a mild dependence on the choice of $z$ for the radial model emission. Choosing a constant $z$ of $3.8$ ($3.5$) ($3.3$) leads to a central value for $c$ of $\sim$$0.38$ ($0.25$) ($0.17$).  We do not consider this source of uncertainty further since it is subdominant compared to other sources of uncertainty in our analyses.

\section{Radial mode decays}\label{sec:radial_decay}

While the semi-static axion strings are protected from decay by the topology of the configuration, the radiated radial modes, with production rate given in~\eqref{eq:Gamma_s}, will promptly decay to lighter states.  In particular, the radial modes may decay to pairs of axions or to pairs of SM states.  The amount of visible energy injected by the string network is proportional to the branching ratio 
\es{}{
{\mathcal B} \equiv {\Gamma_{s \to {\rm SM}} \over  \Gamma_{s \to {\rm SM}} + \Gamma_{s \to aa}} \,,
}
with  $\Gamma_{s \to {\rm SM}}$ ($\Gamma_{s \to aa}$) the decay rate to SM final states (axion final states).  This quantity depends on the UV completion of the theory. We consider KSVZ-type~\cite{Kim:1979if,Shifman:1979if} axion-like-particle scenarios for illustration. (Note that the simplest implementations of the DFSZ~\cite{Dine:1981rt,Zhitnitsky:1980tq} scenario give a coupling of the axion to gluons, which we want to avoid since we are considering axion-like particles and not the QCD axion.)

In the KSVZ-type scenario, the Lagrangian is given by
\begin{align}
    {\cal{L}} = |\partial\Phi|^2 - V(\Phi) + \bar{Q}i\slashed{D} Q - (y_Q \bar{Q}_L Q_R \Phi + {\rm h.c.}),
    \label{eq:Q_largrangian}
\end{align}
with $Q_L, Q_R$ forming a vector-like fermion.  That fermion is charged under $SU(2)_L \times U(1)_Y$ but is a singlet under $SU(3)_c$, since we are focusing on axion-like-particles and not the QCD axion.  After spontaneous symmetry breaking the vector-like fermion acquires a mass $m_Q = y_Q f_a / \sqrt{2}$.

The vector-like quarks must be able to decay as otherwise the population produced in the early Universe, both thermally prior to the PQ phase transition and from the evolution of the string network, would become non-relativistic at later times and over-close the Universe.  New interactions beyond those in~\eqref{eq:Q_largrangian} are needed for the vector-like quarks to decay (see~\cite{DiLuzio:2020wdo} for a review).  In this regard, for specific choices of $SU(2)_L \times U(1)_Y$ charges for the vector-like quarks, additional interactions with the SM are possible.  For example, dimension four operators of the form $\bar f Q H$, where $f$ are SM fermions, $Q$ are the vector-like quarks, and $H$ is the SM Higgs, may lead to heavy quark decay.  
Dimension-five operators may also be responsible for such decays~\cite{DiLuzio:2020wdo}.  For our purposes, the specific forms of such operators are not important; all that is required is that if a heavy quark is produced it will eventually decay to SM final states.

Without loss of generality let us assume that under a $U(1)_{\rm PQ}$ transformation $\Phi \to e^{i \alpha} \Phi$, $Q_L \to e^{i \alpha} Q_L$, and $Q_R \to Q_R$ for constant $\alpha$. This transformation leaves the Lagrangian in~\eqref{eq:Q_largrangian} invariant and allows us to construct operators that respect the PQ symmetry but that induce PQ fermion decay using $Q_R$.

As an illustration, let us consider the possible dimension-four operators that give rise to heavy quark decay. We chose the convention for the weak-isospin such that $U(1)_{\rm EM}$ is generated by $Q= Y+T_3$, with $T_3 = {1 \over 2} \sigma_3$ the third generator of $SU(2)_L$ and $\sigma_3$ the third Pauli matrix; the SM Higgs $H$ then has weak hypercharge $Y=\frac{1}{2}$, right-handed leptons have $Y=1$, and left-handed leptons have $Y=-\frac{1}{2}$. 
If the KSVZ fermions 
have $SU(3)\times SU(2)\times U(1)_Y$ quantum numbers $(1, 1, Y)$ then their Yukawa couplings must be to left-handed SM leptons via ($\widetilde{H}_i=\epsilon_{i j} H^{\dagger j}$)
\begin{equation}
\mathcal{L}=-y \bar{L}^i Q_R H_i+\text { h.c. } \quad \text {or} \quad \mathcal{L}=-y \bar{L}^i Q_R \widetilde{H}_i+\text { h.c. } \,,
\end{equation}
enforcing $Y = -1$ or $Y = 0$ for the hypercharge of $Q_R$.  If $Y = 0$, though, the axion does not have non-trivial interactions with gauge fields at late times.
If instead the quantum numbers are $(1, 2, Y)$ then the Yukawa couplings must be to right-handed SM leptons via
\begin{equation}
\mathcal{L}=-y \bar{Q}_L^i \psi_R H_i+\text { h.c. } \quad \text {or} \quad \mathcal{L}=-y \bar{Q}_L^i \psi_R \widetilde{H}_i+\text { h.c. } \,,
\end{equation}
$Y = \frac{3}{2}$ or $Y = \frac{1}{2}$ for the hypercharge of $Q_L$. 

Recall that $m_s = \sqrt{2 \lambda_\Phi} f_a$ while $m_Q = y_Q f_a / \sqrt{2}$. For the radial modes to decay to KSVZ fermions far away from the string cores we need $2 m_Q < m_s$, which implies $y_Q < \sqrt{\lambda_\Phi}$. If this inequality is not satisfied then radial modes that propagate sufficiently far from the string cores will not be able to decay to KSVZ fermions and must decay directly to axions or to SM final states, as we discuss below. On the other hand, near the string cores, the KSVZ fermion masses are greatly reduced and indeed the fermion masses vanish at the core centers. Thus, the radial modes are always able to decay to $Q$ pairs near the string cores; these KSVZ fermions are then kinematically trapped near the string cores until they decay to SM final states.

We now consider several scenarios for the decays of the radial mode.
Since the radial mode always decays to a pair of axions, we consider one SM channel at a time to compute the branching ratio into the SM.
The expressions given below can be appropriately modified when multiple channels involving SM final states exist at the same time.

\paragraph{Decay into PQ fermions.} Let us start by considering the case $y_Q < \sqrt{\lambda_\Phi}$.  Then, the decay rate of radial modes to KSVZ fermions is 
\es{}{
\Gamma_{s \rightarrow \bar{Q}Q} &= N_L y_Q^2\frac{m_s}{16\pi}\left(1- \frac{4m_Q^2}{m_s^2}\right)^{\frac{3}{2}} \\
&= N_L y_Q^2\frac{\sqrt{2 \lambda_\Phi} f_a}{16\pi}\left(1- {y_Q^2 \over \lambda_\Phi}\right)^{\frac{3}{2}} \,,
}
where $N_L = 1$ if the fermions are $SU(2)_L$ singlets and $N_L = 2$ if they are doublets. This decay channel contributes to $\Gamma_{s \rightarrow {\rm SM}}$, since the KSVZ fermions decay completely to SM final states. 
On the other hand, because of the kinetic term, the radial mode may decay to axions at tree level, such that
\es{}{
    \Gamma_{s\to a a} = \frac{1}{32\pi}\frac{m_s^3}{f_a^2} = \frac{(2\lambda_\Phi)^{3/2}}{32\pi}f_a \,.
}
Thus, the branching ratio of the radial mode to SM final states is
\es{}{
{\mathcal B} = {N_L y^2 \left(1 - y^2\right)^{3/2} \over 1 + N_L y^2 \left(1 - y^2\right)^{3/2} } \,, \qquad y \equiv {|y_Q| \over \sqrt{\lambda_\Phi}} < 1 \,.
}
For $N_L = 1$ ($N_L = 2$) the branching ratio may be as large as ${\mathcal B} \approx 0.16$ (${\mathcal B} \approx 0.27$). On the other hand, if the Yukawa coupling to KSVZ fermions is small relative to the PQ self-coupling, then the branching ratio is suppressed; for example, ${\mathcal B} \approx 2 \times 10^{-4}$ for $y = 10^{-2}$ and $N_L = 2$.

\paragraph{Decay into electroweak gauge bosons.} In addition to the tree-level decays of the radial mode to KSVZ fermions and axions, there are one-loop decays to SM gauge bosons that may also be relevant. 
Taking the vector-like quarks to be in the fundamental representation of $SU(2)_L$ ($N_L = 2$), for example, the decay rate to $W$-bosons is given by~\es{}{
    \Gamma_{s \to WW} = \frac{\alpha_2^2}{72\pi^3}\frac{m_s^3}{f_a^2} = \frac{\alpha_2^2}{72\pi^3} (2\lambda_\Phi)^{3/2}f_a \,.
}
Assuming $N_L = 2$ we thus find ${\mathcal B} \approx 2 \times 10^{-5}$, where we use the value of the weak fine structure constant $\alpha_2(m_s \approx 10^{16} \, \, {\rm GeV}) \approx 0.02$ at energy scales of order the grand unified theory (GUT) scale.  The field $s$ can decay also to $ZZ$, $\gamma \gamma$, and $\gamma Z$, though these processes are roughly a factor of five smaller in total than the decay rate to $WW$.  See App.~\ref{KSVZ} for more details.

\paragraph{Decay into SM Higgs.}
The UV Lagrangian may also contain the renormalizable terms connecting the PQ field with the Higgs field through the potential:\footnote{The $\lambda_{H\Phi}$ quartic term contributes to the electroweak hierarchy problem by adding a mass term for the Higgs field, with mass parameter of order $f_a$, but this theory already has a hierarchy problem of the same order so the addition of this term does not make the hierarchy problem qualitatively worse.  In~\eqref{eq:VHPhi} we leave off additional, bare mass terms for $H$ in the UV that are needed to drive the Higgs mass parameter towards zero in the IR.  }
\es{eq:VHPhi}{
V(H,\Phi) =&\lambda_H \left[ |H|^2 - {\tilde \mu_H^2 \over 2\lambda_H} \right]^2 + \lambda_{\Phi} \left[ |\Phi|^2 - {f_a^2 \over 2} \right]^2  \\
&+ \lambda_{H \Phi} |H|^2 \left[ |\Phi|^2 - {f_a^2 \over 2} \right] \,.
} 
Here, $H$ is the SM Higgs doublet, $\lambda_H$ is the Higgs quartic, and $-\tilde \mu_H^2$ is related to the Higgs mass parameter.  The stability of the electroweak-symmetry-breaking vacuum requires
\es{}{
\lambda_{H\Phi}^2 < 4 \lambda_H \lambda_\Phi \,.
}
With this requirement, one can see by integrating out the radial mode that the Higgs acquires a VEV $\langle |H|^2 \rangle = v_{\rm EW}^2 / 2$, with $v_{\rm EW}^2 = \mu_H^2 / \lambda_H$ and $\mu_H^2 \equiv {4 \lambda_H \lambda_\Phi \over  4 \lambda_H  \lambda_\Phi - \lambda_{H\Phi}^2} \tilde \mu_H^2 $.

Expanding $\Phi$ about its VEV we see that the radial acquires an interaction with the Higgs field,
\es{}{
{\mathcal L} \supset \lambda_{H \Phi} |H|^2 |\Phi|^2 = \lambda_{H \Phi} f_a s |H|^2 + \cdots \,,
}
which allows the radial mode to decay at tree-level to Higgs pairs.
Since the radial mode mass is well above the Higgs mass, the decay rate $s$ to Higgs pairs is given by 
\es{}{
    \Gamma_{s \to HH} = {1 \over 4 \pi \sqrt{2} } {\lambda_{H \Phi}^2 \over \sqrt{\lambda_\Phi}} f_a < {1 \over   \pi\sqrt{2}} {\lambda_{H} \sqrt{\lambda_\Phi} } f_a  \,,
    \label{eq:s_to_HH_width}
}
such that ${\mathcal B} = 1 / [1 +\lambda_\Phi^2 / (2 \lambda_{H \Phi}^2)]$ and ${\mathcal B} < 1 / [ 1 + \lambda_\Phi / (8 \lambda_H)]$.  As we discuss further below, the Higgs quartic famously runs to small and potentially even negative values at high energy scales.  On the other hand, $\lambda_H$ does receive threshold corrections near the PQ scale that push it further positive in the UV (see, {\it e.g.},~\cite{Elias-Miro:2012eoi}).  For definiteness, let us assume that $\lambda_H = 0.01$ at the PQ scale, such that ${\mathcal B} \lesssim 0.1$, 
though smaller, positive values for $\lambda_H$ do not qualitatively change the branching ratio. (For example, ${\mathcal B} \lesssim 0.01$ if $\lambda_H = 10^{-3}$ at the PQ scale.)  On the other hand, if $\lambda_H$ is negative at the PQ scale this leads to runaway behavior for the axion strings, 
and so we do not consider that possibility further. 
 
In summary, depending on the Lagrangian parameters the branching ratio to SM final states may be as large as ${\mathcal B} \sim 0.3$ or as small as ${\mathcal B} \sim 10^{-5}$.  On the other hand, there are many ways of achieving a large branching ratio: for example, if the radial mode is kinematically allowed to decay to PQ fermions then it will generically do so with a large branching ratio unless the Yukawa coupling to the fermions is small.  
The branching ratio ${\mathcal B} \sim 10^{-5}$ is irreducible if we insist on the axion coupling to $W$-bosons; it can be a factor of a few smaller still if all other $s$-decay channels to SM final states are removed and the axion only couples to hypercharge and not to $SU(2)_L$.

Lastly, we note that even if $y_Q > \sqrt{\lambda_Q}$, such that the radial mode is not kinematically allowed to decay to KSVZ fermions asymptotically far away from the string, the radial mode could still decay to fermions close to the string. In the presence of a background field $\Phi$ the KSVZ fermions have a mass $\sim y_Q |\Phi|$; asymptotically far away from the strings $\Phi$ goes to its VEV $f_a / \sqrt{2}$. However, at the string core $\Phi = 0$, with the magnitude of $\Phi$ rising to $f_a$ over a distance of order $m_s^{-1}$ from the string core.  However, we do not consider such decays further here and rather bracket the possible branching ratio to SM final states as being within the range ${\mathcal B} \in (10^{-5},0.3)$.

 \section{Axion-Higgs strings}
 \label{sec:axion-Higgs strings}

The quartic coupling between the PQ scalar and the SM Higgs field -- the term parameterized by $\lambda_{H \Phi}$ in~\eqref{eq:VHPhi} -- is generically present in KSVZ models. This term could be present in the UV but otherwise, it is generated under the renormalization group through the KSVZ fermions at lower energy scales.
We show in this section that the presence of this coupling leads to non-trivial, classical Higgs field profiles surrounding the strings. Below, we refer to these strings as axion-Higgs strings and to the Higgs profiles as ``Higgs sheaths." These Higgs sheaths may have a number of important implications, but for our purposes, they provide efficient sources of SM radiation from the cosmologically-evolving axion-Higgs string network.  In this section, we study the axion-Higgs strings and perform dedicated simulations to study their dynamics and the radiated Higgs fields that are shed during their cosmological evolution. 
 
 Axion-Higgs strings were studied previously in~\cite{Abe:2020ure} in the context of the DFSZ model, where they were called electroweak axion strings.  However, the DFSZ electroweak axion strings are fundamentally different from the axion-Higgs strings that we discuss in this section.  In particular, the Higgs sheaths we discuss have no non-trivial winding around the string cores, while the electroweak axion strings exhibit non-trivial winding in the context of the two-Higgs doublet model (2HDM). In the DFSZ model, there are two Higgs doublets $H_1$ and $H_2$ that have non-trivial interactions with the complex PQ scalar $\Phi$, including those of the form
 \es{}{
 V_{\rm mix} \supset \kappa {\Phi^\dagger}^2 H_1 H_2 + {\rm h.c.} \,,
 }
where $\kappa$ is a coupling constant.  If we let the PQ charge of $\Phi$ be unity and the PQ charges of $H_1$ and $H_2$ be $X_1$ and $X_2$, respectively, then PQ invariance requires $2 - X_1 - X_2 = 0$.
Axion string solutions in this scenario may be of the form, for infinite, straight strings: $\Phi = f_a e^{i \theta}  g(r) / \sqrt{2}$, $H_1 = v_1 e^{i \theta } (0, h_1(r) )^T$, $H_2 = v_2 e^{i \theta } (0, h_2(r) )^T$~\cite{Abe:2020ure}.   
However, while the tension associated with the $\Phi$ profile is $\propto f_a^2$, the tension associated with the $H_1$ and $H_2$ profiles is significantly smaller, of order the Higgs field VEVs $v_{1,2}^2$. 
Given this reason, we do not consider the DFSZ scenario in further detail in this work.\footnote{Recall that since at least one of $H_1$ and $H_2$ is PQ charged, some of the SM quarks would also carry PQ charges. Hence, the axion would acquire tree-level derivative interactions with those SM quarks, and in turn, with gluons through the $a G \tilde G$ operator.
Thus, owing to the QCD-generated mass, $a$ is no longer a candidate for an axion-like particle.
This is another reason why we focus on a KSVZ-type model.}
 
 In contrast to the electroweak axion strings scenario discussed in~\cite{Abe:2020ure}, the Higgs sheath solutions that we present here contribute to the tension at order $f_a^2$.
 The Higgs sheath solutions appear in KSVZ-type scenarios, with single Higgs multiplets, with the Lagrangian as in~\eqref{eq:Q_largrangian} and~\eqref{eq:VHPhi}. As we discuss below, the $\lambda_{\Phi H}$ term generates non-trivial solutions for the Higgs field exterior to the strings with no winding.
 
 Below, we first discuss the Higgs sheath profiles for infinite, straight strings, and then we confirm the semi-analytic expectations for the Higgs profiles using numerical simulations of the axion-Higgs cosmology.
 
 \subsection{Semi-analytic solutions for infinitely straight axion-Higgs strings}
 
Consider an infinitely straight string in the ${\bf \hat z}$ direction in cylindrical coordinates $(r, \theta, z)$,  
in the theory consisting of the PQ scalar $\Phi$ and a single complex scalar field $H$, with potential given in~\eqref{eq:Q_largrangian} and~\eqref{eq:VHPhi}. In reality, the Higgs is an $SU(2)$ doublet, but by gauge symmetry, it is sufficient to work with a singlet Higgs field when computing the contribution to the string tension.

In cylindrical coordinates, the ansatz for an infinite, straight axion string along ${\bf \hat z}$ is 
\es{}{
\Phi = {f_a \over \sqrt{2}} g(r) e^{i\theta} \,,
}
 for some function $g(r)$.
We hypothesize that in the presence of the $\lambda_{H\Phi }$ interaction the Higgs field acquires a non-trivial profile that we may write as 
\es{}{
H ={f_a \over \sqrt{2}} h(r)
}
for a real function $h(r)$.
Then, in Minkowski space, a static solution $(g,h)$ obeys the equations of motion 
\es{eq:h}{h''+\frac{1}{r}h'+\tilde{\mu}_H^2 h - \frac{1}{2}\lambda_{H\Phi}f_a^2h(g^2-1)  -\lambda_H f_a^2h^3=0}
and
\es{eq:static_higgs_PQ}{g''+\frac{1}{r}g'-\frac{1}{r^2}g-\lambda_\Phi g^3 f_a^2 +\lambda_\Phi f_a^2 g - \frac{1}{2}\lambda_{H\Phi}h^2 g f_a^2=0 \,.}
Here primes denote derivatives with respect to $r$.
Writing the total energy density associated with the string solution as $\rho_{\rm tot} = \rho_H + \rho_\Phi + \rho_\mathrm{int}$, the sub-components of the energy density associated with the different fields are
\es{eq:higgs_string_eom}{
\rho_H &= \frac{1}{2}(h')^2f_a^2 -\frac{1}{2}\tilde{\mu}_H^2h^2f_a^2+\frac{1}{4}\lambda_H h^4f_a^4,\\
\rho_\Phi &=  \frac{1}{2}(g')^2f_a^2 +\frac{1}{2}\frac{g^2}{r^2}f_a^2 + \frac{\lambda_\Phi}{4} (g^2 -1)^2f_a^4,\\
\rho_\mathrm{int} &=  \frac{1}{4}\lambda_{H\Phi} h^2 \left (g^2 - 1\right)f_a^4 \,. \\
} 

The tension is defined as the total energy per unit length of the string  
\begin{align}
\mu_\mathrm{tot}  =2\pi\int _0 ^{\infty}  dr\,r \rho_\mathrm{tot}(r) \,.
\end{align}
Note that unlike for the PQ field $\Phi$, which has non-trivial winding around the string, there is no topological protection for the Higgs profile.  The stability of the Higgs profile can rather be determined by the following consideration. A trivial solution to the equation of motion~\eqref{eq:h} is given by $h = 0$ everywhere. We denote the tension corresponding to that solution as $\mu_{h = 0}$.\footnote{More precisely, we define $\mu_{h = 0}$ as the energy density per unit string length of the string solution with $h = 0$ fixed minus the energy density per unit string length of the Universe with $h =0$ and $g = 1$ everywhere.} 
If $\mu_{\rm tot} < \mu_{h= 0}$, with $\mu_{\rm tot}$ including the non-trivial Higgs field profile, then the Higgs sheath is stable.

As already discussed, stability of the PQ and electroweak vacua requires $0<\lambda_{H\Phi}<\sqrt{4\lambda_\Phi\lambda_H}$, and $\lambda_H >0$.
The PQ and Higgs VEVs are then 
\begin{equation}
\begin{aligned}
&\langle |\Phi|^2 \rangle = {v_\Phi^2 \over 2} \,, \qquad v_\Phi^2=f_a^2 - {\lambda_{H \Phi}  \over 2 \lambda_\Phi} \frac{\mu_H^2}{\lambda_H} \approx f_a^2 \\
&\langle |H|^2 \rangle = {v_{\rm EW}^2 \over 2} \,, \qquad v_{\rm EW}^2= {\mu_H^2 \over \lambda_H} \,.
\end{aligned}
\end{equation}
We use this notation below in describing the behavior of the field profiles.  In practice, we may approximate $v_{\rm EW} \approx 0$, since $f_a \gg v_{\rm EW}$.  The radial mode mass is $m_s^2 = 2\lambda_\Phi f_a^2$.  With the approximation $v_{\rm EW} \approx 0$, and assuming all of the dimensionless coupling constants ($\lambda_H$, $\lambda_{H \Phi}$, $\lambda_\Phi$) are order unity, there is only one dimensionful scale, which is $f_a$. At small $r$, the equations of motion enforce $g(0) = 0$ in addition to $h'(0) = 0$.  The quantity $h(0)$ cannot be computed analytically; it must be computed numerically.  However, the fact that the equations of motion do not depend on any dimensionful parameters ensures that $h(0) \sim 1$, such that $H(0) \sim f_a$. Asymptotically far from the string core $h \to 0$ and $g \to 1$. The fields $\Phi$ and $H$ thus have field excursions of order $f_a$ over distance scales of order $f_a^{-1}$, since $f_a$ is the only dimensionful parameter in the problem. This implies that the contributions to the tension from both $H$ and $\Phi$ are expected to be of order $f_a^2$.  Note that both $g$ and $h$ approach their asymptotic values at $r \to \infty$ through terms the fall of with $r$ as $1/r^2$, which implies that both the radial mode and the Higgs field have IR-finite contributions to the tension, unlike the contribution from the axion field, which is logarithmically-divergent in the IR.

It is important to contrast the Higgs solution above with that found in the DFSZ electroweak axion string. In the latter case, the Higgs fields have non-trivial winding, which implies that regularity at $r = 0$ forces $H_1(0) = H_2(0) = 0$. On the other hand, $H_{1,2}(r = \infty) \sim v_{1,2}$. Thus, the Higgs fields in the electroweak axion strings only have field excursions of order their VEVs $v_{1,2}$. It is precisely because our Higgs field does not wind that it is able to have $H(0) \sim f_a$ and thus contribute substantially to the string tension. 

\begin{figure}[!tb]
\begin{center}
\includegraphics[width=0.9\linewidth]{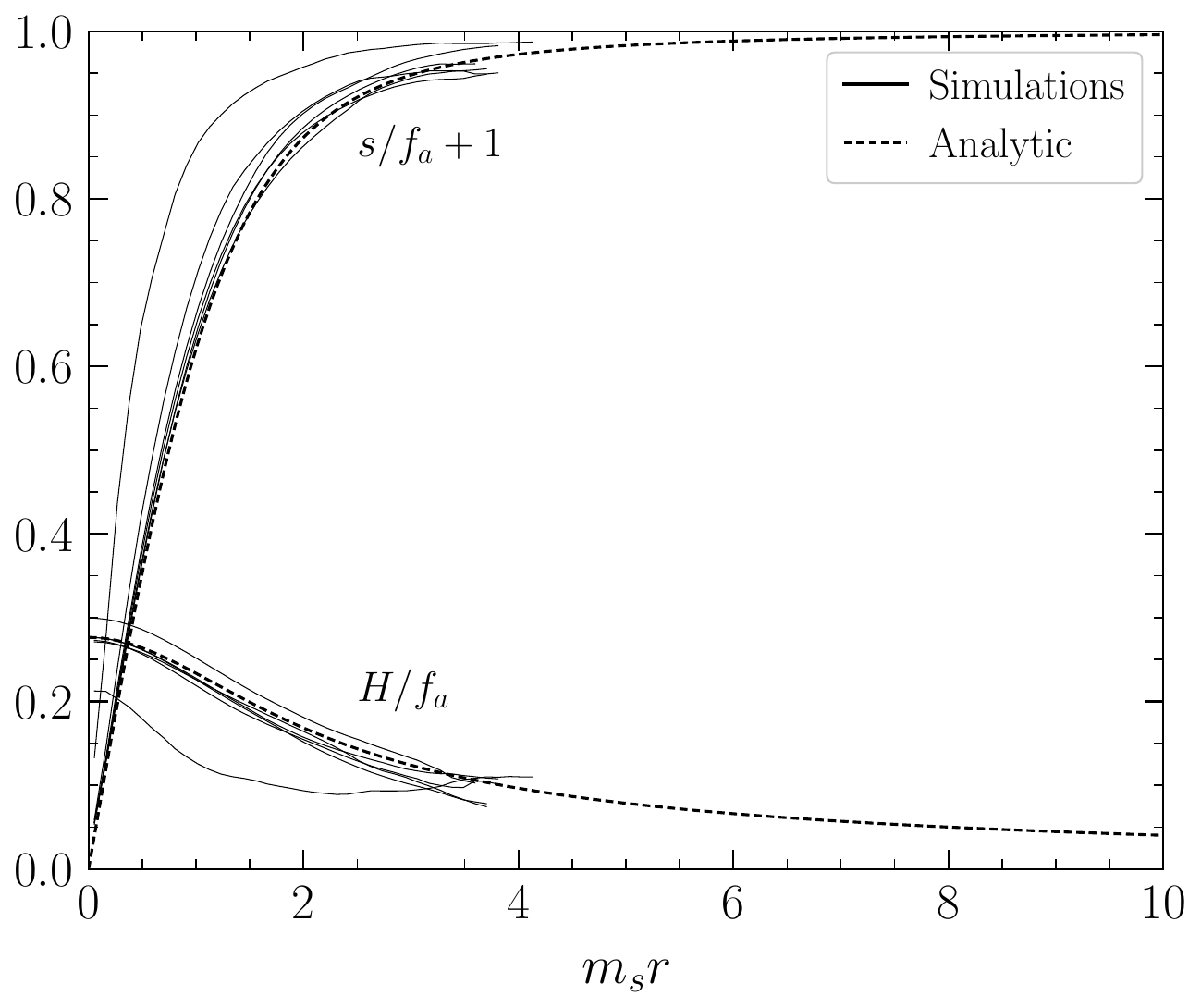} 
\caption{Higgs and radial mode profiles around six string segments extracted from the AMR simulations (solid) at $\log(m_s/H)=5.79$. The profiles are computed from a small sub-volume around the string location, by  averaging the field values in bins of distance to the  string. The measured profiles are compared to the infinitely-long, static string solution (dashed), which is found by numerically solving~\eqref{eq:static_higgs_PQ} for $(\lambda_\Phi, \lambda_H, \lambda_{H\Phi}) = (1, 4, 2)$. 
}
\label{fig:higgs_prof}
\end{center}
\end{figure}

In Fig.~\ref{fig:higgs_prof} we illustrate the radial mode and Higgs field profiles found for the infinite, straight string by numerically integrating~\eqref{eq:h} and~\eqref{eq:static_higgs_PQ} using a $4^{\rm th}$ order collocation method.  We take the Higgs VEV to be zero since physically it is much less than $f_a$, and we make the choices 
$(\lambda_\Phi, \lambda_H, \lambda_{H\Phi}) = (1, 4, 2)$.  Note that $|\Phi| = 0$ at the string core, as required in order to remove the singularity associated with the axion winding, while at large distances from the string, the PQ mode asymptotes to its VEV $|\Phi| = f_a / \sqrt{2}$. The Higgs field has a non-zero value at the string core 
($|H|/f_a \approx 0.3$), since it has trivial winding, and it asymptotes to zero infinitely far from the string. 

\begin{figure}[!tb]
\begin{center}
\includegraphics[width=0.9
\linewidth]{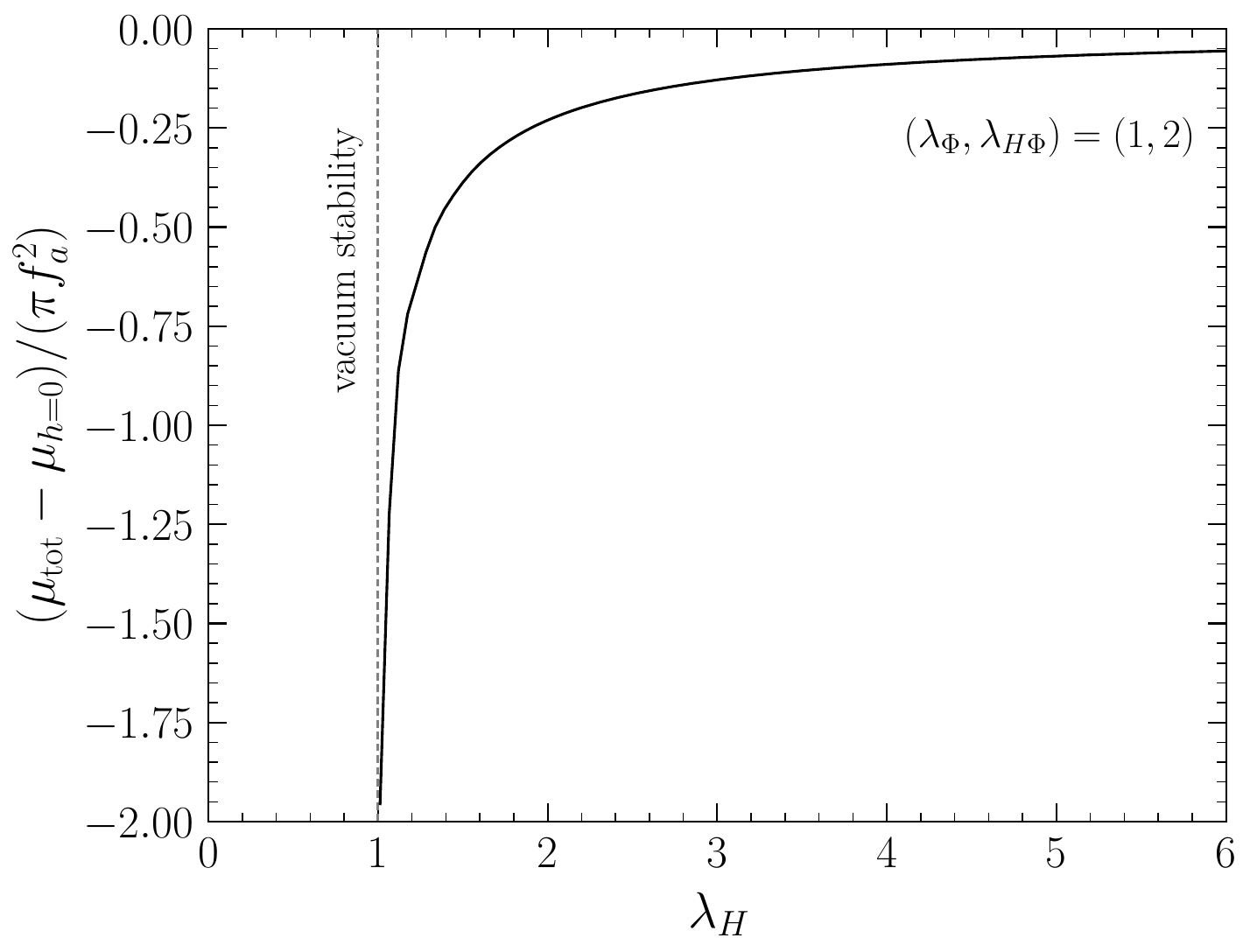}
\caption{
The difference in tension between the infinitely straight Higgs sheath solution and the solution with no non-trivial Higgs field profile ($h = 0$).  We illustrate this difference as a function of $\lambda_H$ with the other dimensionless coupling constants fixed to the indicated values. Negative values indicate that the Higgs sheath profile is stable. 
}\label{fig:tension_h}
\end{center}
\end{figure}

Let us now verify that the Higgs sheaths are stable by computing the tension of these configurations and comparing to the solution with $H =0$ everywhere.  In Fig.~\ref{fig:tension_h} we illustrate $\mu_{\rm tot} - \mu_{h= 0}$ as a function of $\lambda_H$ for the choices 
$(\lambda_\Phi, \lambda_{H\Phi})=(1,2)$.   Note that vacuum stability requires $\lambda_H > \lambda^2_{H\Phi} / (4 \lambda_\Phi) = 1$ 
in this case, which is indicated.  For all $\lambda_H$ the differences in tension are negative, suggesting that the Higgs sheaths represent the energetically preferred solutions and are stable to decay. 

A network of axion strings with Higgs sheaths may be expected to evolve by emitting classical axion, Higgs, and radial mode radiation.  By energy conservation, the total rate of energy loss to radiation by the string network is equal to the time derivative of the energy difference between the evolving string network and the free string network (see, {\it e.g.},~\cite{Gorghetto:2018myk}). The axion emission rate should be mostly unaffected by the Higgs profiles since the Higgs profiles extend over a distance $\sim$$m_s^{-1}$ while the axion emission dominantly comes from longer wavelengths.  The Higgs emission rate should then arise from the IR-finite part of the string tension $\mu \sim \pi f_a^2$. More precisely, assuming the Higgs emission arises from energy conservation associated with the IR-finite part of the string tension we expect 
\es{eq:Gamma_H_ansatz}{
\Gamma_H = \left( 8 H^3 \xi \pi f_a^2\right) f(\lambda_H, \lambda_\Phi, \lambda_{H \Phi}) \,,
}
where $f(\lambda_H, \lambda_\Phi, \lambda_{H \Phi}) $ is a function of the dimensionless coupling constants.
An analytic derivation of $f(\lambda_H, \lambda_\Phi, \lambda_{H \Phi}) $ appears difficult, due to the non-linear nature of the equations of motion.  In the following sub-section, we numerically calculate $f$ for specific choices of the coupling constants by performing AMR simulations.

\subsection{AMR simulations of axion-Higgs strings}

\begin{figure*}[!htb]
    \centering
    \includegraphics[width=1\textwidth]{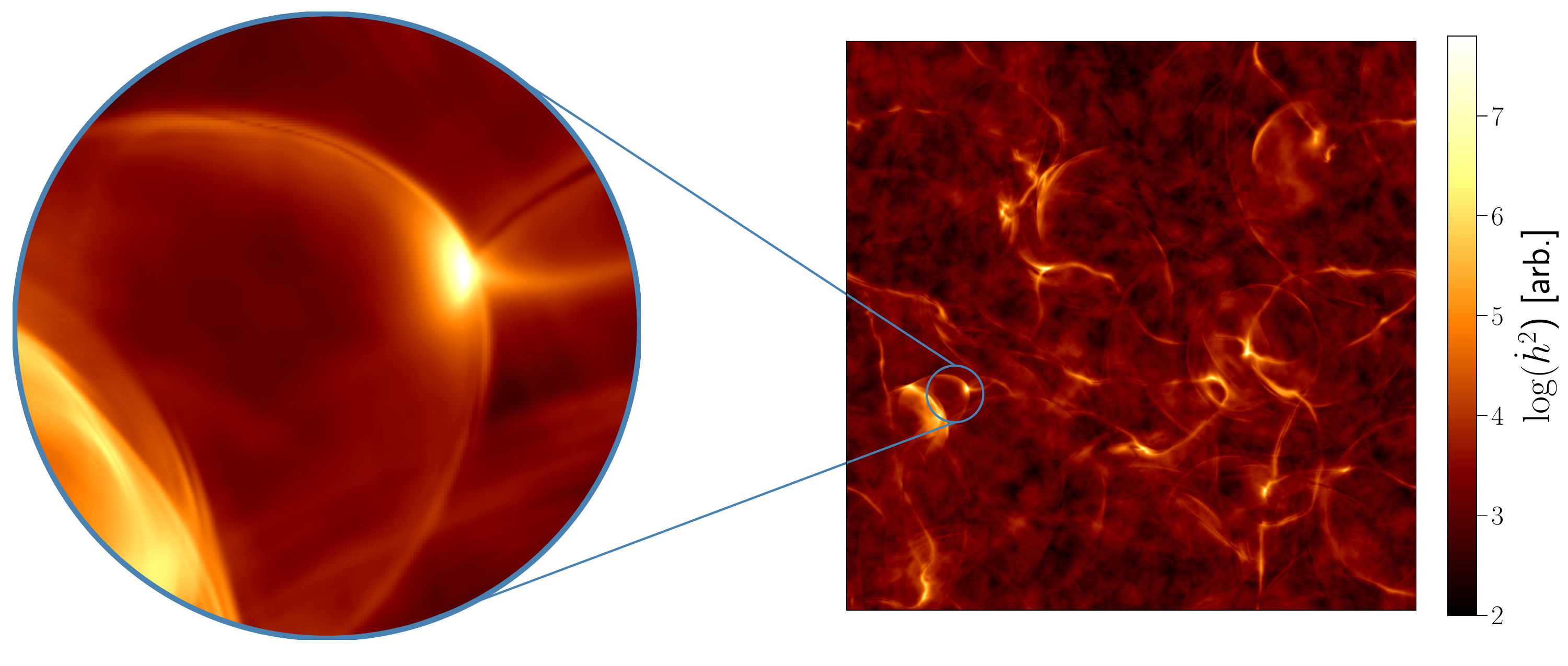}
      \includegraphics[width=1\textwidth]{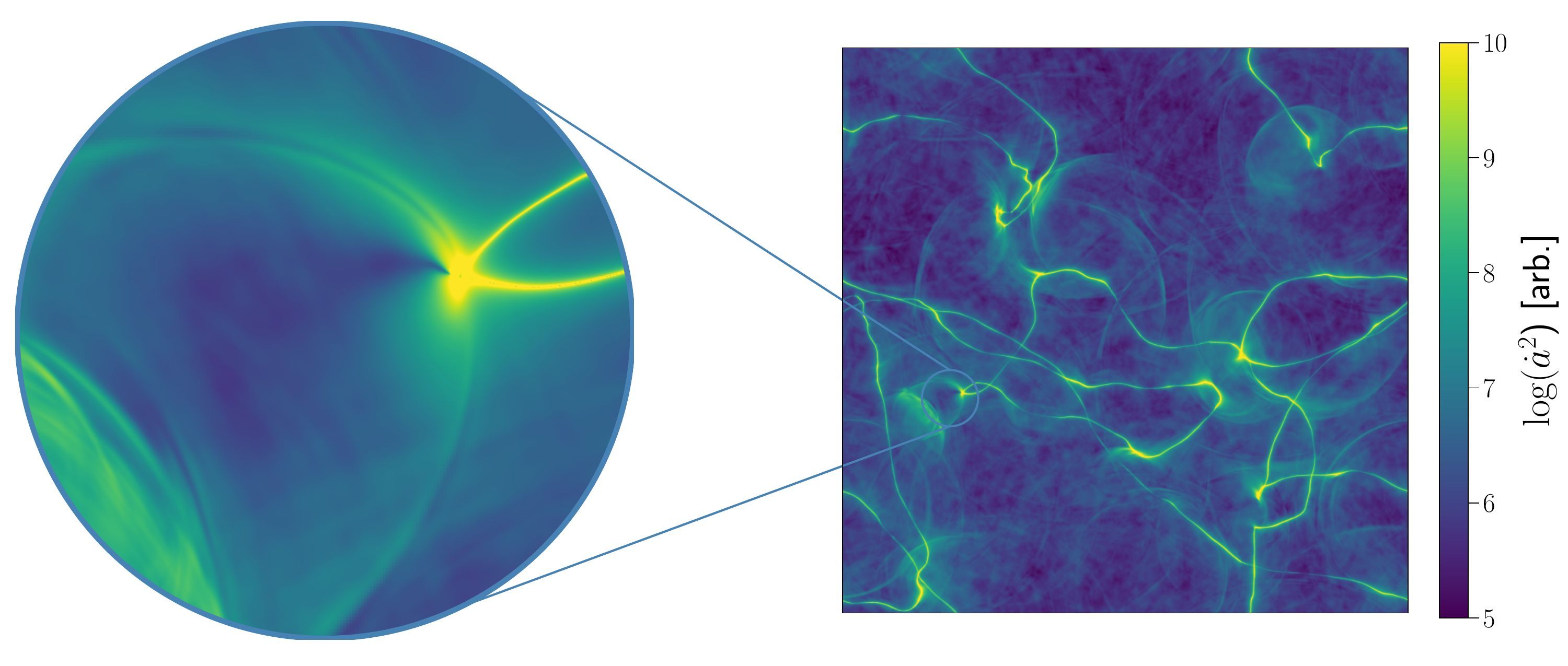}
    \caption{As in Fig.~\ref{fig:RadialModeEmission} but for the simulations including Higgs fields. (Top panels) 2D projection of the Higgs energy $\dot h^2$ towards the end of our fiducial 3D simulation around $\log(m_s/H)\sim 7.0$. The full simulation box, spanning $\sim$1.65 Hubble lengths, is shown on the right with a detailed view shown on the left. (Bottom panels) The same state of the string network but illustrated for the axion energy density $\dot a^2$ instead of that of the Higgs.  Animations available \href{https://goo.by/qHk9d}{here}.}
    \label{fig:higgs_emission_proj.png}
\end{figure*}

We verify the development of axion-Higgs strings and the subsequent classical radiation of Higgs modes through AMR lattice simulations of the coupled equations of motion in the early Universe. 
The simulation setup is similar to that in Sec.~\ref{sec:prompt_radial_mode_emission} except that it uses an adaptive mesh; details are described in App.~\ref{supl:AxionHiggsSims}.  We use an AMR grid instead of a static lattice grid in order to access a larger dynamical range.  Note that we simulate the coupled Higgs-PQ system on an adaptive lattice with comoving side length $L=46/(R_1H_1)$ up to $\log(m_s/H) = 7.2$, with $R_1$ and $H_1$ as defined in Sec.~\ref{sec:prompt_radial_mode_emission}.  We start the simulation with thermal initial conditions for both the Higgs field and the axion.  Our fiducial choice for the couplings is $\lambda_\Phi=1$, $\lambda_{H\Phi}=0.3$, and $\lambda_H=0.05$, as discussed further below, though we also consider variations to this fiducial choice.  A 2D projection of the simulation volume, for an example simulation, is shown in Fig.~\ref{fig:higgs_emission_proj.png} at $\log(m_s / H) \sim 7$ for both the Higgs  radiation energy density and the axion radiation energy density. Higgs sheaths and Higgs emission regions are clearly visible surrounding axion strings. 

The goal of the simulations is to measure the function $f$ characterizing the Higgs emission rate that appears in~\eqref{eq:Gamma_H_ansatz}. First, let us consider reasonable choices for the parameters $\lambda_H, \lambda_\Phi, \lambda_{H \Phi}$. Without any prior for $\lambda_\Phi$, we simply take $\lambda_\Phi = 1$ for illustrative purposes.  The value we take for $\lambda_H$ is more subtle since it is known that $\lambda_H$ runs to small and potentially even negative values at high energy scales, assuming no new heavy physics (see, {\it e.g.},~\cite{Degrassi:2012ry}).  We insist that $\lambda_H$ is positive for consistency.  In fact, the PQ field itself should give a threshold correction to the Higgs quartic at energy scales of order $f_a$, which is the relevant scale for considering the Lagrangian when solving the classical equations of motion at distances of order $f_a^{-1}$ from the string core. That threshold correction could push the Higgs quartic to values around $\lambda_H \sim 5 \times 10^{-2}$, depending on the PQ scale and on the top quark mass~\cite{Elias_Mir__2012}. For definiteness, let us then take $\lambda_H = 5 \times 10^{-2}$.\footnote{More precisely, a constant value of $\lambda_H$ may not capture the full dynamics, since  if we associate the radial direction from the string with the renormalization group scale, the value of $\lambda_H$ should drop by its threshold correction when starting at the string core and traveling to distances much larger than $m_s^{-1}$.  We adopt a constant value of $\lambda_H$ here for simplicity.}  We may then vary $\lambda_{H \Phi}$ from small values all the way to the vacuum stability limit.  

\begin{figure}[!t]
\begin{center}
\includegraphics[width=0.9\linewidth]{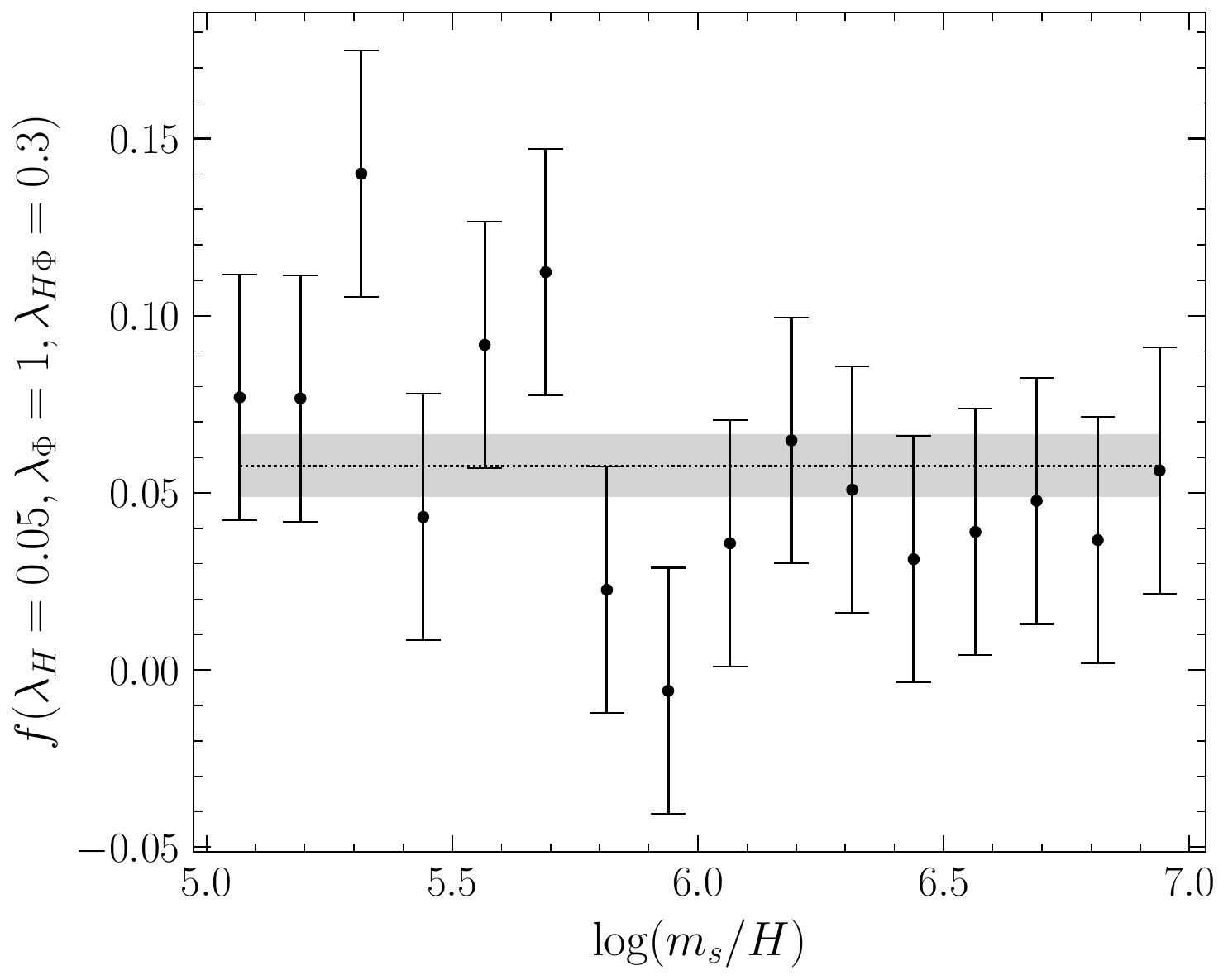}
\caption{
Direct Higgs emission $f(\lambda_H, \lambda_\Phi, \lambda_{H\Phi}) = \Gamma_H / (8 H^3 \xi \pi f_a^2)$ (data points) at our fiducial parameter set including linear fit (dotted line) with corresponding $1\sigma$ uncertainty band (grey band).}
\label{fig:gamma_h}
\end{center}
\end{figure}

\begin{figure}[!t]
\begin{center}
\includegraphics[width=0.9\linewidth]{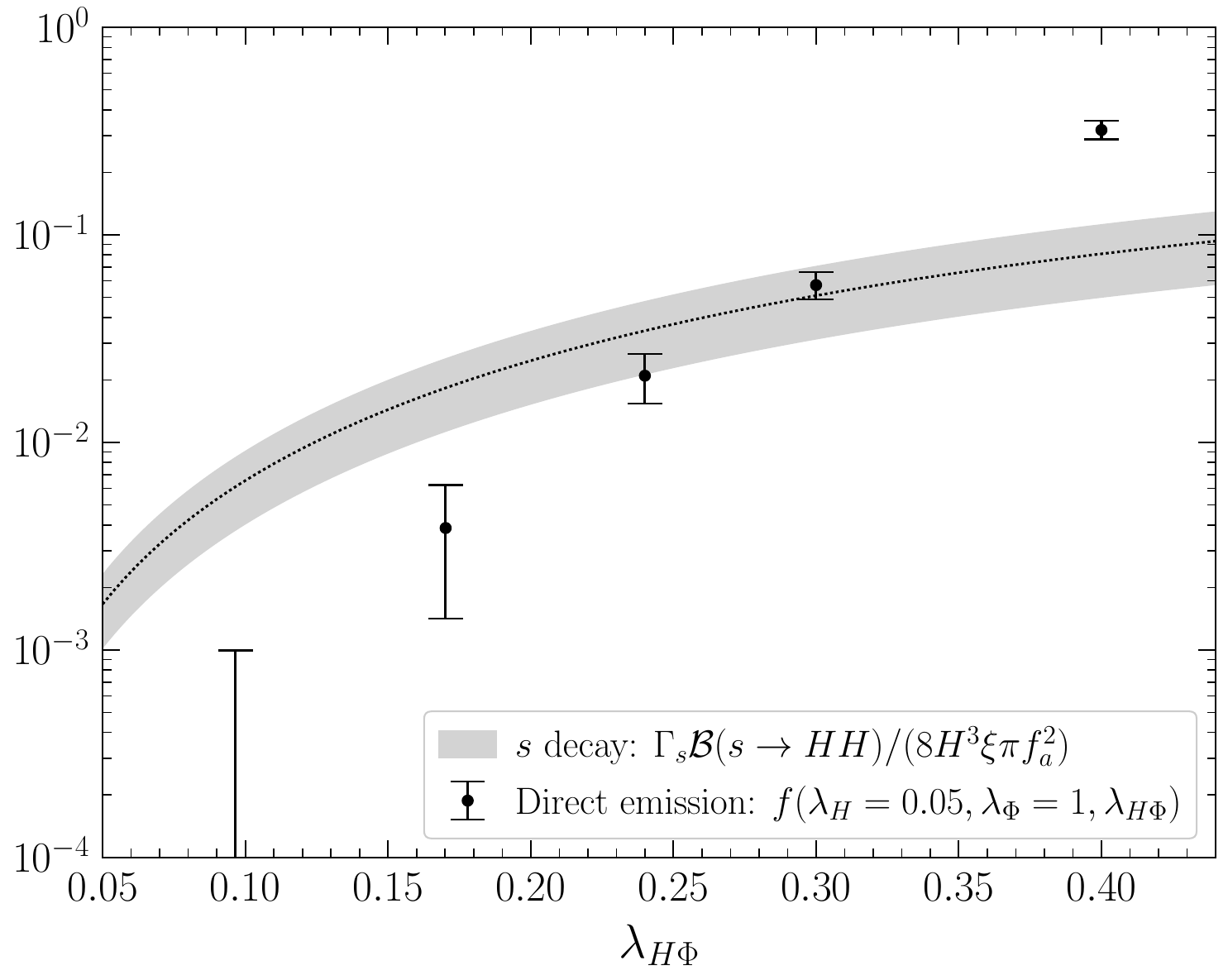}
\includegraphics[width=0.9\linewidth]{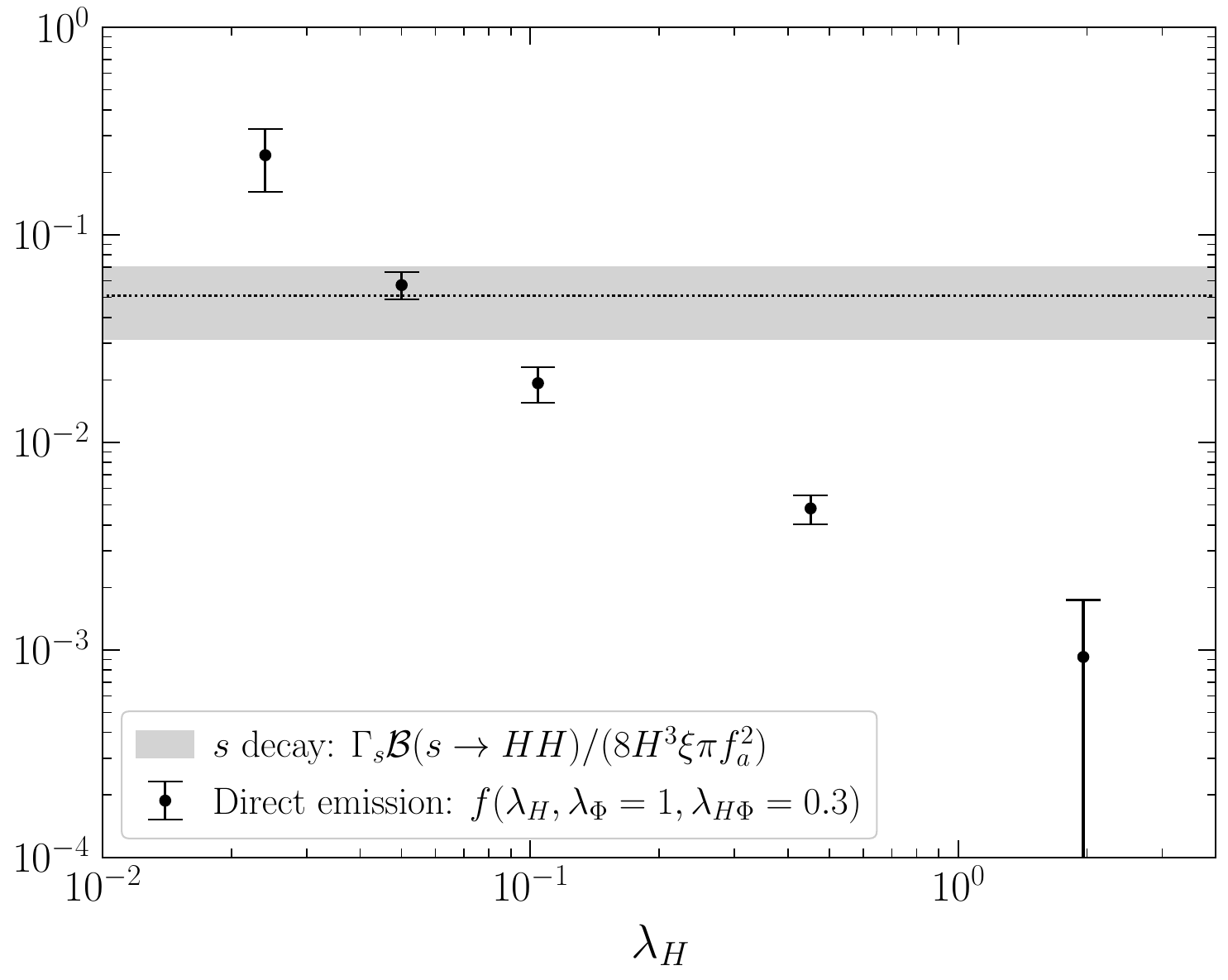}
\caption{
Comparison between direct Higgs emission, $f(\lambda_H, \lambda_\Phi, \lambda_{H\Phi}) = \Gamma_H / (8 H^3 \xi \pi f_a^2)$, and Higgs production through radial mode decay, $\Gamma_s \mathcal{B}(s\rightarrow HH) / (8 H^3 \xi \pi f_a^2)$ with $\mathcal{B}(s\rightarrow HH) = 1 / [1 +\lambda_\Phi^2 / (2 \lambda_{H \Phi}^2)]$, as a function of $\lambda_{H\phi}$ (top) and $\lambda_H$ (bottom). The respective other parameter has been fixed to our fiducial choice, $\lambda_{H\Phi}=0.3$ and $\lambda_{H}=0.05$. The production mode through $s$ decay has been extracted from the simulation without Higgs feedback in Sec.~\ref{sec:prompt_radial_mode_emission} where the grey band corresponds to the $1\sigma$ uncertainty.}
\label{fig:emission_h}
\end{center}
\end{figure}

 For each simulation we compute $\Gamma_H$ through an analogous procedure to that used in Sec.~\ref{sec:prompt_radial_mode_emission}, restricting to $\log(m_s / H) > 5$.  In particular, we extract $\Gamma_H$ analogously to $\Gamma_s$ and $\Gamma_a$ using~\eqref{eq:rho_X_sim} with $X(x)=h(x)$ and~\eqref{eq:Gamma_X_sim} with $z=4$.  In Fig.~\ref{fig:gamma_h} we illustrate the example data points for $\Gamma_H$ for our fiducial simulation, along with the resulting linear fit to extract $f$, as defined in~\eqref{eq:Gamma_H_ansatz}. 
In Fig.~\ref{fig:emission_h} we show $f(\lambda_H,\lambda_\Phi,\Lambda_{H \Phi}) = \Gamma_H / (8 H^3 \xi \pi f_a^2)$ as a function of $\lambda_{H\Phi}$ and $\lambda_{H}$ for a sequence of five simulations. All other couplings are fixed to our fiducial choice. 

Note that there are two distinct ways that the axion strings may create Higgs radiation: (i) the strings may radiate high-energy radial modes, which decay quantum mechanically to Higgs pairs with branching ratio ${\mathcal B} \approx 2 \lambda_{H \Phi}^2 / \lambda_\Phi^2$, in the limit $\lambda_{H\Phi} \ll \lambda_\Phi$;
and (ii) the strings directly radiate classical Higgs radiation, with rate given in~\eqref{eq:Gamma_H_ansatz}. 
Both contributions should be accounted for when computing the energy injection due to axion strings. In Fig.~\ref{fig:emission_h} we show, in addition to $f(\lambda_H,\lambda_\Phi,\lambda_{H \Phi})$, the radial-mode emission rate times the branching ratio ${\mathcal B}$ of the radial mode to Higgs particles, though for the purpose of illustration, we neglect the back-reaction of the Higgs field on the PQ field when calculating $\Gamma_s$. That is, in Fig.~\ref{fig:emission_h} we use the PQ-only simulation results when computing $\Gamma_s$.  Still, this comparison suggests that while the direct emission of Higgs particles may dominate in certain regions of parameter space, the Higgs sheaths do not parametrically increase the energy-injection relative to what would naively be estimated based off of radial-mode emission alone.

\section{Observational constraints on axion strings from SM radiation}
\label{sec:obs_cons}

We now consider the observational constraints that arise from radial-mode-induced radiation into SM final states for string networks that survive until at least the epoch of BBN.  Note that while we frame the discussion in terms of radial mode emission and decay, the following arguments also apply to high-energy Higgs emission from Higgs sheaths.  A summary of all of the upper limits on $f_a$ derived in this section is provided in Tab.~\ref{tab:summary}.

\begin{table}[]
\centering
\renewcommand{\arraystretch}{1.27}
\begin{tabularx}{0.366\textwidth}{|c|c c c c|}
\hline
probe & $z$ &  $m_a^{\rm max}$ [eV] & $\xi(z_*)$ & $\sqrt{{\mathcal B}} f_a^{\rm max}$ [GeV]      \\ \hline
BBN & $\sim$$10^{6}$ & $10^{-23}$ & 25 & $10^{14}$ \\
CMB & $\sim$$600$ & $10^{-29}$ & 30 & $2.2 \times 10^{12}$ 
\\
$\gamma$-ray & 0 &  $10^{-33}$ & 30 & $9.2 \times 10^{11}$
   \\ \hline
\end{tabularx}
\caption{A summary of the observational constraints derived in this work on axion strings from the high-energy SM radiation they emit in the scaling solution. The BBN and CMB constraints arise from primordial energy injection at these epochs, while the gamma-ray constraint is from present-day gamma-ray searches. We provide the approximate redshift of the constraint, the maximum axion mass $m_a^{\rm max}$ for the constraint to apply (though considering domain wall formation more non-trivial constraints could apply at higher axion masses), and the upper bound on the decay constant $f_a^{\rm max}$ for a given branching ratio ${\mathcal B}$ of the radial mode to SM final states.  More formally all of the upper bounds apply to $f_a / \sqrt{ {0.33 \over c} {1 \over {\mathcal B}} {25 \over \xi(z_*) }}$, but for this table we fix $c = 0.33$, which parameterizes the energy injection into radial modes, and the number of strings-per-Hubble volume $\xi(z_*)$ at the values given in the table.   Note that these limits are quoted in terms of radial mode production and decay but also apply, with the appropriate modifications, to the scenario in which the strings directly produce high energy Higgs radiation, as discussed in Sec.~\ref{sec:axion-Higgs strings}. 
}
\label{tab:summary}
\end{table}

As discussed in Sec.~\ref{sec:radial_decay}, radial modes decay to SM final states with model-dependent branching ratio ${\mathcal B}$. The SM particles, which may be, {\it e.g.}, Higgs boson pairs or heavy gauge bosons, then subsequently undergo a sequence of prompt decays to produce a spectrum of SM final states, with characteristic energy scale given by the radial mode mass $m_s$.  At the epochs of BBN and CMB decoupling those SM particles rapidly deposit their energy in the primordial plasma through scattering processes. Note that this also applies to neutrino final states. High-energy neutrinos ({\it e.g.}, $E \sim 10^{12}$~GeV) scatter off the primordial plasma at rates much faster than Hubble at the epochs of BBN and reionization (see~\cite{Formaggio:2012cpf}, for example, for a discussion of the high-energy neutrino cross-sections). Thus, for the purpose of the following discussions we do not differentiate between neutrino and non-neutrino SM final states.

We derive constraints on the axion decay constant associated with axion strings at three different cosmological epochs: (i) BBN, (ii) the dark ages between CMB decoupling and reionization, and (iii) today (redshift $z = 0$).  All of these constraints arise from injecting additional energy into the Universe from the string network through radial mode decay into SM final states. At a given redshift $z$ the energy injected per unit time per unit volume from radial mode decay is, referring to~\eqref{eq:Gamma_s}, \es{eq:energy_inj_string}{
\left. {\D E(z) \over \D t \D V} \right|_{ \rm string} = 8 \pi c\, f_a^2 \xi(z) H^3(z) {\mathcal B}  \,.
}  
The redshift dependence of~\eqref{eq:energy_inj_string} may be made more explicit by recalling that 
\es{eq:hubble}{
H^3 = H_0^3 \left[ \Omega_{\Lambda} + \Omega_{\rm m} (1+z)^3 + \Omega_{\rm rad} (1+z)^4\right]^{3/2} \,,
}
with $\Omega_\Lambda$ ($\Omega_{\rm m}$) ($\Omega_{\rm rad}$) the present day relative abundances of the cosmological constant (matter) (radiation) relative to the critical density $\rho_c$.

The expression in~\eqref{eq:energy_inj_string} is closely related to those for energy injection from DM decay and annihilation.  Moreover, constraints exist already from the epochs of BBN, CMB decoupling, and today on annihilating and decaying DM models. We may thus reinterpret these constraints in the context of radial mode radiation from strings.  The energy deposited per unit volume per unit time from DM decay and annihilation is, respectively,
\es{eq:energy_inj_DM}{
\left. {\D E(z) \over \D t \D V} \right|_{ \rm DM \, \, decay} &= \rho_{c, 0} \Omega_{\rm DM} (1+z)^3 \Gamma_{\rm DM \to {\rm SM}}  \,, \\
\left. {\D E(z) \over \D t \D V} \right|_{ \rm DM \, \, ann.} &= \rho_{c, 0}^2 \Omega_{\rm DM}^2 (1+z)^6 {\langle \sigma v \rangle \over m_{\rm DM}}  \,,
} 
where in the top line $\Gamma_{\rm DM \to {\rm SM}}$ is the DM decay rate to SM final states, $\Omega_{\rm DM}$ is the energy density fraction in DM today, and in the bottom line the DM with mass $m_{\rm DM}$ annihilates to the SM with velocity-averaged cross-section $\langle \sigma v \rangle$.

\subsection{Constraints from BBN}

During the radiation-dominated epoch, the energy deposition for string-induced radial mode decay and for DM annihilation scale the same with redshift, up to the logarithmic dependence of $\xi$ on $z$. Thus, we may determine the upper limit on $f_a$ by identifying
\es{}{
\left( {f_a \over M_{\rm pl}}\right)^2 = {9 \over 8 \pi c} {\Omega_{\rm DM}^2 \over \Omega_{\rm rad}^{3/2}} {1 \over {\mathcal B} \xi(z_*)} \left[ {\langle \sigma v \rangle \over m_{\rm DM}} H_0 M_{\rm pl}^2 \right] \,,
}
where $\xi(z_*)$ is the value at the epoch given by redshift $z_*$ where the DM annihilation constraint is evaluated.  Note that above $M_{\rm pl} \approx 2.4 \times 10^{18}$ GeV is the reduced Planck mass.

We now consider the constraints on axion strings from BBN by reinterpreting the BBN constraints on DM annihilation.  DM annihilation constraints during the epoch of BBN arise from two different mechanisms related to (i) hadronic energy injection, and (ii) photonic and leptonic energy injection. Hadronic energy injection may increase the neutron-to-proton ratio, which in turn increases the primordial $^{4}{\rm He}$ mass fraction~\cite{Reno:1987qw,Kawasaki:2004qu,Hisano:2009rc,Hisano:2008ti,Jedamzik:2009uy}. Electromagnetic energy injection, on the other hand, may photo-dissociate nuclei~\cite{Kawasaki:1994sc,Cyburt:2002uv,Kawasaki:2004qu,Hisano:2009rc,Jedamzik:2009uy}. For example, photo-dissociation of $^4{\rm He}$ may lead to the overproduction of $^3{\rm He}$.  The hadronic energy injection constraints and the electromagnetic energy injection constraints scale differently with DM mass: for masses $m_{\rm DM}$ well above a GeV, the electromagnetic constraints are for fixed $\langle \sigma v \rangle / m_{\rm DM}$, since they are constraints on the total injected energy, while the hadronic constraints scale as $\langle \sigma v \rangle / m_{\rm DM}^{3/2}$ since they are proportional to the number of injected nucleons~\cite{Henning:2012rm}. 
This implies that for very large DM masses the electromagnetic constraints are more powerful; thus, the electromagnetic constraints are the leading ones to use when constraining heavy radial mode decay.  

We adopt the $S$-wave DM annihilation constraint from~\cite{Hisano:2009rc} on the $^3{\rm He} / {\rm D}$ ratio, which states 
\es{}{
{\langle \sigma v \rangle \over m_{\rm DM}} H_0 M_{\rm pl}^2 \lesssim 2.4 \times 10^{-13} (0.5 / \epsilon_{\rm vis}) \,,
}
where $\epsilon_{\rm vis}$ is the fraction of annihilation energy that goes into photons and $e^{\pm}$.  For both $W^+W^-$ and $hh$ final states, $\epsilon_{\rm vis} \approx 1/2$, which is the value we adopt.\footnote{Note that ultra-high energy hadronic particles will rapidly -- on time scales much faster than Hubble -- cascade through scattering processes to produce low-energy particles, including photons and $e^{\pm}$. It is thus possible that our choice of $\epsilon_{\rm vis} = 0.5$ is conservative, since really $\epsilon_{\rm vis}$ should reflect the fraction of energy injected into visible final states after the cascade processes.}  Note that photo-dissociation only becomes efficient when $T \lesssim 0.3$ keV~\cite{Hisano:2009rc}, which implies that the BBN constraint requires $m_a \lesssim 2 \times 10^{-23}$ eV.  When the mass is less than this critical value, the decay constant is constrained to be less than:
\es{eq:f_a_BBN}{
f_a \lesssim 1 \times 10^{14} \, \, {\rm GeV} \sqrt{ {0.33 \over c} {1 \over {\mathcal B}} {25 \over \xi(z_*) }} \,.
}
Note that above we use $c \approx 0.33$ from Sec.~\ref{sec:PQ_sim_results}.
Recall from Sec.~\ref{sec:existing} that at this same epoch, constraints exist from not overproducing the observed value of $N_{\rm eff}$ from the axion radiation emitted by the string network, constraining $f_a \lesssim 9 \times 10^{14}$ GeV. The constraint in~\eqref{eq:f_a_BBN} is stronger than the $N_{\rm eff}$ constraint at this epoch for ${\mathcal B} \gtrsim 0.01$.

\subsection{Constraints from the CMB}

At redshifts $z \lesssim 3000$, energy injected into the SM plasma by annihilating or decaying DM, or radial mode decay from strings, changes the ionization history of the ordinary matter. These changes, in turn, change the CMB angular power spectrum, which is accurately measured and modeled under standard cosmology.  CMB constraints on annihilating and decaying DM have been extensively studied (see, {\it e.g.},~\cite{Kanzaki:2009hf,Slatyer:2009yq,Slatyer:2012yq,Galli:2013dna,Slatyer:2015jla,Slatyer:2015kla,Liu:2016cnk,Kawasaki:2021etm}).   
Recently, Ref.~\cite{Kawasaki:2021etm} found that for 
 $W^+W^-$ final annihilation states\footnote{ Note that $hh$ final states are expected to be similar since they have comparable visible energy deposition fractions.} the annihilation cross-section is constrained by the CMB power spectrum to be smaller than 
\es{}{
{\langle \sigma v \rangle \over m_{\rm DM}} H_0 M_{\rm pl}^2 \lesssim 8.8 \times 10^{-16} \,.
}
These constraints predominantly arise from energy injection at redshifts $z \approx 600$.  Note that the CMB constraints, as in the case of BBN, are a function of $\langle \sigma v \rangle / m_{\rm DM}$ since they constrain the total injected energy.

It is less straightforward to reinterpret the CMB angular power spectrum constraints in terms of strings.
This is because the injected energy from string emission does not redshift the same way as either energy injection from DM annihilation or decay, as the Universe goes through matter radiation equality into the epoch of matter domination.  Below, we carefully compute the upper limit on $f_a$ by performing a dedicated CMB power spectrum analysis for the specific form of the redshift-dependent energy injection appropriate for string emission. First, though, let us approximate the upper limit by assuming that the energy deposition happens instantaneously at $z_{\rm inj} = 600$. Then, we may translate the CMB constraints on DM annihilation\footnote{Using instead the limits from DM decay leads to compatible results.} to constraints on radial mode decay by equating~\eqref{eq:energy_inj_string} and $\left. {\D E \over \D t \D V} \right|_{\rm DM \, \, ann}$  from ~\eqref{eq:energy_inj_DM} at $z_{\rm inj}$.  This leads to the relation
\es{}{
\left( {f_a \over M_{\rm pl}}\right)^2 = {9 \over 8 \pi \, c} {\Omega_{\rm DM}^2 (1+z_{\rm inj})^{3/2} \over \Omega_{\rm m}^{3/2} {\mathcal B} \xi(z_{\rm inj})} \left[ {\langle \sigma v \rangle \over m_{\rm DM}} H_0 M_{\rm pl}^2 \right] \,,
}
where both $f_a$ and $\langle \sigma v \rangle$ represent the upper limits.  This then implies that
\es{eq:fa_CMB_guess}{
f_a \lesssim 1.06 \times 10^{12} \, \, {\rm GeV} \sqrt{{0.33 \over c} {1 \over {\mathcal B}} {30 \over \xi(z_{\rm inj})}    } \,.
}
Performing this calculation more carefully leads to a similar result, as we discuss below. 

Before describing our simulation framework for energy injection from axion strings, let us briefly comment on a crucial assumption in the above estimate. We assume that, apart from the redshift dependence, the energy-injection signal from axion strings has the same phenomenology as that from DM decay and annihilation. On the other hand, we know that in detail this assumption cannot be true, since the axion strings deposit their energy in narrow cylinders around the string cores, while DM annihilation and decay processes deposit energy in a relatively smooth fashion throughout the entire Hubble volume of interest. On small angular scales, we thus expect to see differences between the morphology of the DM-induced signals and the string-induced signals, for example as manifest through the angular power spectrum of the CMB. 

Roughly speaking, the co-moving separation between strings at $z_{\rm inj}$ is $( R(z_{\rm inj}) H(z_{\rm inj}) \sqrt{\xi(z_{\rm inj})})^{-1}$.  From this estimate we may compute the typical angular separation between strings $\theta$ by dividing the length scale above by the co-moving distance from today to the redshift $z_{\rm inj}$; 
then, identifying the angular multipole number $\ell$ through $\theta \sim \pi / \ell$, we estimate that only for $\ell \gtrsim 800 (z_{\rm inj}/600)^{1/2} \sqrt{\xi(z_{\rm inj}) /30}$ should the morphology of the strings signal differ from that of DM annihilation and decay.
The analyses we describe below make use of Planck data with $\ell \lesssim 2500$; thus, the high-$\ell$ modeling we perform likely underestimates the true anisotropy. We suspect that this implies the limits presented here are conservative, but this should be checked in the future with a dedicated analysis that accounts for the anisotropies on small angular scales arising from strings depositing their energy locally.  On the other hand, we verify that limiting the analysis to $\ell \leq 800$, corresponding to $z_{\rm inj}=600$, does not qualitatively change the sensitivity of the search, since the string-induced energy-injection signal predominantly appears at low $\ell$; in particular, we check that limiting $\ell \leq 800$ actually makes the upper limit stronger by $\sim$20\%, though at a level expected from statistical fluctuations alone.

\subsubsection{Dedicated CMB bound on axion strings}

To go beyond the approximation in~\eqref{eq:fa_CMB_guess} we perform a dedicated analysis that accounts for the unique redshift dependence of the axion-string-induced energy injection. 
We use \texttt{CLASS} to compute the CMB anisotropy with energy injection from radial mode decay, using the formalism described in \cite{Lucca:2019rxf}.  The steps of the computation are as follows: (i) the recombination histories of the matter temperature and ionization fraction of H are computed using a modified version of \texttt{RECFAST}~\cite{2011ascl.soft06026S}, which allows for an exotic energy injection; (ii) the evolution of matter and metric perturbations is then determined following cosmological perturbation theory, by solving (in Fourier space) the coupled linearized Einstein  plus fluid equations for the photons, baryons, DM, and neutrinos; (iii) finally, a line-of-sight integral is used to compute the angular power spectra for the temperature (TT), E-modes (EE), and their cross-spectrum (TE).

We now discuss the recombination modeling in more detail. In the following we  assume the rate of energy injection from radial mode decays is given by~\eqref{eq:energy_inj_string},  where the prefactor $c$ is extracted from our simulations, and we take a constant value of $\xi$ for simplicity.
As the stable radial mode decay products (high-energy photons, electrons, and positrons) cool, they deposit their energy in various channels, including into the ionization of $\mathrm{H}$ and $\mathrm{He}$, Ly-$\alpha$ excitations, free-streaming continuum photons, and heating of the intergalactic medium. These processes increase the ionization fraction of H. 
\texttt{CLASS} has built-in options to include exotic energy injection from, {\it e.g.}, DM annihilation/decay. 
We perform straightforward modifications to the thermodynamics module to include energy injection from radial mode decay from axion strings.

To obtain the fraction of energy deposited into each channel for $s \to WW$ decay,\footnote{Given the similar energy injection signatures for $WW$, $ZZ$, and $hh$ production at high energies, we only use the $WW$ final state in the analysis that follows.} we use \texttt{DarkHistory} \cite{Liu:2019bbm}, which models the evolution of the ionization fraction of H and the gas temperature and the cooling of decay products. Importantly, it includes the back-reaction of changes in the ionization fraction and gas temperature on the various energy-loss mechanisms.
As \texttt{DarkHistory} requires an initial electron and photon energy spectrum, but limits the mass of the annihilating particle to $10^5$ GeV and the highest energy bin to $\sim$5.4 TeV for both the electron and photon, we assume a box function for these spectra spanning a single energy bin centered at $1$ TeV with width $\sim$60 GeV for the electron and $\sim$80 GeV for the photon,
with the integrated energy obtained from \texttt{HDMSpectra}~\cite{Bauer:2020jay}, which computes decay spectra for GUT-scale masses. We verify that our results are insensitive to the width and mean energy of these box spectra, and thus, to a good approximation, depend only on the total energy injected; for example, injecting GeV energy particles instead of TeV energy particles leads to test statistic changes less than $\sim$10\%. As a consistency check, we also verify that using the deposition fractions from \texttt{DarkHistory} in \texttt{CLASS}  retrieves the electron ionization fraction computed by \texttt{DarkHistory} to the 10\% level. The deposition fractions calculated by \texttt{DarkHistory} are not expected to be accurate below this level. 

We set constraints on the injected energy using Planck 2018 CMB data~\cite{Aghanim:2018eyx}. We use \texttt{cobaya}~\cite{Torrado:2020dgo} to interface between \texttt{CLASS} and the Planck 2018 likelihood code \texttt{Plik} (described in detail in \cite{Planck:2019nip}). In particular, we use the likelihood \texttt{planck\_2018\_highl\_plik.[TT|TTTEEE]\_lite}, a version of the Planck 2018 high-$\ell$ $TT+EE+TE$ binned likelihood which is marginalized (in the Bayesian sense) over $47$ nuisance parameters that model the foreground. 

As noted in \cite{Padmanabhan:2005es}, modifications to the $TT$ spectrum from energy injection are almost degenerate with the primordial scalar spectral index $n_s$ and amplitude $A_s$. This degeneracy is broken by the polarization information in the $EE$ and $TE$ spectra. To account for the degeneracy we profile (in the frequentist sense) over the $\Lambda \text{CDM}$ parameters $(h, \Omega_b, \Omega_{\text{cdm}}, A_s, n_s)$, fixing all other cosmological parameters to their best-fit values from the Planck 2018 $TT+EE+TE+ \mathrm{low} \ell + \mathrm{low} E + \mathrm{lensing}$ data analysis.\footnote{ \url{https://wiki.cosmos.esa.int/planck-legacy-archive/images/b/be/Baseline_params_table_2018_68pc.pdf}} 
When profiling, we also fix the deposition fractions to their values calculated with the Planck 2018 best-fit  parameters for ease of computation. Note that for simplicity we do not profile over the optical depth $\tau$ and instead fix the reionization history to the default model of \texttt{DarkHistory}. As reionization only affects redshifts $z<20$, our constraint would only differ negligibly under small perturbations to the reionization history.

\begin{figure}[!t]
    \centering
    \includegraphics[width=1\columnwidth]{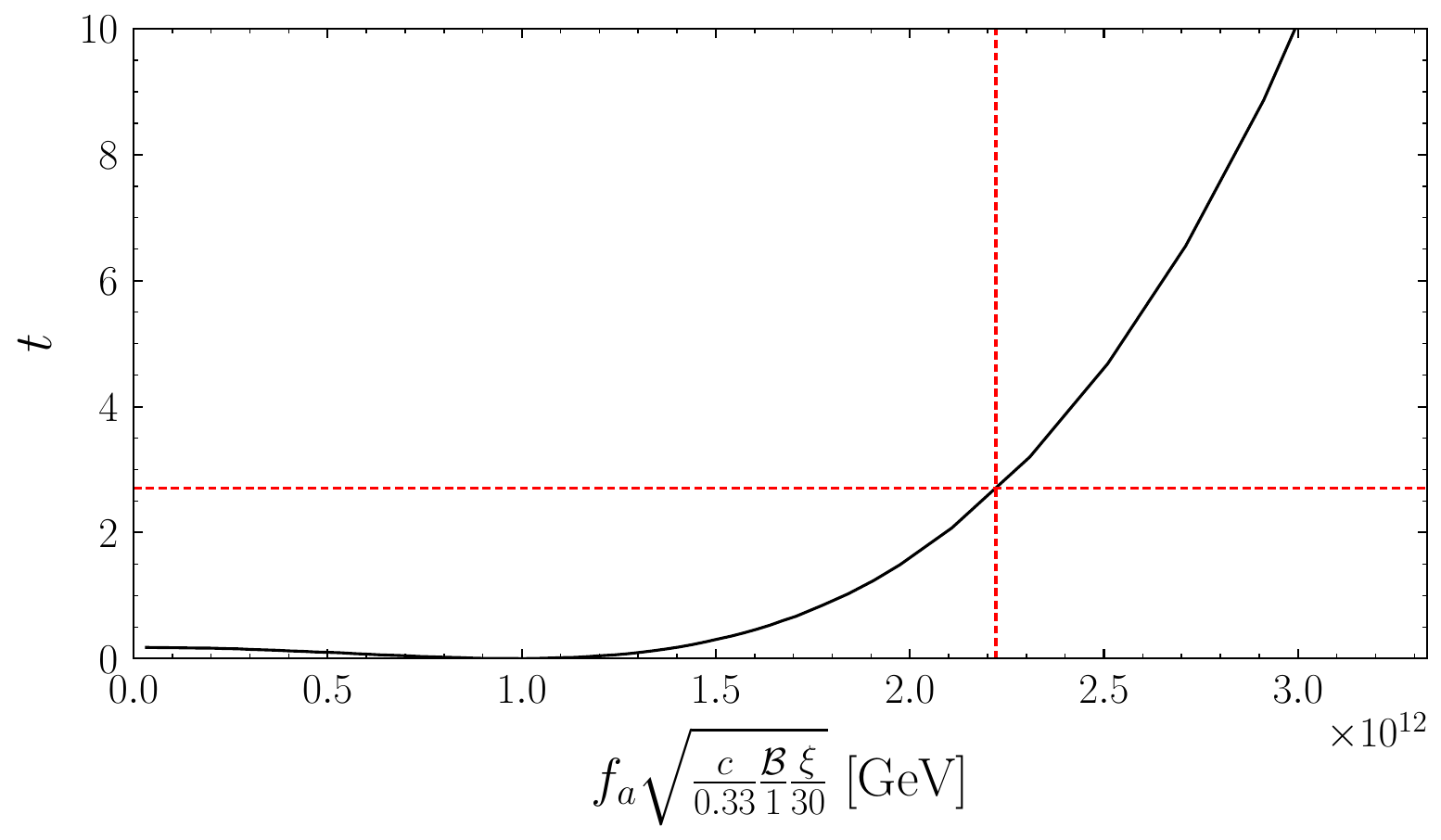} 
    \caption{The test statistic, defined in~\eqref{eq:t_fa}, for the profiled Planck 2018 TT + EE + TE + lensing likelihood for the CMB anisotropy with radial mode energy injection. The $95\%$ upper limit is shown in red. 
    }
    \label{fig:TS_cmb_injection}
\end{figure}

More precisely, we construct the profile likelihood ratio 
\es{}{
\lambda(f_a) = { p( {\bf d} | \{    { \hat{\hat{\bm \theta}}}_{\rm nuis} , f_a \} ) \over  p( {\bf d} | \{    { {\hat{\bm \theta}}}_{\rm nuis} , \hat f_a \}} \,,
}
where $p$ is the Planck partially-marginalized likelihood, given the data ${\bf d}$, the 5 $\Lambda$CDM nuisance parameters ${\bm \theta}_{\rm nuis}$, and the signal parameter $f_a$.  The quantities $\{    {\bm {\hat \theta}}_{\rm nuis} , \hat f_a \}$ represent those which maximize the marginalized likelihood, while $\hat{\hat{\bm \theta}}$ denotes the nuisance parameters that maximize the likelihood at fixed $f_a$. We then compute the test statistic 
\es{eq:t_fa}{
t(f_a) = - 2\log \lambda(f_a) \,,
}
which is illustrated in Fig.~\ref{fig:TS_cmb_injection}. We invoke Wilks' theorem and set the one-sided 95\% upper limit as the value of $f_a > \hat f_a$ for which $t(f_a) \approx 2.71$ (see~\cite{Safdi:2022xkm} for details).  Note that this analysis is formally a hybrid Bayesian-frequentist analysis, since the likelihood $p$ has been marginalized over the 47 foreground nuisance parameters. Additionally, note we find no evidence in favor of the axion model, given that the test statistic difference between the best-fit point $\hat f_a$ and the null hypothesis ($f_a = 0$) is much less than unity.
We find the 95\% one-sided upper limit to be
\es{eq:CMB_bound_CLASS_rad_dom}{
f_a \le 2.2 \times 10^{12} \, \, {\rm GeV} \sqrt{{0.33 \over c} {1 \over {\mathcal B}} {30 \over \xi(z_{\rm inj})}    } \,,
}
Comparing to~\eqref{eq:fa_CMB_guess} we see that this upper limit is similar to the naive estimate based off of translating the DM annihilation limit.  

There are two caveats related to the bound in~\eqref{eq:CMB_bound_CLASS_rad_dom} that are important to consider related to $\xi$. First, we note that the energy injection does not take place instantaneously at $z_{\rm inj} \sim 600$ but rather over a range of redshift values near this characteristic redshift. Thus, taking a constant number of strings-per-Hubble patch is not completely correct, but in practice since $\xi$ varies logarithmically with time we verify that this approximation is valid to the precision quoted. Second, and more importantly, the time specified by $z_{\rm inj}$ is within the epoch of matter domination, and the string network has a difference scaling solution during matter domination than during radiation domination, as we discuss in App.~\ref{strings:matter_domm}. 
 In particular, if we assume the string network follows the radiation-epoch scaling solution until $z_{\rm in}$ then we expect a characteristic value $\xi(z_{\rm inj}) \sim 30$, while using the matter-dominated scaling solution in App.~\ref{strings:matter_domm} we would infer $\xi(z_{\rm inj}) \sim 7$.
 In practice, we expect $\xi(z_{\rm inj})$ to be between these two values, since $z = 600$ is only slightly below matter-radiation equality. Computing $\xi(z_{\rm inj})$ directly through simulations is difficult because of the large $\log(m_s / H)$ values where matter-radiation equality occurs. Thus, we simply note that the pre-factor in~\eqref{eq:CMB_bound_CLASS_rad_dom}, accounting for the $\xi(z_{\rm inj})$ dependence as well, may be as large as  $\sim 4.5 \times 10^{12}$ GeV if we use the lower bound $\xi(z_{\rm inj}) \gtrsim 7$.

\begin{figure*}[!htb]
\centering
\includegraphics[width=\textwidth]{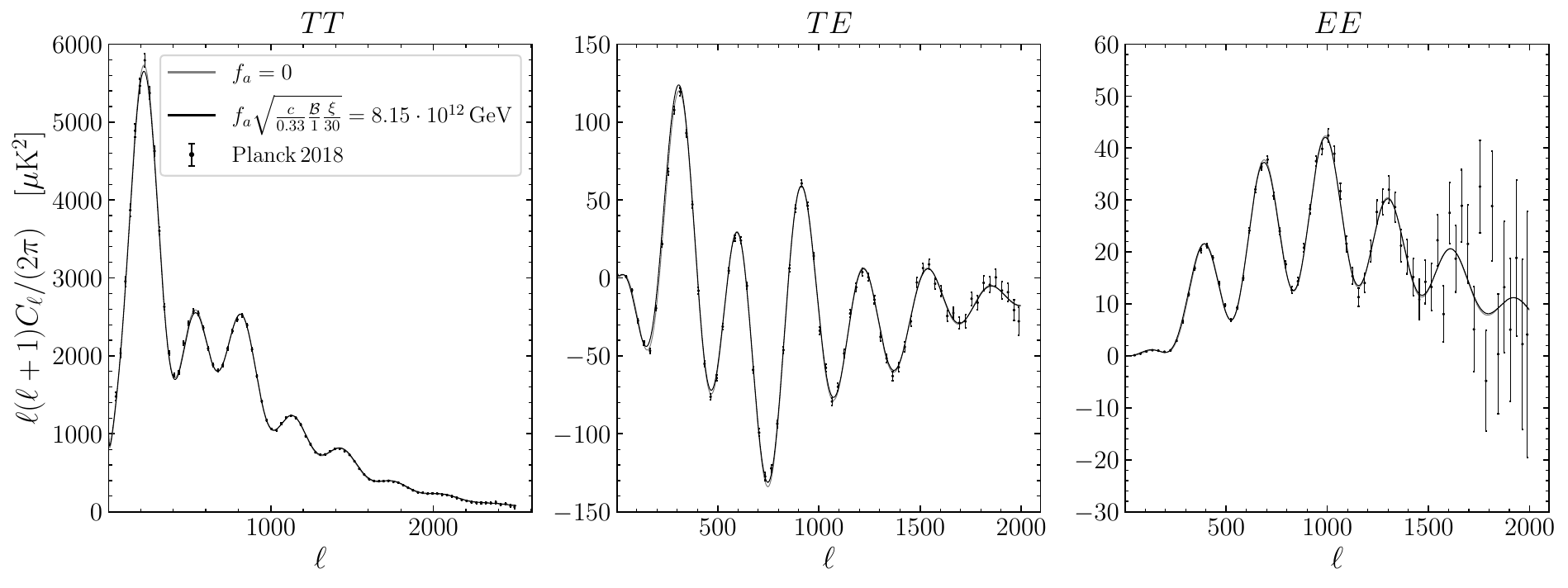}
\caption{
We illustrate the binned CMB anisotropy data ($TT$, $TE$, and $EE$) from the Planck 2018 Data Release on top of the best-fit model for the CMB anisotropy without any energy injection computed from \texttt{CLASS} ($f_a = 0$). We also show the best-fit model, profiled over nuisance parameters, with $f_a \sqrt{ {c \over 0.33} { {\mathcal B} \over 1} {\xi \over 30 }} = 8.15 \times 10^{12}$ GeV. 
}
\label{fig:cmb_ps_from_axion_string_inj}
\end{figure*}

To help unpack this analysis in Fig.\ref{fig:cmb_ps_from_axion_string_inj} we illustrate the angular power spectra for an energy injection signal from radial mode decay corresponding to $f_a \sqrt{ {c \over 0.33} { {\mathcal B} \over 1} {\xi \over 30 }} = 8.15 \times 10^{12}$ GeV.
We observe a suppression of the power spectrum which is stronger at smaller scales.

\subsection{Constraints from present-day gamma rays}

Additional constraints appear if one assumes that the network persists until $z = 0$ ($m_a \lesssim 10^{-33}$ eV). 
In this case, we may reinterpret the results of searches for extragalactic DM decay.
 High-energy gamma-rays and $e^{\pm}$ from DM decay are reprocessed to lower energy gamma-rays through a cascade of electron-positron pair production and inverse Compton scattering off of background radiation fields, and for high $m_s$, well above the PeV scale, the spectrum of reprocessed gamma-rays observed on Earth approaches a universal spectrum that peaks, in terms of the flux $E_\gamma^2 {d\Phi_\gamma \over dE_\gamma}$ (units of GeV/cm$^2$/s/sr), between 10 and 100 GeV~\cite{Murase:2012xs}.\footnote{Note that we approximate the sub-100 GeV emission as isotropic, since the extragalactic gamma-ray background limits do not incorporate spatial information; a potentially stronger and more accurate analysis, however, would incorporate the expected anisotropy. }    
 Ref.~\cite{Cohen:2016uyg} constrained $\tau \gtrsim 1.3 \times 10^{27}$ s for DM decay to $b \bar b$ for $m_{\rm DM} > 10^{11}$ GeV; considering that the upper limit only depends on the  energy injected into non-neutrino species, we may infer that the upper limit for $hh$ or $W^+W^-$ final states would be the same to within 10\%. Ref.~\cite{Blanco:2018esa} used more aggressive modeling of the extragalactic gamma-ray background to constrain $\tau \gtrsim  10^{28}$ s for $m_{\rm DM} \gtrsim 10^9$ GeV, for both $W^+W^-$ and $hh$ final states, with the limits insensitive at the less than 10\% level to $m_{\rm DM}$ for $m_{\rm DM} \gtrsim 10^9$ GeV~\cite{Cohen:2016uyg}.  Since the energy injection is dominated by decays with $z \ll 1$, we may translate these limits to limits on $f_a$ through  the relation
 \es{}{
 \left( {f_a \over M_{\rm pl}}\right)^2 = {3 \over 8 \pi\, c} {\Omega_{\rm DM} \over {\mathcal B} \xi(z=0)} {\Gamma_{{\rm DM} \to {\rm SM}} \over H_0} \,. 
 }
 where $\xi(z=0)$ is the value at $z = 0$.  Using the lifetime bound from~\cite{Blanco:2018esa} this then implies
 \es{}{
f_a \lesssim 9.2 \times 10^{11} \, \, {\rm GeV} \sqrt{{0.33 \over c} {1 \over {\mathcal B}} {30 \over \xi(z_{\rm inj})}    } \,.
}
This upper limit is marginally stronger than the CMB upper limit in~\eqref{eq:CMB_bound_CLASS_rad_dom}, though --- as we discuss further in App.~\ref{strings:matter_domm}, the number of strings-per-Hubble is likely more comparable to $\xi(z=0) \sim 7$ at this epoch.

\section{Discussion}
\label{sec:discuss}

In this work, we set strong constraints on axion-like particle strings that survive to temperatures at or below that of BBN.
We study the effects of primordial energy injection from the decays of massive radial modes, released during the evolution of the axion string network. The strength of the derived upper limits depends on how long the network persists and on the branching ratio of the radial modes to SM final states. For relatively generic branching ratios ${\mathcal B} \sim 0.1$ to SM final states, the upper limit from BBN (CMB) is around $f_a \lesssim 3 \times 10^{14}$ GeV ($f_a \lesssim 5 \times 10^{12}$ GeV).  These upper limits rely crucially on understanding how the axion-string network sheds energy into radial modes and directly into SM final states, which we study through a combination of analytic arguments and dedicated numerical simulations.

QCD axion string simulations predict that the QCD axion decay constant should be within a factor of a few of $10^{11}$ GeV in order to produce the correct DM abundance if the PQ symmetry is broken after inflation~\cite{Gorghetto:2020qws,Buschmann:2021sdq}, assuming a standard cosmological history.  It is thus also well motivated to consider axion-like particle strings with similar decay constants, as these axion-like particles may accompany the QCD axion in some realization of the axiverse paradigm. Interestingly, this region of parameter space should be probed by the next generation of CMB experiments~\cite{Chang:2022tzj}, making CMB probes an exciting future possible discovery channel for axion-like particles.

Lastly, we note that while this work focuses on global strings it is possible that some of the results may be relevant to local strings, such as strings in the Abelian-Higgs model.  The Abelian-Higgs model is obtained simply by taking the same Lagrangian used in this work to produce axion strings and gauging the $U(1)_{\rm PQ}$ symmetry with an abelian gauge field (see, {\it e.g.},~\cite{Vilenkin:2000jqa} for a review). After spontaneous symmetry breaking both the radial mode and the gauge field are heavy, with masses of order $f_a$.  The previously massless axion is now ``eaten" by the massive gauge field, such that there are no light degrees of freedom.

It is typically assumed that local string networks evolve by emitting gravitational-wave radiation, with the direct production of massive modes exponentially suppressed and thus not relevant for dynamics~\cite{Olum:1999sg}.  On the other hand, the question of whether heavy-state emission is truly exponentially suppressed is unresolved, with some works claiming that it is relevant for the network dynamics~\cite{Vincent:1997cx,Hindmarsh:2008dw,Hindmarsh:2017qff,Blanco-Pillado:2023sap}.  Our work concludes that for global strings the production of heavy states is only logarithmically suppressed relative to the production of massless states.  This may suggest that in the local string scenario the strings are able to produce heavy states with a similar efficiency as in the global case.  Heavy-mode production from local strings would have a number of important implications, including making such strings susceptible to energy injection constraints along the lines of those discussed in this work. Dedicated local-string cosmological simulations are needed, however, to understand to what extent the results found here for global strings carry over to local strings.

\section*{Acknowledgements}

{\it
We thank Joshua Foster, Anson Hook, Hongwan Liu, Andrew Long,  and Tracy Slatyer for helpful conversations.
M.B. was supported by the DOE under Award Number
DESC0007968.
J.B., Y.P., and B.R.S are supported in part by the DOE Early Career Grant DESC0019225.
S.K. is supported in part by the U.S. National Science Foundation (NSF) grant PHY-1915314 and the DOE contract DE-AC02-05CH11231.
This research used resources of the National Energy Research Scientific Computing Center (NERSC), a U.S. Department of Energy Office of Science User Facility located at Lawrence Berkeley National Laboratory, operated under Contract No. DE-AC02-05CH11231 using NERSC award HEP-ERCAP0023978.
}

\appendix
\section{KSVZ radial mode decay widths}
\label{KSVZ}

In this Appendix we compute the decay widths of the radial mode into various final states.
We focus on a KSVZ-type UV completion, for reasons explained in the main text.

The Lagrangian for the PQ field and the KSVZ fermions is given by,
\begin{align}
    {\cal{L}} = |\partial\Phi|^2 - V(\Phi) + \bar{Q}i\slashed{D} Q - (y_Q \bar{Q}_L Q_R \Phi + {\rm h.c.}),
\end{align}
with 
\begin{align}
    V(\Phi) = \lambda_\Phi (|\Phi|^2  - f_a^2/2)^2.
\end{align}
Here $Q_L,Q_R$ form a vector-like fermion which we take to be neutral under $SU(3)_c$.
They can, however, be charged under $U(1)_Y$ and $SU(2)_L$, where we denote $N_L=1(2)$ if they are singlets (doublets) under $SU(2)_L$.
In particular, we now consider the case where the KSVZ fermions are charged under $SU(3)_c \times SU(2)_L \times U(1)_Y$ as $(1, 2, 1/2)$.
We can parameterize the radial ($s$) and the axion ($a$) mode as,
\begin{align}
    \Phi = \frac{1}{\sqrt{2}}(f_a + s)e^{ia/f_a}.
\end{align}
Below the PQ breaking scale, $f_a$, we can write the interactions with the fermion as,
\begin{align}
    \mathcal{L} \supset \frac{y_Q}{\sqrt{2}}(f_a + s) e^{ia/f_a} \bar{Q}_L Q_R + {\rm h.c.}.
\end{align}
To remove the axion from this interaction, we may perform the standard anomalous rotation $Q_L\rightarrow e^{ia/(2f_a)} Q_L$ and $Q_R\rightarrow e^{-ia/(2f_a)} Q_R$.
After this rotation, the effective theory is given by
\begin{equation}
\begin{aligned}
\mathcal{L}\supset & \frac{1}{2}(\partial s)^2 + \frac{1}{2}(\partial a)^2 + \frac{1}{f_a} s (\partial a)^2 + \frac{1}{2}\frac{s^2}{f_a^2}(\partial a)^2 \\
& - \frac{\lambda_\Phi}{4} (4 f_a^2 s^2 + 4 f_a s^3 + s^4) \\
& + \bar{Q}i\slashed{D} Q - \left[\frac{y_Q}{\sqrt{2}}\bar{Q}_L Q_R (f_a + s) + {\rm h.c.}\right] \\
& + \frac{\alpha_2}{8\pi f_a}a W \tilde{W} + \frac{\alpha_Y}{8\pi f_a}a B \tilde{B}.  
\end{aligned}
\end{equation}
Thus the mass of the radial mode is given by $m_s = \sqrt{2\lambda_\Phi} f_a$.
From the above we can compute the relevant decay widths,
\begin{equation}
    \begin{aligned}
            \Gamma(s\rightarrow a a) & = \frac{1}{32\pi}\frac{m_s^3}{f_a^2}, 
            \\
            \Gamma (s \rightarrow \bar{Q}Q) & = N_L y_Q^2\frac{\sqrt{2 \lambda_\Phi} f_a}{16\pi}\left(1- {y_Q^2 \over \lambda_\Phi}\right)^{\frac{3}{2}},\\
            \Gamma(s\rightarrow W^+W^-) & = \frac{\alpha_2^2}{144\pi^3}\frac{m_s^3}{f_a^2},\\
            \Gamma(s\rightarrow ZZ) & = \frac{\alpha_2^2}{576\pi^3}\frac{m_s^3}{f_a^2}\frac{(c_w^4+s_w^4)^2}{c_w^4},\\
            \Gamma(s\rightarrow \gamma\gamma) & = \frac{\alpha_{\rm em}^2}{144\pi^3}\frac{m_s^3}{f_a^2},\\
            \Gamma(s\rightarrow \gamma Z) & = \frac{\alpha_2^2}{576\pi^3}\frac{m_s^3}{f_a^2}\frac{(c_w^2-s_w^2)^2s_w^2}{c_w^2} \,,
    \end{aligned}
\end{equation}
where $c_w \equiv \cos \theta_w$, $s_w \equiv \sin \theta_w$, $\theta_w$ is the Weinberg angle, and $\alpha_w \equiv g_2^2 / (4 \pi)$, with $g_2$ the coupling constant of $SU(2)_L$.

\section{Pre-evolution and adiabatic regime}
\label{sec:pre-evolution}
We follow the procedure described in~\cite{Gorghetto:2018myk} to avoid transient radial-mode excitations at small values of $\log(m_s/H)$. Instead of starting the simulation from a thermal initial state and having strings form dynamically by explicitly simulating the PQ phase transition we start the simulation in this procedure after the PQ phase transition from a pre-evolved initial state that already contains strings. This pre-evolved initial state is generated by evolving a thermal initial state within a modified physics scenario where strings have a constant width. This width can be tuned to match the string width at the intended starting time of the actual simulation, $\eta_i\approx 2.3$ ($\log(m_s/H)=2)$. Additionally, this scenario contains a moderate amount of Hubble friction, which allows us to evolve strings for far longer such that they have sufficient time to de-excite. This is achieved by changing the relation $R/R_1 = \sqrt{t/t_1}$ to $R/R_1 = t/t_1$ and forcing $m_s\propto 1/R$. These modified equations of motion read
\es{}{
\psi_i'' + \frac{3}{\eta}\psi_i' - \frac{\bar \nabla^2 \psi_i}{\eta^2\eta_i} + \frac{\eta_i^2}{\eta^2}\psi_i \left(|\psi|^2 - 1\right) &= 0,
}
where the dimensionless fields $\psi_i=\psi_{1},\psi_2$ are given by $\Phi=(\psi_1 + i \psi_2)f_a/\sqrt{2}$. 

The initial state for the pre-evolution stage is a thermal state with wavenumbers up to a certain threshold in each spatial direction, see~\cite{Buschmann:2019icd} for more details. We perform a total of nine different simulations that differ in the wavenumber threshold: Two initial states are statistically independent realizations using the first 13 wavenumbers, the other initial states are based on the first 10, 15, 18, 20, 25, 30, and 35 wavenumbers, respectively. Due to the existence of an attractor solution~\cite{Gorghetto:2018myk} the impact of our choice of threshold is marginal at sufficiently large $\log(m_s/H)$. The pre-evolution simulation is performed with $512^3$ grid cells as the resolution is not an issue here due to the constant string width. The simulation starts at $\eta=1$ and ends when the total string length is close to the attractor solution at $\eta_i$, $\xi(\eta_i)\approx 0.18$. 

Furthermore, as the underlying physics changes instantaneously when starting the main simulation from a pre-evolved state, we introduce a short adiabatic period between the two regimes. We do this by computing $\psi_{1,2}''$ in both scenarios and combining them, $\psi^{(1)''}_{1,2}(1-f) + f\psi^{(2)''}_{1,2}$, with a logistical function $f=1/(1+\exp[-10(\eta-2.8)])$.

\section{Axion-Higgs simulations}
\label{supl:AxionHiggsSims}
The equations of motion for the coupled PQ-Higgs system are derived analogously to the PQ-only case and read
\begin{equation}
    \begin{aligned}
        h_{i}^{\prime \prime}+\frac{2}{\eta} h_{i}^{\prime}-\bar{\nabla}^{2} h_{i} +\\
        h_{i}\eta^{2}\left(4\lambda_H |h|^2+\lambda_{H\Phi}|\psi|^2-2\mu_H^2\right)&=0,
    \end{aligned}
\end{equation}
along with 
\begin{equation}
    \begin{aligned}
        \psi_i^{\prime \prime}+\frac{2}{\eta} \psi_i^{\prime}-\bar{\nabla}^2 \psi_i + \psi_i\eta^2\left(\lambda_{H\Phi}|h|^2 + |\psi|^2-1\right) =0.
    \end{aligned}
\end{equation}
Here, given the  $\sim 12$ order of magnitude hierarchy between the SM Higgs VEV and $f_a$, we set the Higgs VEV far away from strings to zero.

While we chose a simulation volume that is slightly larger than that in our PQ-only simulation it is nevertheless not large enough to avoid the necessity of pre-evolving the initial state to mitigate the effects of transient oscillations of the string cores.  The corresponding pre-evolution equations of motion are derived analogously to the PQ-only simulation described in App.~\ref{sec:pre-evolution} and read
\begin{equation}
\begin{aligned}
h_i'' + \frac{3}{\eta}h_i' - \frac{\bar \nabla^2 h_i}{\eta^2\eta_i} +\\ \frac{\eta_i^2}{\eta^2}h_i \left(4\lambda_H |h|^2+\lambda_{H\Phi}|\psi|^2-2\mu_H^2\right) &= 0,
\end{aligned}
\end{equation}
\begin{equation}
\begin{aligned}
\psi_i'' + \frac{3}{\eta}\psi_i' - \frac{\bar \nabla^2 \psi_i}{\eta^2\eta_i} + \frac{\eta_i^2}{\eta^2}\psi_i \left(\lambda_{H\Phi}|h|^2 + |\psi|^2-1\right) &= 0.
\end{aligned}
\end{equation}
Our pre-evolution procedure is identical to that described in App.~\ref{sec:pre-evolution} but with an increased static grid size of $1024^3$ cells to accommodate the larger volume. The thermal initial state includes the first 18 wavenumbers in each spatial direction where the Higgs field is generated analogously to the PQ field but without an effective mass (see~\cite{Buschmann:2019icd} for more details).

We use the final state of the pre-evolution as our initial state for the main simulation starting at $\log(m_s/H)=2$. This state is taken as our coarse level but an extra refinement level with $\Delta x\rightarrow \Delta x/2$ is introduced whenever the string core width would be resolved by less than four grid sites on the finest level. This means by the end of the simulation we will have three refinement levels on top of the coarse level. A static lattice simulation would have needed $8192^3$ cells to match this dynamic range. In which part of our simulation volume the refinement level is placed is based on two criteria: (i) the location of string cores, and (ii) a data-driven convergence criterion. The refined region is re-adjusted frequently every $\Delta\eta=0.12$.

String cores are identified using the procedure outlined in~\cite{Fleury:2015aca}. We ensure the refinement region around string cores is large enough that even a string segment moving with the speed of light will always be at least an entire string width away from any coarse-fine boundary until the grid is readjusted.
This criterion ensures that the strings themselves are properly resolved at all times, however, the emission leaving the string may not be. To guarantee this emission is resolved appropriately as well we additionally employ a data-driven method that estimates the convergence at individual grid cells. 

The basic idea behind the data-driven technique is to independently evolve a cell at two different resolutions for a short period of time. The difference between the results informs us about the size of the numerical truncation error due to the finite resolution. When this difference gets too large it means that numerical convergence is locally bad at the current resolution and refinement is needed. In practice, this is done easily within an AMR framework as we are already evolving the field at different resolutions. That is, at the end of every time step and before level synchronization we can compare the results of, let us say, the coarse level and the first refinement level. If this difference $\Delta X$ exceeds a threshold $\tau$ the area around this cell will be covered by the second refinement level during the next regrid. In order to identify problematic cells on the coarse level, however, an even coarser level with half the resolution of the coarse level is required. This setup is known as a \textsl{self-shadow hierarchy}. The parameter $\tau$ is chosen empirically and we find $\tau=10^{-3}$ to work well for $X=\psi_i, h_i$, and $\tau=10^{-3}/\Delta \eta_\ell$ for $X=\psi_i^{\prime}, h_i^{\prime}$. 

Our integration scheme is the same as that of our PQ-only simulation. The simulations are performed on the NERSC Perlmutter GPU cluster and utilize 256 NVIDIA A100 GPUs and 64 AMD EPYC 7763 CPUs for about an hour per run.

\section{String density in matter-domination}
\label{strings:matter_domm}
To study the evolution of the string density at times after matter-radiation equality, we simulate the string network in a matter-dominated cosmology. The simulations are such that the radial mode acquires its broken VEV and strings form when the Universe is already matter-dominated. Of course, this does not correspond to the physical scenario of string network formation at the PQ phase transition in a radiation-dominated epoch, and the subsequent evolution of the Universe through the epoch of matter-radiation equality. However, the choice of initial conditions in the simulation is not important as here we are only interested in the scaling regime during the matter-dominated era. 

Neglecting radiation energy density, we have at late times $\Omega_{\rm m} + \Omega_\Lambda \approx 1$, and the energy densities of matter and cosmological constant are equal when the scale factor is $\eta_\mathrm{eq} = (\Omega_{\rm m}/\Omega_\Lambda)^{\frac{1}{3}}$. We perform two AMR simulations: 
(i) a matter-only simulation where we impose $\eta_\mathrm{eq}\to \infty$ and which ends at $\log(m_s/H) = 5.78$, 
and  (ii) a simulation entering the cosmological constant-dominated epoch 
with $\eta_\mathrm{eq} = 20$, which ends at $\log(m_s/H) = 4.71$. 
We use the same AMR simulation setup as for our axion-Higgs simulation in App.~\ref{supl:AxionHiggsSims}. However, as we are not interested in measuring any emission spectrum we can safely skip the pre-evolution procedure. Instead, we simulate explicitly through the PQ phase transition by starting at $\eta=0.1$ from a thermal initial state with the first 9 
wavenumbers included. 

Both simulations evolve the following equations of motion
\es{eq:pq_eom_with_cc_matter}{
&[\eta^4 +\eta_\mathrm{eq.}^{3}\eta]\psi_{i}^{\prime \prime}+[4\eta^{3}+2.5\eta_\mathrm{eq.}^{3}] \psi_{i}^{\prime} \\&-[1+\eta_\mathrm{eq.}^{3}]\left(\bar{\nabla}^{2}\psi_i
+\lambda \psi_{i}\left[\eta^{2}\left(|\psi|^2-1\right)+\frac{T_{1}^{2}}{3 f_{a}^{2}}\right]\right) \\&=0 \,,
}
where the thermal term ensures the PQ symmetry breaks early on in the simulation.  These are the Euler-Lagrange equations of the Lagrangian~\eqref{eq:PQ} in an FRW metric with Hubble parameter given by~\eqref{eq:hubble} and $\Omega_\mathrm{rad} =0$. We also define the dimensionless fields $\psi_i$ as in App.~\ref{sec:pre-evolution}.    

\begin{figure}[!htb]
\begin{center}
\includegraphics[width=0.9
\linewidth]{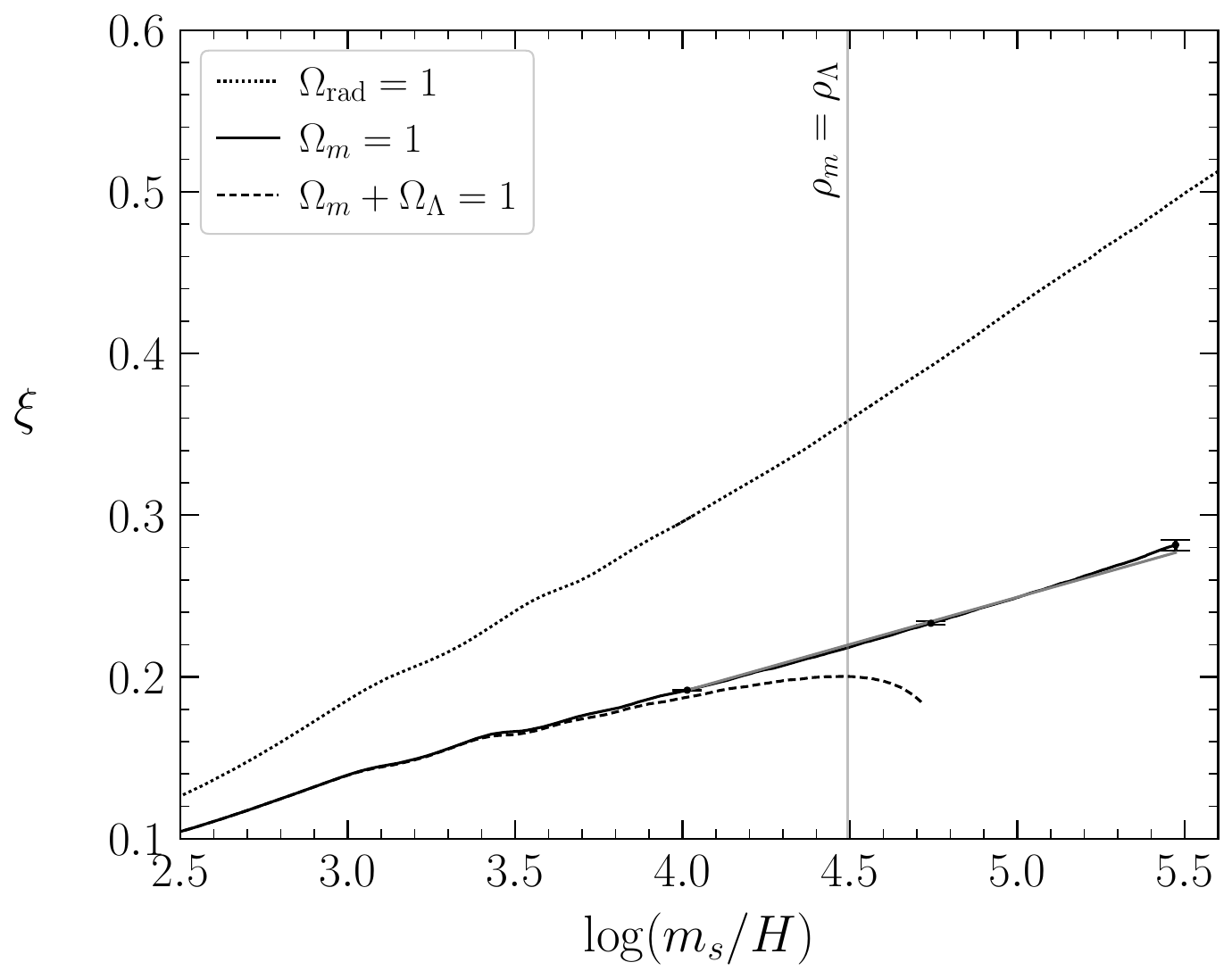}
\caption{Evolution of the string length per horizon $\xi$ in the matter-dominated (black solid) and radiation-dominated (black dotted) scenarios, and during a transition between matter-dominated and cosmological constant-dominated epochs (dashed). For the matter-dominated case, the model $\xi = c_0+c_1\log$ is fit to the $\xi$ data. 
}
\label{fig:xi_cosmo_comparison.pdf}
\end{center}
\end{figure}

\begin{figure}[!htb]
\begin{center}
\includegraphics[width=0.9
\linewidth]{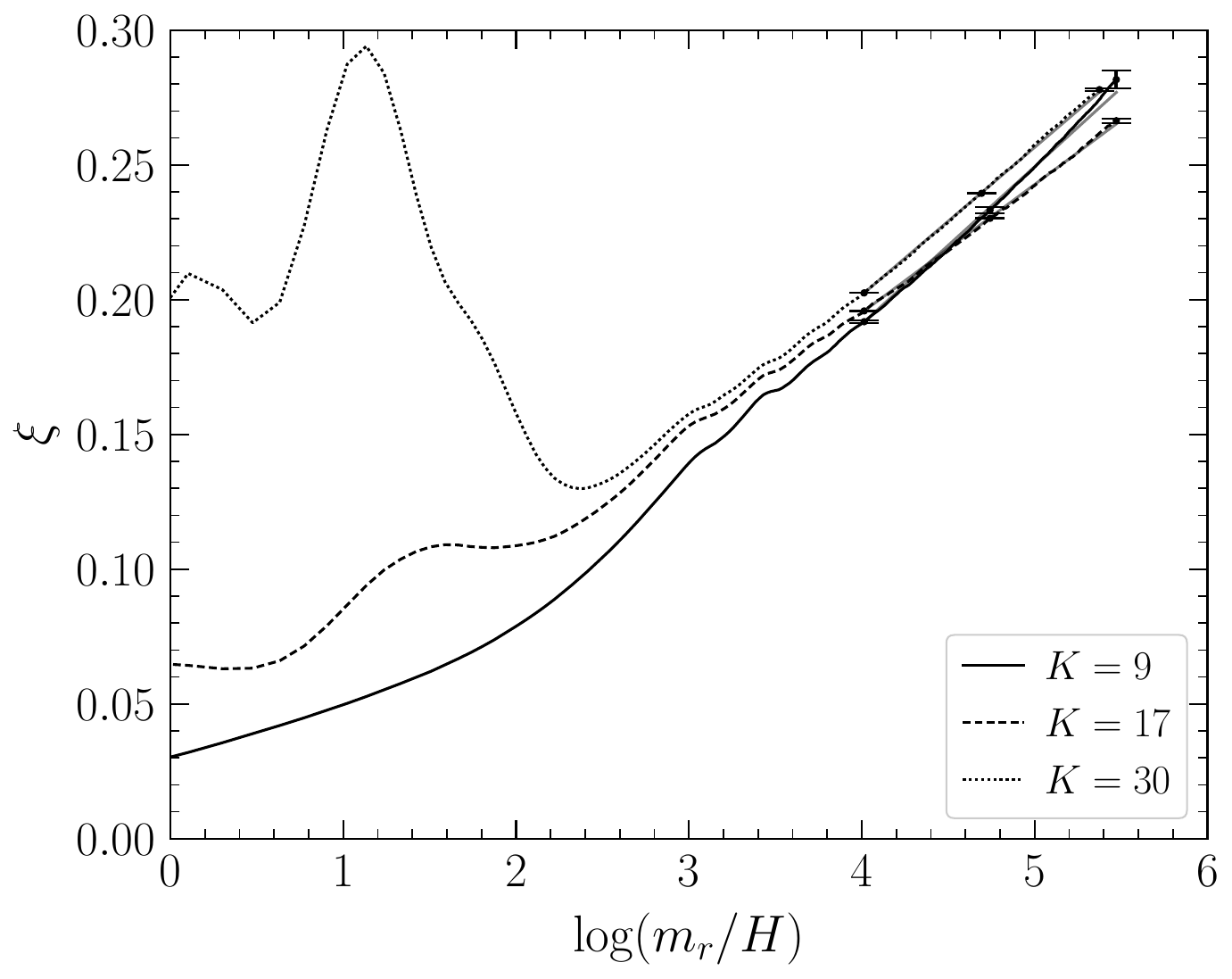}
\caption{String length per horizon in the matter-dominated scenario, varying the number $K$ of wavenumbers included in the initial thermal state. 
}
\label{fig:attractor_soln_test.pdf}
\end{center}
\end{figure}

Denoting the total string length inside a cube of side-length $D$ by $\ell(D)$, the string length per horizon is then defined by
\begin{align}
\xi \equiv \lim _{D \rightarrow \infty} \frac{\ell(D) t^2}{D^{3}} \,.
\label{eq:xi_def}
\end{align}
To evaluate Eq.~\eqref{eq:xi_def}, we can write the physical time $t$ as
\begin{align}
t = \frac{2}{3}f_a^{-1}\sqrt{1+\eta_\mathrm{eq}^3}\sinh^{-1}\left(\left(\frac{\eta}{\eta_\mathrm{eq}}\right)^{\frac{3} {2}}\right)\,,
\end{align}
using our convention for the Hubble scale for the simulations described in this Appendix,
\es{}{
H = f_a \sqrt{\Omega_{\rm m}/\eta^3 + \Omega_{\Lambda}} \,,
}
where by definition $H=f_a$ at $\eta=1$.
The string length per horizon is compared between the various cosmological scenarios in Fig.~\ref{fig:xi_cosmo_comparison.pdf}. 
For the matter-dominated simulation, we follow the method of Ref.~\cite{Buschmann:2021sdq} to fit $\xi = c_0 + c_1 \log(m_s / H)$ in the interval $\log(m_s/H) = [4,5.48]$
to find $c_1 = 0.0584 \pm 0.0013$. In the scenario where the Universe transitions from matter to cosmological constant domination, $\xi$ begins to drop exponentially, as expected, when the energy densities in matter and cosmological constant are equal.
The string length for the radiation-dominated scenario in Fig.~\ref{fig:xi_cosmo_comparison.pdf} is extracted from the AMR simulation discussed in ~\cite{Buschmann:2021sdq}, which was performed with $2048^3$ grid-cells.

In Fig.~\ref{fig:attractor_soln_test.pdf} we vary the number $K$ of wavenumbers included in the initial thermal state for the matter-dominated simulation. The proximity of $\xi$ at large values of $\log(m_s/H)$ as we vary $ K = 9, 17, 30$ indicates that, like in the radiation-dominated era, the string network approaches an attractor solution in the matter-dominated era as well.  The attractor solution implies that at late times the number of strings per Hubble patch is the same regardless of the initial condition. In our case, we find $c_1 =  0.0584 \pm 0.0013$,  $0.0476 \pm 0.0003$, and $0.0548 \pm 0.0003$, for the simulations with initial conditions $K=9,17,30$, respectively. Note that the error bars quoted above and shown in Fig.~\ref{fig:attractor_soln_test.pdf}  are statistical.  We may infer a systematic uncertainty from the variance in $c_1$ between the simulations with varying initial mode numbers, leading to the estimate $c_1 = 0.0584 \pm 0.0013_{\rm stat} \pm 0.0055_{\rm sys}$.  Note, also, that in reality the network evolves smoothly from the radiation dominated scaling solution to the matter dominated scaling solution around matter-radiation equality; depending on the cosmological epoch of interest, the network may not be well approximated by the scaling solution in either epoch but could take on intermediate values. 

\section{Single loop spectrum}
\label{sec:single_loop}

We simulate a single circular collapsing string with instantaneous radius $R$ to study the spectral shape of the resulting instantaneous axion emission spectrum; we find evidence that the instantaneous emission spectrum scales with axion momentum $k$ as $1/k$ between the characteristic frequency $k\sim 2\pi/R$ and $k\sim m_s$, supporting the consistent conclusions reached in~\cite{Saurabh:2020pqe}. 

Our simulation setup for this test is mostly the same as that of our axion-Higgs simulations but with an artificial initial state that creates a singular and perfectly circular string loop. To achieve this initial condition we generate the initial fields at rest with
\es{}{
\Phi_1(i,j,k)&=1-2/(1+e^{-0.05[|\hat x-N/2|-N/4]})\\
\Phi_2(i,j,k)&=\sin(2\pi k/N) \,,
}
in index space $\{ijk\}$ over the three Cartesian dimensions, with $N$ cells in each direction ({\it e.g.}, $i = 0,1, \cdots, N-1$) and $\hat x=(i, j, k)^T$. The result is a circular string loop with a radius of $N/4$ and approximately periodic boundary conditions. This construction has the advantage that the initial radius $R_0$ can be controlled by choosing a box length $L$. The exact details of the initial state such as the overall field amplitude are not crucial as we start our simulation in the unbroken phase at $\eta=0.1$. By explicitly simulating the PQ phase transition the string is generated dynamically.

\begin{figure}[!t]
\begin{center}
\includegraphics[width=0.9
\linewidth]{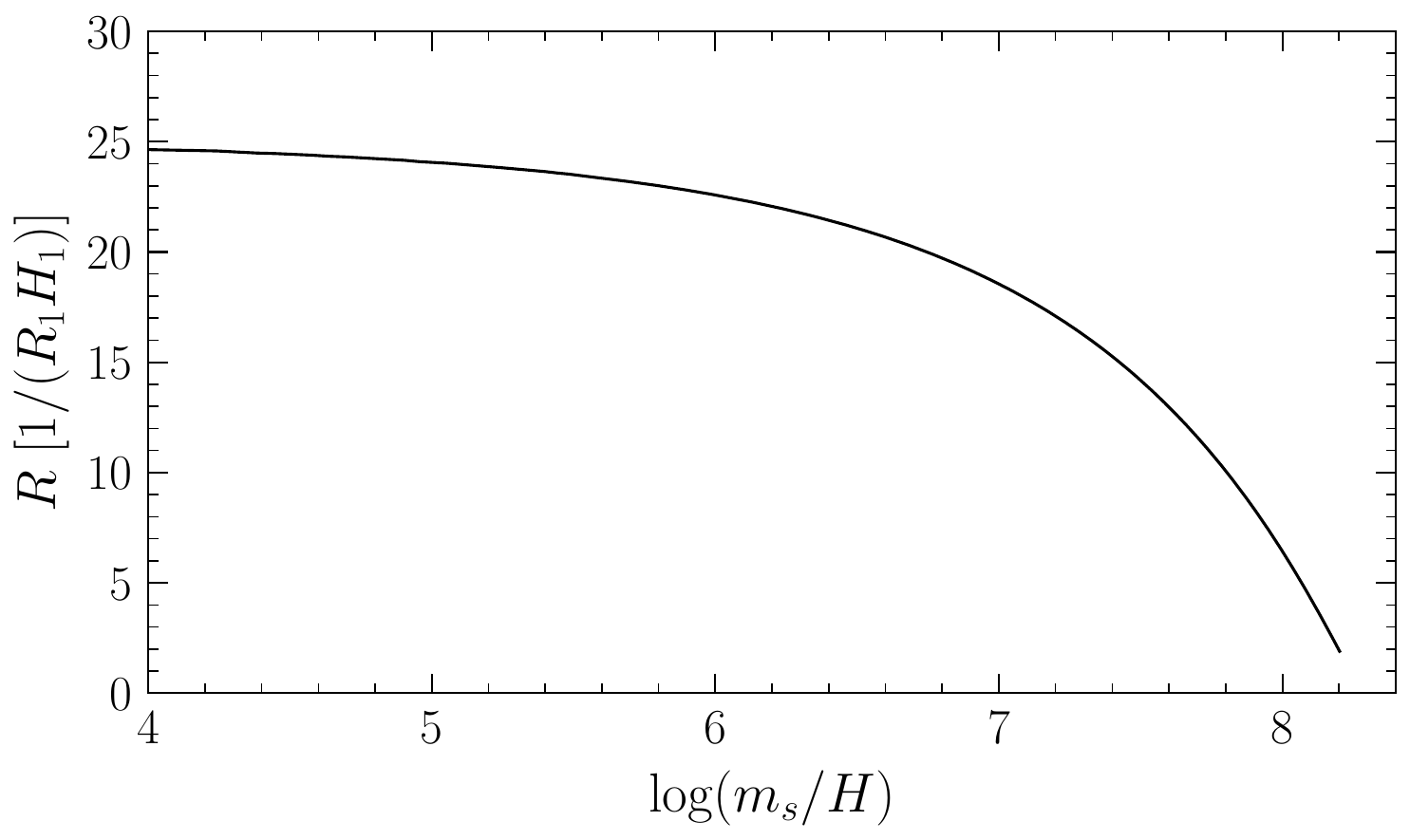}
\caption{Evolution of the string loop radius $R$ as a function of $\log(m_s/H)$ as measured in our simulation.
}
\label{fig:SingleLoopR}
\end{center}
\end{figure}

\begin{figure}[!t]
\begin{center}
\includegraphics[width=0.9
\linewidth]{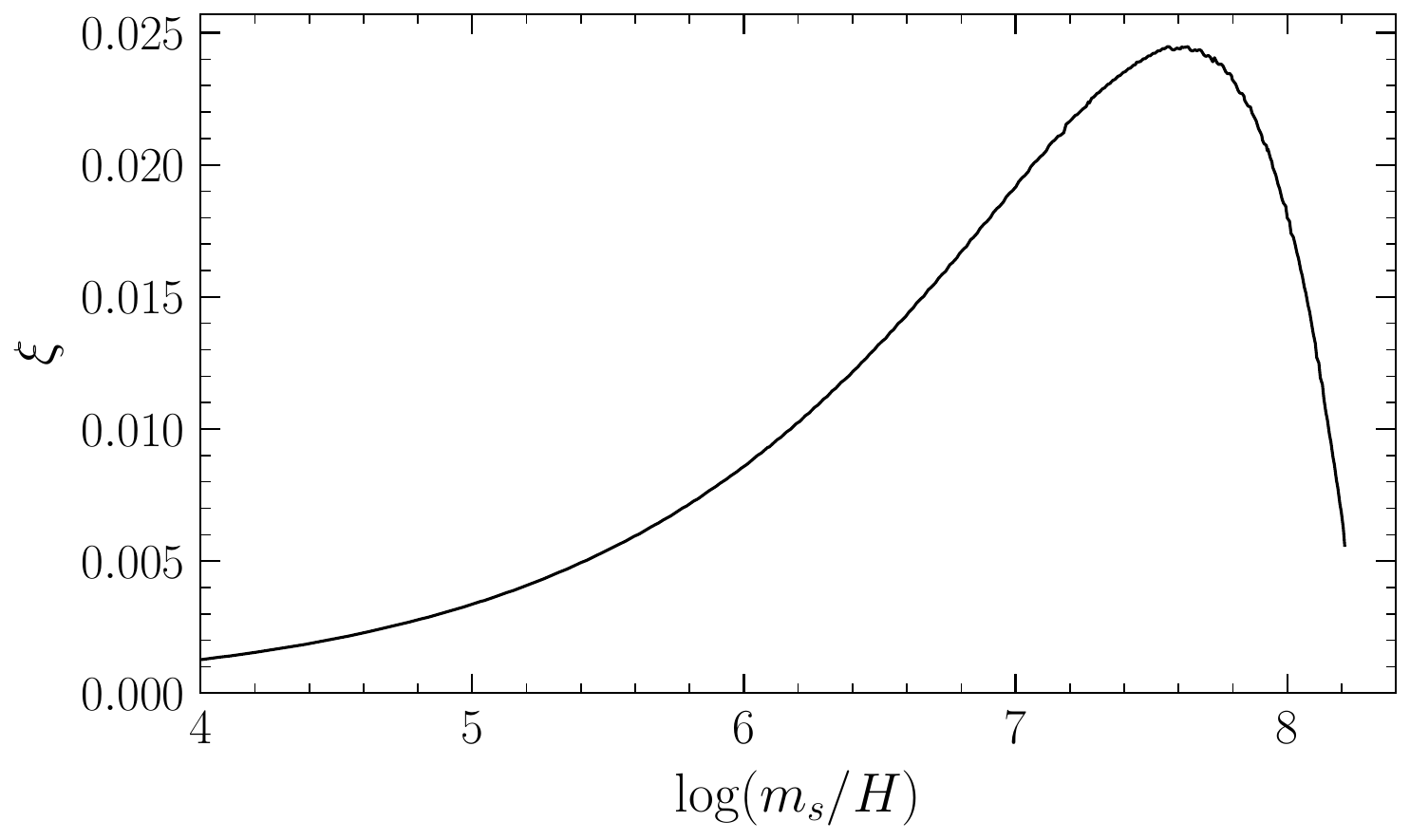}
\caption{String length $\xi$ as a function of $\log(m_s/H)$ for a circular decaying string as measured in our simulation. 
}
\label{fig:SingleLoopXi}
\end{center}
\end{figure}

\begin{figure}[!t]
\begin{center}
\includegraphics[width=0.9
\linewidth]{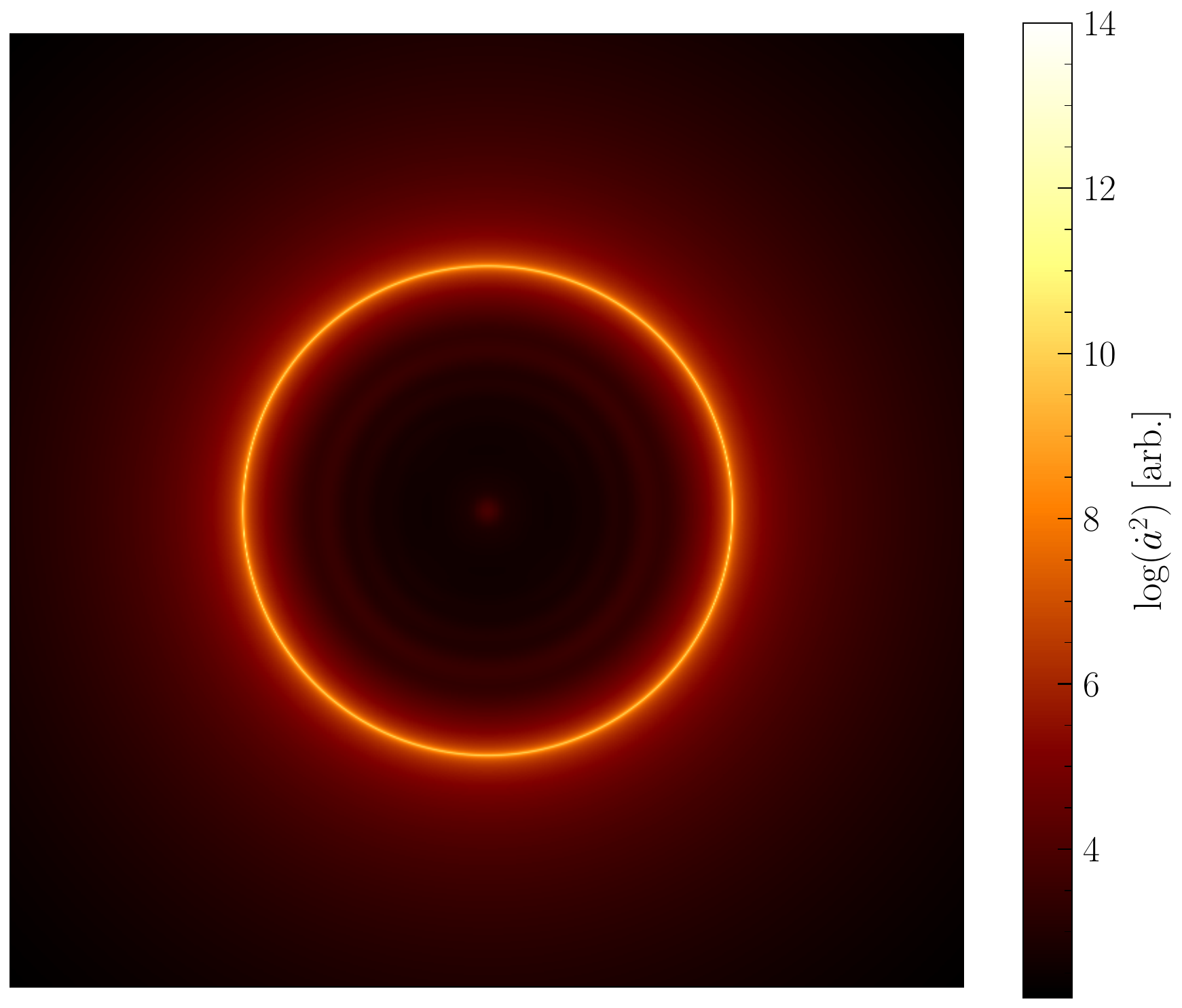}
\caption{2D projection of the axion energy density of a circular decaying string at $\log(m_s/H)\sim 7.6$. The radius of the string loop at this snapshot is is $R\sim 12.8/(R_1H_1)$, \textsl{i.e.} about 0.34 Hubble lengths. 
}
\label{fig:SingleLoop}
\end{center}
\end{figure}

Since we study a single string we can use a relatively aggressive AMR setup. Our coarse level consists of merely $1024^3$ grid sites, but this is compensated by five refinement levels at the end of our simulation at $\eta\sim50$ ($\log(m_s/H)\sim8.2$) to maintain at least four grid sites per string core width. On a static lattice such a simulation would require a grid of $32,768^3$ grid sites. We chose $L=100 / (R_1H_1)$ such that the initial string radius is $R_0=25 / (R_1H_1)$. The evolution of $R$ is shown in Fig.~\ref{fig:SingleLoopR} with corresponding string length $\xi$ in Fig.~\ref{fig:SingleLoopXi}. We perform the simulation on the NERSC Perlmutter GPU cluster and utilize 256 NVIDIA A100 GPUs and 64 AMD EPYC 7763 CPUs. An illustration of the axion energy density is shown in Fig.~\ref{fig:SingleLoop}. 

\begin{figure}[!t]
\begin{center}
\includegraphics[width=0.9
\linewidth]{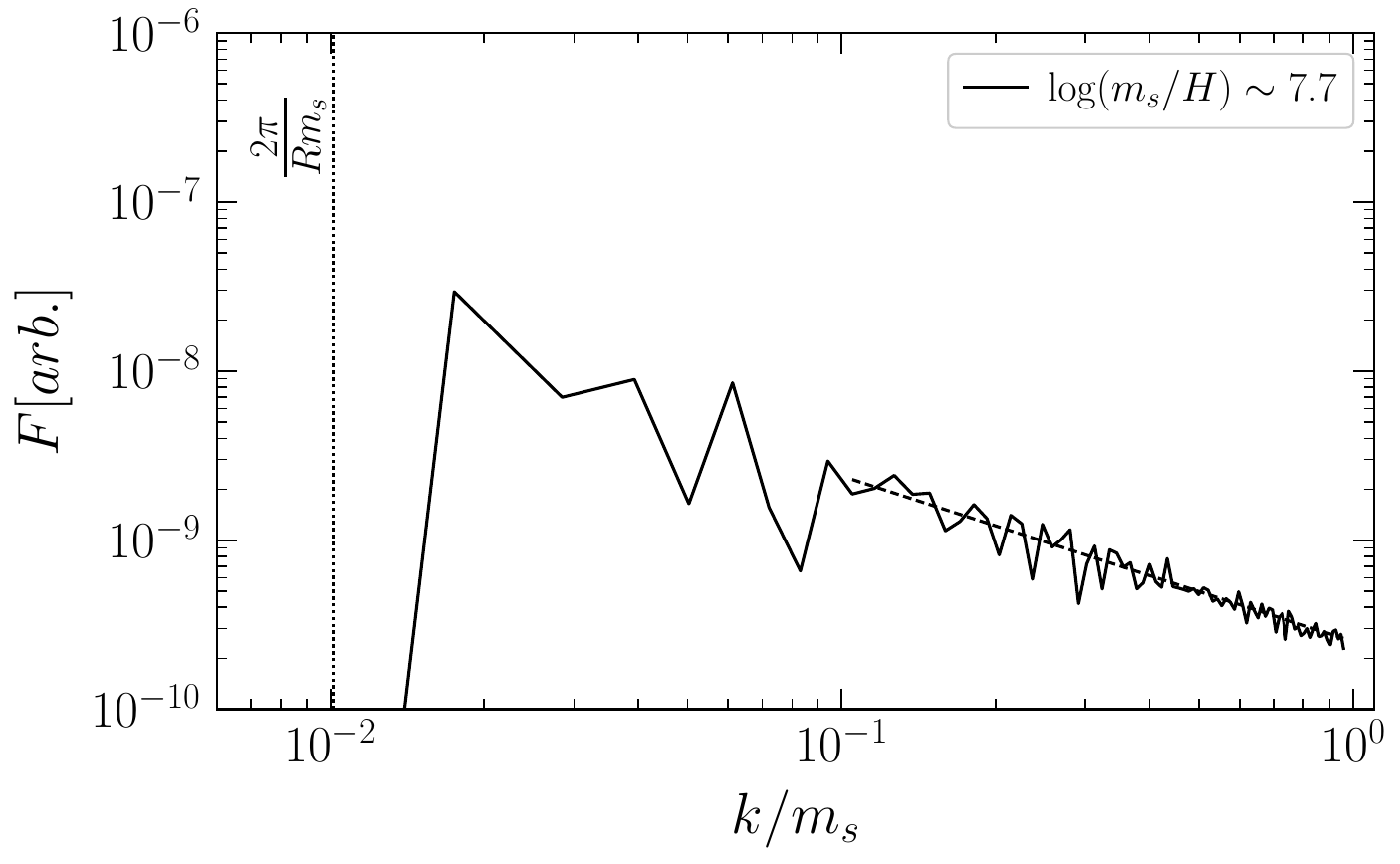}
\caption{Instantaneous axion emissions spectrum $F$ for a collapsing circular string loop. We perform a power law fit to the regime between $k\sim 0.1 m_s$ and $k\sim m_s$ (dashed line).
}
\label{fig:SingleLoopF}
\end{center}
\end{figure}

\begin{figure}[!t]
\begin{center}
\includegraphics[width=0.9
\linewidth]{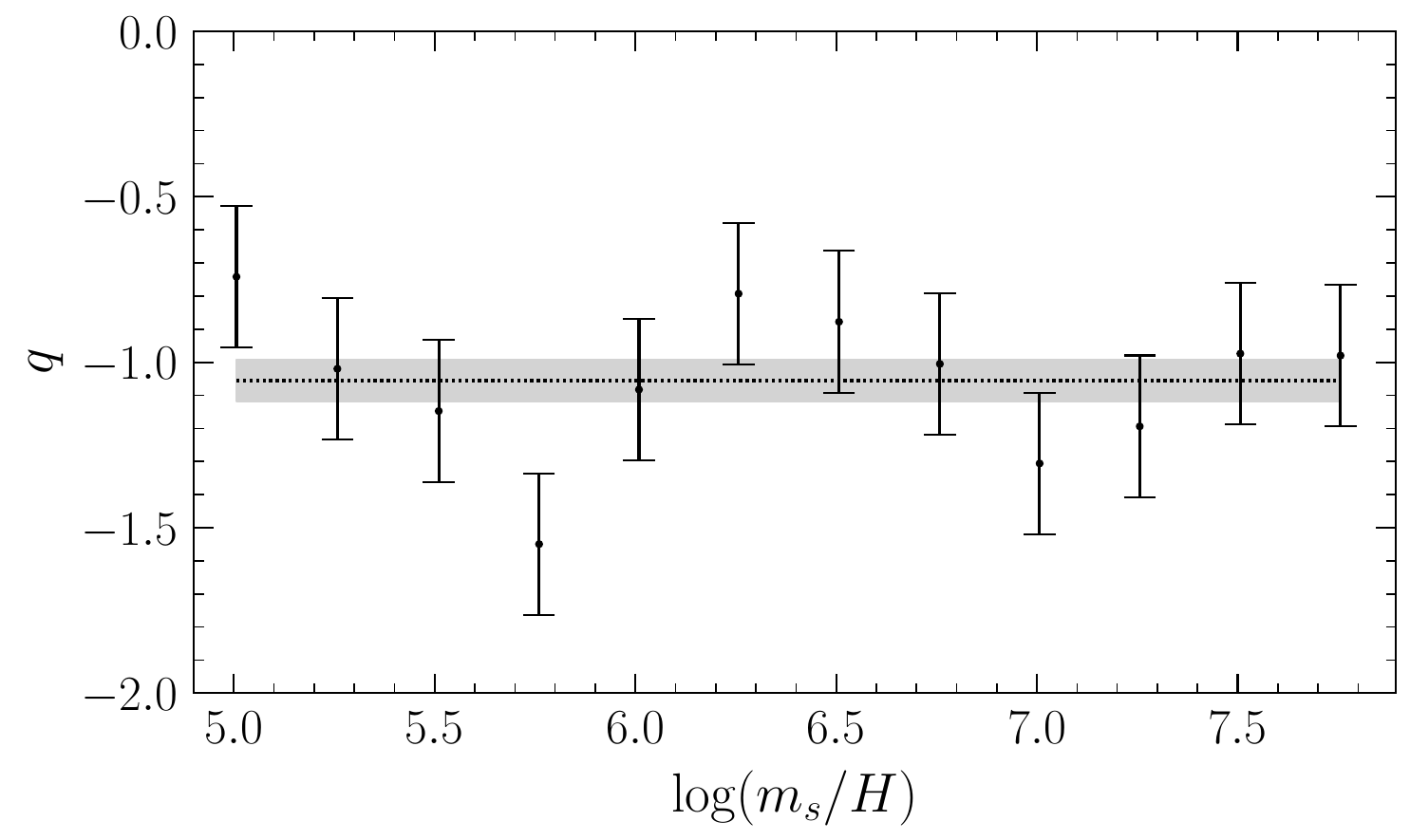}
\caption{Power law index $q$ from fits to the instantaneous axion emission spectra $F$ from a collapsing circular axion string at different $\log(m_s/H)$. We perform a linear fit to those indices (dotted line) with $1\sigma$ uncertainty indicated by the grey band. The fit yields $q=1.06 \pm 0.06$.}
\label{fig:SingleLoopIndex}
\end{center}
\end{figure}

We compute the instantaneous axion emission spectrum $F\propto (1/R^3)\frac{d}{dt}(R^3\partial \rho_a/\partial k$) with $\partial \rho_a/\partial k$ the time-dependent differential axion energy density spectrum. Numerically, we obtain the time-derivative via finite differences uniform in $\log(m_s/H)$ with $\Delta\log(m_s/H)\sim0.25$ (see, \textsl{e.g.}, \cite{Gorghetto:2018myk, Buschmann:2021sdq}). We fit a power-law model $F\sim 1/k^q$ to the instantaneous spectrum between $k\sim 4\pi/R$ and $k\sim m_s$. We do not, however, extend the fit below $k\sim 0.1m_s$ in case of large $R$. An example spectrum and the corresponding fit are presented in Fig.~\ref{fig:SingleLoopF}. The power-law index $q$ for each fit is shown in Fig.~\ref{fig:SingleLoopIndex}. We perform a linear fit to these indices, analogously to the fit of $\Gamma_s/(8H^3\xi\pi f_a^2)$, and find $q=1.06 \pm 0.06$. This result supports our claim of a $1/k$ scaling in this regime to within $\sim$6\% accuracy.

In Sec.~\ref{sec:general_expectations} we argue that sub-horizon-size string loops are distributed as $dn_\ell / d \ell \sim {1 \over \ell}$. On the other hand, above we find evidence that string loops of radius $R$ emit axions with instantaneous spectra $F \propto 1/ k$ between, roughly, $2 \pi / R$ and $m_s$. Let us now combine these two points to argue that the network as a whole should emit axions with instantaneous spectrum $F \propto 1/k$.

The $F \propto 1/k$ scaling of the full network can be understood analytically in the following way.
We can write the axion emission spectra by including the contributions from all loops of size $\ell$ as,
\es{eq:F_loop}{
F\propto \int d \ell {d n_\ell \over d \ell} F_\ell(k)
}
The factor $d n_\ell /d \ell$ governs the number density of loops as a function of their size $\ell$, and as argued and seen in simulations in~\cite{Buschmann:2021sdq}, we expect $d n_\ell / d \ell \propto 1/\ell$.
The spectral function $F_\ell(k)$ governs the instantaneous emission spectrum from a single loop with size $\ell$ and is normalized, without loss of generality, via $\int dk  k F_\ell(k)=1$.
As an illustration, we first focus on the case where each loop of size $\ell$ emits axions at a single frequency $k \sim 1/\ell$, {\it i.e.}, $F_\ell(k) = \delta(k-1/\ell)$.
We then arrive at,
\es{}{
F\propto \int {\D \ell \over \ell}  \delta\left(k-{1 \over \ell}\right) \propto {1 \over k} \,.
}
We now consider the more realistic scenario seen above where each loop with size $\ell$ exhibits a conformal emission spectrum with lower and upper frequency cut offs of $1/\ell$ and $m_s$, respectively.
In other words, for ${1 \over \ell} \leq k \leq m_s$
\es{}{
F_\ell(k) = {1 \over \log(m_s \ell)}{1\over k},
}
where the pre-factor is fixed by normalization condition.
Substituting this expression into~\eqref{eq:F_loop}, we find
\es{}{
F \propto \int_{1/k}^{1/m_s} {d \ell \over \ell} {1 \over \log(m_s \ell)}{1\over k} \propto {\log(\log(m_s/k))\over k}.
}
Therefore, we still expect to have a conformal spectra, as seen in the simulations above, with small doubly-logarithmic corrections.

\bibliography{Bibliography}

\end{document}